\documentclass{article}
\usepackage[utf8]{inputenc}
\usepackage{import}

\usepackage[left=3cm, right=3cm, top=3cm, bottom=3cm]{geometry}

\usepackage{graphicx}
\usepackage{hyperref}
\usepackage{subfig}

\title{The ISTI Rapid Response on Exploring Cloud Computing 2018}
\author{Editors: Carleton Coffrin, James Arnold, Stephan Eidenbenz}
\date{August 2018}

\begin{document}

\maketitle

\setcounter{tocdepth}{2}
\tableofcontents

\clearpage
\section{Background}


The Information Science \& Technology Institute (ISTI), with partnering entities at Los Alamos National Laboratory (LANL), regularly organizes and executes Rapid Response efforts in mission-relevant growth areas within information science. This report describes 18 projects conducted at LANL in response to a Rapid Response Familiarization call titled ``Exploring Cloud Computing 2018.'' 

These projects focused on demonstrations for leveraging Cloud Computing for scientific computation. Demonstrations ranged from deploying LANL-developed software in the cloud environment to leveraging established cloud-based analytics work flows (e.g., pandas, CAFFE, PyTorch, TensorFlow, and SageMaker) for processing datasets of interest to LANL missions.  All of the demonstrations were conducted in Amazon Web Services (AWS), a leading cloud computing service provider.  Each project was given a budget of one thousand USD for AWS compute resources and was encouraged to leverage the breadth of features available in the AWS platform, such as,
\begin{itemize}
\item 
Infrastructure automation (e.g., cloud formation / auto-scaling)
\item 
Distributed Computing Abstractions (e.g., MapReduce / Hadoop / Spark / Kinesis)
\item
Serverless computation (e.g., Lambda / Container Services)
\item 
Databases (e.g., Relational Databases / DynamoDB / Neptune)
\item Leveraging the spot-instance market for affordable computation
\end{itemize}
Section \ref{sec:iaas} provides a brief overview of all of the AWS services that were used by the projects.

The bulk of this document is the experience reports from these 18 projects, each of which briefly introduces the computational task of interest and reports on the challenges and successes of using cloud computing to complete that task.  Sections \ref{sec:cc}--\ref{sec:deploy} organize the projects reports in to the following three topic areas,
\begin{itemize}
\item Cluster Computing -- These projects investigated how cloud computing services can be utilized to supplement or augment typical cluster computing applications (e.g., MPI and large batch processing tasks).
\item Machine Learning -- These projects explored how cloud computing services can be utilized for a variety of machine learning tasks.  Many of these projects leveraged AWS's built-in machine learning tools such as, SageMaker, Rekognition, Polly, and Lex.
\item Software Deployment -- These projects tested the viability of using cloud computing services for delivering LANL capabilities externally and for archiving data.
\end{itemize}
By and large, the projects were successful and collectively these experience reports demonstrate that AWS, and cloud computing in general, can be a valuable computational resource for scientific computation at LANL.

\section{Computational Infrastructure as a Service}
\label{sec:iaas}

The terminology around {\em the cloud} is overloaded and notoriously ambiguous.  In the context of this activity, the term {\em cloud computing} is used to refer to established web-based services such as, Amazon Web Services, Google Cloud, and Microsoft Azure, which provide computational infrastructure as a service.  Specifically, these web services provide on-demand short-term rental of computational tools such as compute servers, data storage, networking, and communications.  The cloud computing model has become a standard practice in the computing industry because it allows companies to rapidly explore a wide variety of computational architectures and quickly scale those architectures with a pay-as-you-go model, which does not require large capital investments in computing infrastructure. 

In the interest of simplicity and convenience, this activity standardizes around AWS, which is one of the oldest and most well established cloud computing providers.  A wide variety of services are available on AWS.  To provide background for the reports presented in Sections \ref{sec:cc}--\ref{sec:deploy}, this section provides a brief introduction to the AWS services referenced in those reports.  A more detailed description of these services can be found at \url{https://aws.amazon.com/}.  The services in this section are presented by their AWS service category (i.e., Computation, Storage, Databases, etc.).

\subsection{Computation}

\paragraph{Elastic Compute Generation 2 (EC2):}
EC2 is the core computational resource. It allows users to rent virtual machines and dedicated hosts on a per-second basis.  EC2 features over 70 different computing configurations providing a variety of processors, memory, local storage, and hardware accessories such as GPUs. 

\paragraph{Lambda:}
Lambda is a serverless computation framework that allows users to execute fast code snippets without the need to start up a dedicated server (i.e., an EC2 instance).  Lambda is ideal for cases where many small state-less computations need to be applied a large amount of data.  Lambda currently supports code snippets for Python, Java, Node.js, .Net, and Go.

\paragraph{Elastic Container Service (ECS):}
ECS is a serverless computation framework that allows users to run Docker containers without the need to start up a dedicated server (i.e., an EC2 instance).  In contrast to Lambda, ECS is ideal for cases where long or persistent computations are required.  Building on the Docker virtualization layer allows ECS to support nearly any software dependencies.

\paragraph{Lightsail:}
Lightsail provides a lightweight and simplified version of EC2, with special features tailored to hosting a standalone web server.  The minimal learning curve of Lightsail makes it ideal for beginner AWS users to become familiar with provisioning and managing virtual machines in the cloud.

\subsection{Storage}

\paragraph{Simple Storage Service (S3):}
S3 is the core persistent data storage service on AWS. It provides users a seemingly infinite and globally accessible data store that balances performance, scalability, and price.  At the top level S3 organizes data into {\em buckets}.  Each bucket behaves as a key value store, where the key is a character string and the value is the data file.

\paragraph{Elastic Block Storage (EBS):}
EBS is a flexible storage device that is mounted directly to an EC2 instance's file system, similar to a commodity hard drive.  In comparison to S3, EBS provides much higher performance at a slightly higher price and is only available within the AWS region where it was created.

\subsection{Databases}

\paragraph{DynamoDB:}
DynamoDB is a serverless NoSQL key-value data store with seemingly infinite capacity and scalablity. Similar to S3, one master DynamoDB service is shared across all AWS regions and has a latency typically below ten milliseconds.

\paragraph{ElastiCache:}
ElastiCache provides fully managed, EC2-hosted, key-value data store (e.g., Redis and Memcached).  These in-memory data stores provide sub-millisecond latency and are ideal for building responsive real-time applications.

\subsection{Networking and Content Delivery}

\paragraph{CloudFront:}
CloudFront provides a fast and secure content delivery service for web-hosting.  CloudFront simplifies the process of delivering content with low latency and high bandwidth across the globe and provides basic threat mitigation tools, for example to protect the web service from DDoS attacks.

\paragraph{Route 53:}
Route 53 is a reliable and scaleable DNS service that makes it easy to route users to web applications hosted at specific IP addresses.

\subsection{Security, Identity, and Compliance}

\paragraph{Identity and Access Management (AMI):}
AMI is AWS's user account and permission management tool.  It allows an AWS account administrator to build user groups and assign fine-grained permissions to those groups.  AMI can limit a users access to specific AWS services and to the data stored in AWS.

\paragraph{Cognito:}
Cognito is a user authentication service for web applications.  It streamlines the process of adding robust sign-up, sign-in, and authentication capability to a web service.

\subsection{Analytics}

\paragraph{Elastic MapReduce (EMR):}
EMR is a managed cluster service that makes it easy to run distributed computing frameworks such as Hadoop, Spark, and Presto.  EMR automatically builds a cluster of EC2 instances, deploys a distributed file system, starts the desired computation and uses S3 to archive the results.

\subsection{Media Services}

\paragraph{Elemental MediaConvert:}
Elemental MediaConvert provides simple file-based video conversion that is scalable.  Media conversion is most often used to take a high resolution video file and make variants that are better suited for different devices, such as smartphones, tablets, and desktop computers.

\paragraph{Elastic Transcoder:}
Elastic Transcoder is an older service providing media conversion functionalities similar to Elemental MediaConvert.  The newer Elemental MediaConvert service is recommended for developing new media workflows in AWS.

\subsection{Augmented Reality and Virtual Reality}

\paragraph{Sumerian:}
Sumerian provides a simple web interface for designing and deploying interactive 3D environments.  Users upload content (e.g., 3D assets, audio, and text files) and build scenes in Sumerian's web interface.  When complete, users can publish their content, and Sumerian will deliver it as a web-based application that is accessible from a variety of devices including desktop computers, mobile devices, and virtual reality headsets (e.g., Oculus Rift and HTC Vive).

\subsection{Management Tools}

\paragraph{CloudFormation:}
CloudFormation provides simple text-based description of an AWS cloud configuration.  A CloudFormation script can be used to automatically provision all of the infrastructure required for any cloud-based application.  CloudFormation makes it easy to replicate a cloud deployment across multiple runs or multiple AWS accounts.

\subsection{Machine Learning}

\paragraph{SageMaker:}
SageMaker is a managed service that streamlines the process of building machine learning workflows.  It provides a flexible framework for designing a machine learning model, training that model on vast mounts of data, and deploying a trained model.

\paragraph{DeepLens:}
DeepLens is a self-contained smart video camera for deploying image-based machine learning models built with SageMaker.  After a video-based machine learning model is designed and trained with SageMaker, the trained model is copied to the DeepLens camera and can be used for real-time computer vision applications in the field. 

\paragraph{Polly:}
Polly is a text-to-speech service that uses deep learning models to synthesize speech that is similar to a human voice.  Polly supports multiple languages and voice styles (e.g., gender and accents).  

\paragraph{Rekognition:}
Rekognition provides image and video analysis tools that streamline the process of image classification, text extraction, sentiment analysis, and facial recognition.

\paragraph{Lex:}
Lex provides tools for building human conversation applications similar to Alexa, Google Home, and Siri.  Lex streamlines the process of speech recognition, speech-to-text conversion, and natural language processing to understand the intent of text.

\subsection{Internet of Things}

\paragraph{Greengrass:}
Greengrass provides a framework for orchestrating computations across the Internet of Things (IoT).  It coordinates local compute, data syncing, and software deployment, effectively extending the AWS cloud abstraction into IoT devices.

\clearpage
\section{Cluster Computing}
\label{sec:cc}

\subsection{Mapping Industrial Seismic Noise in the Continental United States}

\subsubsection{Contributors}

\begin{itemize}
    \item Jonathan MacCarthy, EES-17, jkmacc@lanl.gov
    \item Omar Marcillo, EES-17, omarcillo@lanl.gov
\end{itemize}

\subsubsection{Motivation}

 \begin{figure}[t]
     \begin{center}
     \includegraphics[width=10.25cm]{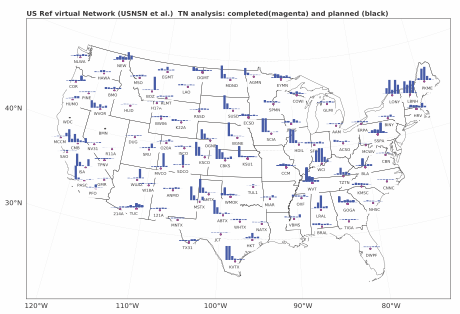}
     \end{center}
     \caption{Harmonic Tonal Noise detections for 60 USNSN channels for the year 2017. The magenta dots show the stations and the bar plots above the histograms of detections with 0.5 frequency bins from 0.5 to 4 Hz with 0.5 bin frequency width. This calculation took 60 hours with a traditional desktop implementation.}
     \label{fig:NSN}
 \end{figure}
 
 \begin{figure}[th!]
     \begin{center}
     \includegraphics[width=8.75cm]{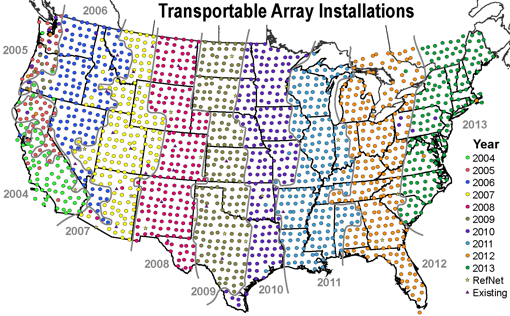}
     \end{center}
     \caption{The color dots show the stations of the US TA totaling over 5,000 channel-years of data.  An HTN calculation on one channel of these data would take approximately 70 days to complete with a traditional desktop implementation.}
     \label{fig:TA}
 \end{figure}

Industrial activity has a significant impact in the local and regional seismic wavefield, as the operation of large machines generates mechanical energy that potentially propagates into the air as acoustic and infrasonic waves and/or into the solid earth as seismic waves. The repetitive nature of industrial processes generates persistent elastic energy that can be observed at distances of several 10’s of kilometers, and its signature is often characterized by both broadband and tonal noise \cite{Marcillo:2018}. Using a sparse seismic network (US Seismic National Network, USNSN) we identified areas in the contiguous US with a high occurrence of harmonic tonal noise (HTN, Figure \ref{fig:NSN}).  In this project, we use AWS to expand processing to include a denser seismic and acoustic sensor network (USArray Transportable Array, TA) to increase the resolution of our initial coarse mapping (Figure \ref{fig:TA}). The TA recorded continuous data in the contiguous US between 2009 and 2017 using broadband sensors with an inter-sensor distance of around 75 km.

The data recorded on the US and TA networks, as well as many other networks, is stored at the Data Management Center of the Incorporated Research Institutes for Seismology (IRIS DMC), and served over HTTP web services through a REST-like API.  This makes it possible to query, acquire, and process the data programmatically. Using a traditional implementation, a single 4-core Mac desktop using threads can process approximately one station-year per clock hour.  At this rate, it would take nearly 70 days to process a single channel (e.g., vertical component of seismic motion) of data across the whole contiguous TA network for one year.  We use an AWS EC2 cluster to parallelize and horizontally scale this calculation to reduce the time-to-result.

\subsubsection{Solution Approach}

\begin{figure}[t]
     \begin{center}
     \includegraphics[width=8.75cm]{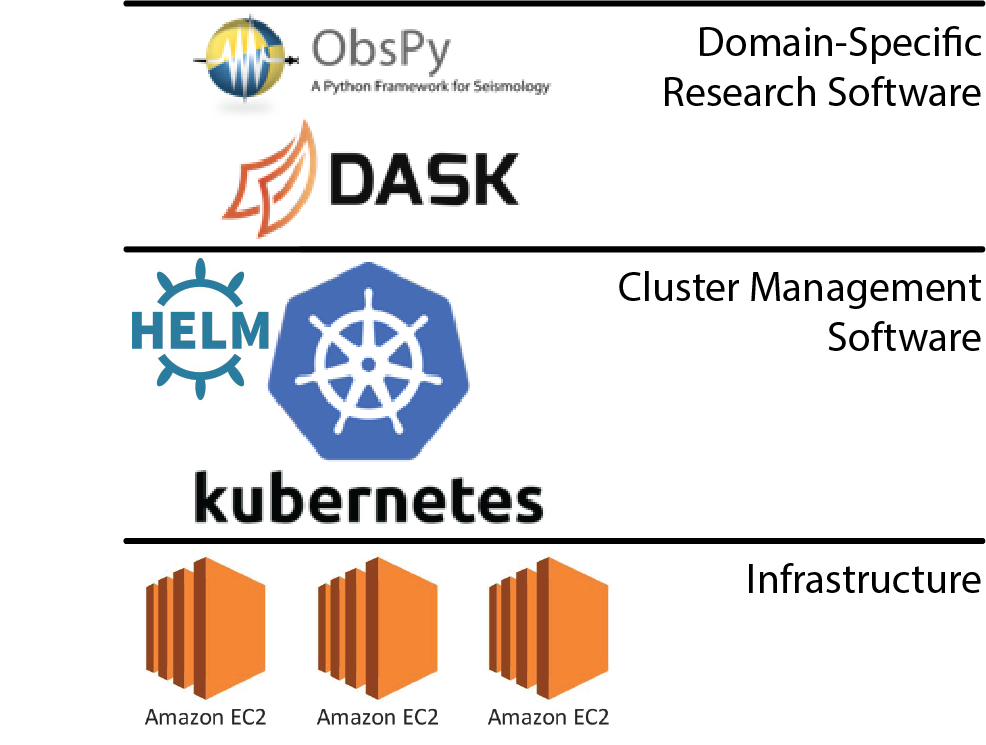}
     \end{center}
     \caption{The three layers of technologies employed in this application.}
     \label{fig:infrastructure}
 \end{figure}

Our solution consists of three layers of technologies: cluster hardware infrastructure, cluster management software, and domain-specific research software (Figure \ref{fig:infrastructure}).  The lowest ``infrastructure" layer is provided by an Amazon EC2 cluster.  We used 10--100 \texttt{t2.large} general computing Debian Linux nodes, and a \texttt{t2.medium} head node for cluster management via Kubernetes.  The cluster was initialized by creating a small Linux EC2 instance with proper IAM permissions to create more instances.  On this administrative instance, command-line tools for managing the cluster were installed.

Kubernetes is cluster management and scheduling software that is available on clusters in all major cloud providers.  Among many other features, it facilitates creation, scaling, monitoring, and resiliency of cluster operations.  Building upon Kubernetes, Helm is an application manager for cloud-native applications on Kubernetes-managed clusters.  Helm coordinates parameterizations between Docker containers installed on different nodes in the Kubernetes cluster, such that they operate in a coordinated manner as the underlying cluster may grow larger or smaller according to computing needs.  The specific Helm application we installed underlies the final domain-specific technology layer.

Dask is a framework that facilitates parallelization of user-defined Python functions (UDFs). This is an important component of our solution, as it allows us to quickly scale-out successful algorithms written by researchers who are not experts in cluster computing (e.g., the author of this report).  The specific UDF employed in our application was a simple function that downloaded a chunk of waveform data from the IRIS DMC using its web services API and ran an HTN detector over the waveform, using ObsPy, the de-facto seismological processing library in Python.  The HTN detector function was embedded into a computation graph representing all desired HTN downloads and calculations for the TA network, split into into six-hour chunks.  Dask executes the computational graph in the most efficient way possible, using all available worker nodes in the cluster.

\subsubsection{Results}

\begin{figure}[t!]
    \begin{center}
    \includegraphics[width=8.75cm]{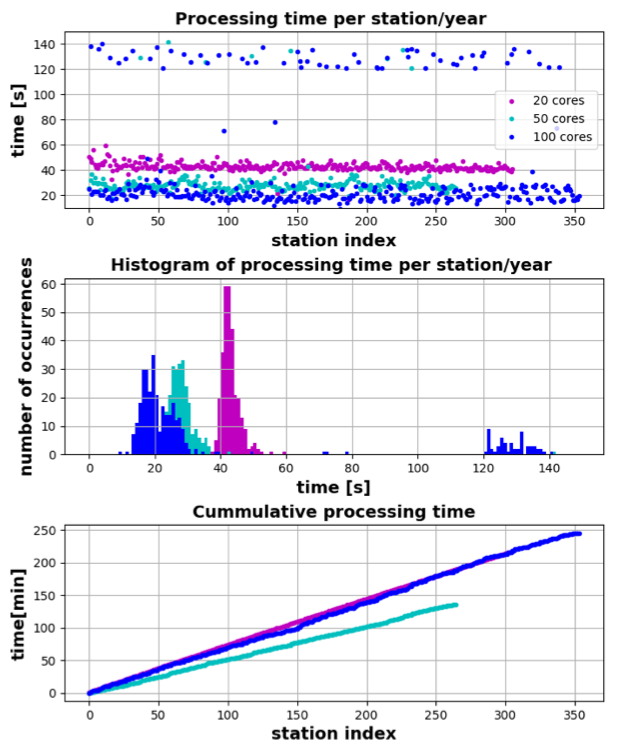}
    \end{center}
    \caption{Per-station timings for cluster sizes of 20, 50, and 100 nodes. Top: individual station processing times.  Middle: histograms of per-station processing time. Bottom: cumulative processing time with station index (a proxy for station name).}
    \label{fig:timings}
\end{figure}

The solution described above was able to successfully scale-out the HTN calculation, decreasing the time it takes to run an HTN detector over a single station for one year from one hour on a standard desktop to under 30 seconds, a 120x speedup.  The HTN detector was run over one year of all TA stations in the contiguous United States in approximately 14 hours, \textit{with no changes in the researcher's HTN detection code}.  Figure \ref{fig:timings} shows per-station and cumulative processing times with three cluster sizes.  For a cluster consisting of 20 compute nodes, we find a median one-year station processing time of approximately 45 seconds (top, middle).  Increasing the node-count to 50 reduces the median one-year station processing time to approximately 30 seconds, a sub-linear scaling.  

Doubling the node count to 100 also lowers the median one-year station processing time to approximately 20 seconds but  results in a large number of 130-second processing times (Figure \ref{fig:timings}, top and middle).  These are delays generated at the IRIS DMC and may be due to the increased request load put on their data services system.  These delays reduce the overall processing efficiency of the 100-node cluster to approximately the same efficiency as that of the 20-node cluster, highlighting the importance of operating within the tolerances of the IRIS DMC capacity to serve data.

\subsubsection{Conclusion}

We feel that protoyping the use of AWS to request and process seismic data in a streaming manner was an extremely worthwhile investment of time and funds.  A summary of the positive results includes the following:

\begin{itemize}
    \item A two order-of-magnitude increase in data processing speed, compared to a traditional desktop implementation.
    \item Rapid deployment of a researcher-developed UDF on a computational cluster with \textit{no changes to the researcher's calculation.}
    \item Utilization of the community-supported IRIS DMC archive in a streaming manner with no need to store, manage, and access downloaded data on-premises.
\end{itemize}

Future work for this application may include:

\begin{itemize}
    \item Make adjustments to our processing algorithm to increase request throughput. 
    \item Make adjustments to cluster topology to better balance per-node request rate.
    \item Use Kubernetes autoscaling configurations to reduce the need to manually manage cluster size and reduce cost.
\end{itemize}
\subsection{Framework for On-Demand Cloud Computation of Theoretical Models of Materials over Discretized Phase Volumes}

\subsubsection{Contributors}

\begin{itemize}
    \item David M. Fobes, A-1, dfobes@lanl.gov
\end{itemize}

\subsubsection{Motivation}

Computationally demanding theoretical calculations of properties of materials, such as electronic or magnetic band structures, used in the analysis of experimental data is an important aspect of modern condensed matter physics. This is particularly relevant when using techniques such as inelastic neutron scattering \cite{doi:10.1081/ASR-100100843} or angle-resolved photoemission spectroscopy (ARPES), where the volume of data measured per experiment continues to increase substantially every year \cite{hill_mulholland_persson_seshadri_wolverton_meredig_2016}. It is therefore not uncommon for experimentalists to now require computing resources for one-off calculations for data analysis. 

An example relevant to LANL is the computation of the magnetic excitation spectra \cite{0022-3727-50-29-293002} in the topologically-protected magnetic phase of MnSi, the so-called skyrmion phase \cite{Muhlbauer915}, which is believed to hold great potential in magnetic memory applications \cite{Fert2017}. The theory behind this phase is based on the original theoretical work performed by Tony Skyrme here at LANL in 1962 \cite{SKYRME1962556} and has been the subject of several recent LDRD-funded projects. I previously developed a program to calculate the magnetic band structure within the skyrmion phase, which was used to analyze inelastic neutron scattering data. 

The calculation was performed over a grid of points filling a partial cylindrical volume taking $\sim$25 days on a dedicated workstation ($\sim$\$10,000), during which the workstation was otherwise unusable. This calculation represents the current typical situation for many experimental materials scientists and represents the primary motivation for this project.

\begin{figure}[h]
\begin{center}
	\includegraphics[width=0.5\textwidth]{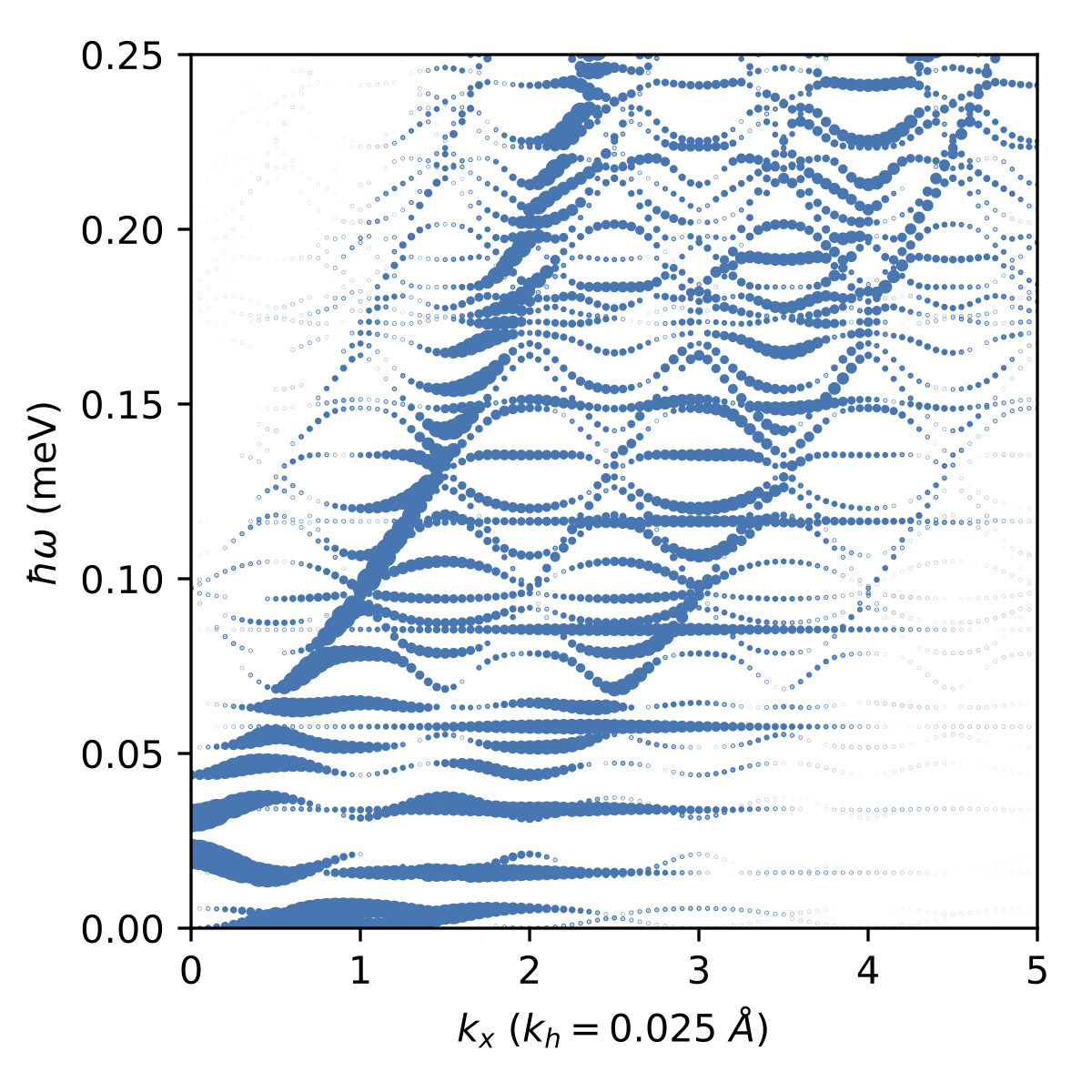}
\end{center}
\caption{Magnetic excitation spectra of the skyrmion lattice in MnSi calculated along the (100) crystallographic direction, in units of magnetic Brillouin zones, or scattering wave vector $k_h$ ($\sim0.025\AA$).}
\label{mnsi}
\end{figure}

\subsubsection{Goals and Approach}

The primary goals of this project were to (i) explore the viability and cost effectiveness of performing materials science computations in the cloud and (ii) create an easy-to-use framework to this end. This problem, and the example I used to test, were chosen because of their suitability to intuitive extension to a cloud environment. These types of problems are characterized by their discrete nature, i.e. the calculations are performed over volumes in phase space at discrete points, and the calculations at any given point do not depend on the results from calculations at any other point. This means that calculations can be easily performed completely independently in an efficient manner, making the mapping to multiple cores on a single machine or multiple instances/nodes in the cloud fairly trivial in concept.

To achieve these goals I chose a small subset of points (see Figure \ref{mnsi}) that could be calculated relatively quickly but would be large enough to test the use of groups of instances of multiple types. 

\subsubsection{Results}

The following details the framework I designed to solve this problem and makes a comparison to the original calculation that inspired this project as a comment on the overall viability and cost-effectiveness of this approach.

\paragraph{Framework Design and Implementation}
\begin{figure}[t]
\begin{center}
\includegraphics[width=0.85\textwidth]{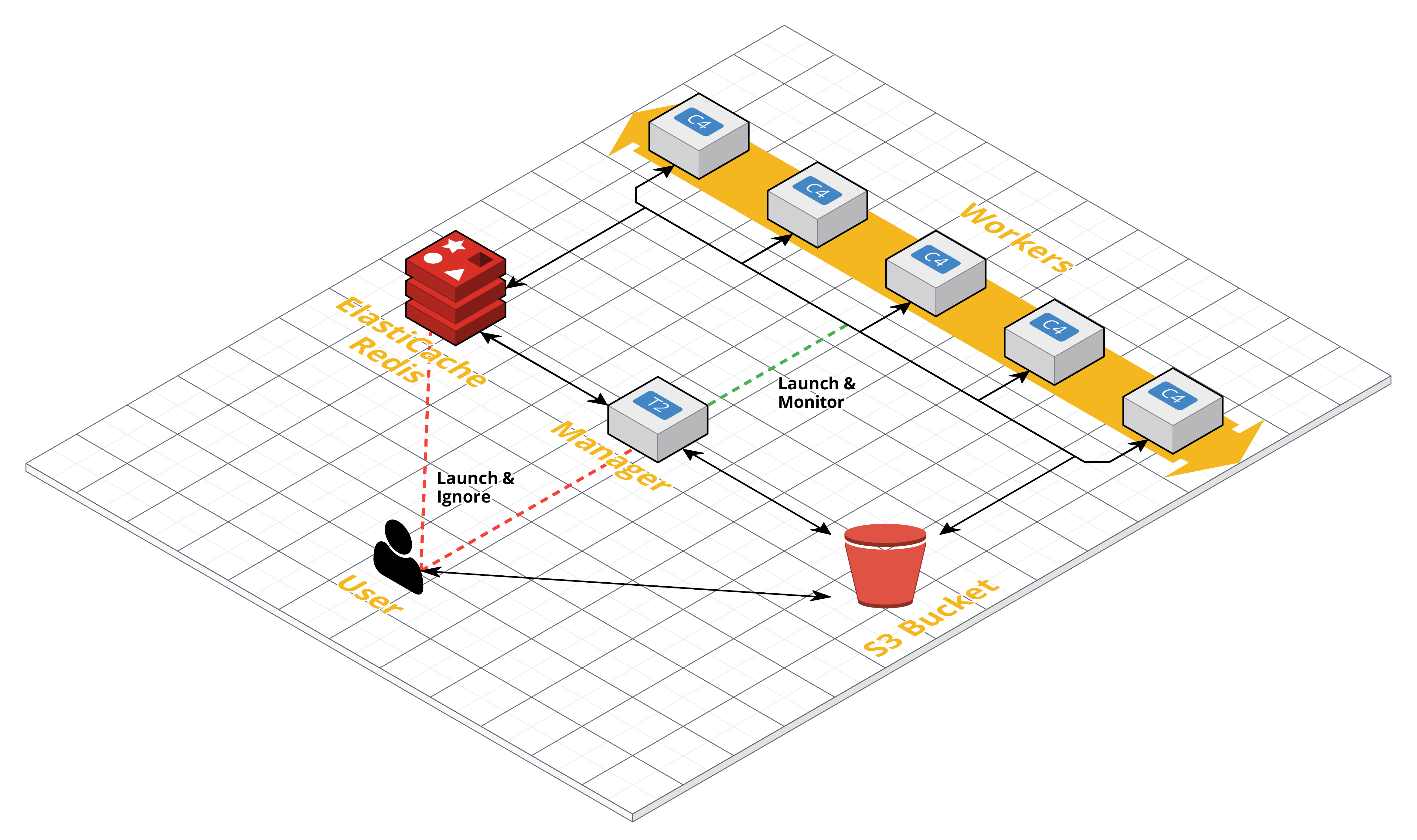}
\end{center}
\caption{Materials cloud compute's AWS resource utilization design.}
\label{fig:aws-framework}
\end{figure}
The framework designed for this project, named Materials Cloud Compute (MCC), utilizes the following AWS services: EC2 (computation), S3 (cloud storage), and ElastiCache (Redis database). Internally, it relies on Python 3.6, with only a limited number of external Python packages.

The audience for this software is only expected to have some experience with Python, mainly for data analysis type tasks, but little to no experience with cloud computing or even parallel computing. With this in mind, I identified the following core requirements for MCC: (i) automated creation of AWS resources, (ii) automated performance of the calculation with no user intervention, (iii) easy uploading and downloading of required files, and (iv) simple price and time estimation tools.

These core requirements resulted in the framework design shown in Figure \ref{fig:aws-framework}, for which the basic workflow is as follows. The User creates the templates for the EC2 instances, the S3 bucket, and the ElastiCache Redis via a small number of functions, plugs in two scripts, i.e. (a) a point generator and (b) a file combiner, and finally uploads the main program and the required templates generated by MCC to S3 with a function. After these preparatory actions, the User launches a Manager instance and waits for the calculation to complete, after which they can download the resulting data file(s).

\begin{figure}
\centering
\raisebox{-0.5\height}{\includegraphics[width=0.3\textwidth]{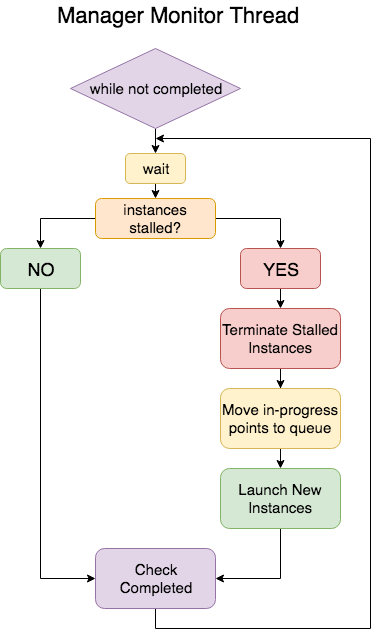}}
\hspace{0.2in}
\raisebox{-0.5\height}{\includegraphics[width=0.3\textwidth]{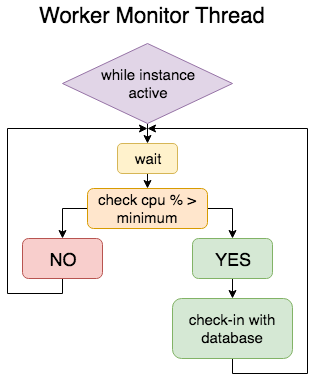}}
\caption{Flowcharts of Manager (left) and Worker (right) monitoring threads.}
\label{monitor-flowchart}
\end{figure}

The Manager instance launches Worker instances to actually perform the calculation, and monitors those workers for stalls (see Fig.~\ref{monitor-flowchart}). The workers poll the Redis server for points, taking ones that are still available (atomically) and running the calculation. They also ``check-in'' with the database, confirming they are not stalled, and eventually self-terminate when no more points remain. When all workers have terminated, the managing instance combines the partial data files, uploads the result to S3, and also self-terminates. The automation that makes MCC easy-to-use depends on the AWS API's EC2 launch parameter `UserData,' a script that will run at the final steps of the launch sequence.

\paragraph{Lessons Learned}
The biggest barrier to implementing the automation behind MCC was determining how to run code on each instance after they are launched. Two solutions are typical, (a) UserData scripts or (b) ssh via paramiko. The first is more attractive because direct communication between instances isn't required. However, UserData by default can only be run at the first launch of an instance, but because MCC uses custom EC2 AMIs, each instance image has already been run once. Fortunately, by altering ``/etc/cloud/cloud.cfg'' to include ``[scripts-user, always],'' UserData scripts can be run upon any launch.

Hyperthreading was a significant issue during testing. With hyperthreading enabled, VCPUs do not represent ``physical'' cores,  but instead 2 VCPUs $\approx$ 1 CPU. In some cases, especially for instances with VCPUs $\gg 2$, I discovered that instances would be more likely to stall if num.~processes = num.~VCPUs. In those cases it was more economical to only run half as many processes.

Proper timing was also a major challenge. EC2 instances do not have uniform initialization times, e.g. for custom AMI creation. While the AWS API offers most information, in some cases a custom solution was necessary. For example, when creating a custom AMI with a UserData script, the only way to determine if UserData completed was to add a Tag to the instance via the AWS CLI at the end of the script.

\paragraph{Comparison to Original Calculation}
The original calculation was performed on a Mac Pro (late 2013) with a 2.7 GHz 12-Core Intel Xeon E5 processor and 64 GB 1866 MHz DDR3 memory, costing $\sim$\$10,000 (now available for $\sim$\$6,500). This calculation consisted of a grid of 9090  points filling a cylindrical-wedge phase volume. The total time of that calculation was approximately 610 hours (25.5 days).

In order to quantify MCC's performance, I performed the same calculation on seven t3.medium EC2 instances (14 VCPUs, Hyperthreading enabled). The only change to the original calculation's code was to remove any Python multiprocessing, in order to run strictly in a single-threaded mode; multithreading was achieved via `subprocess.Popen.' On AWS using MCC, this calculation had a duration of 90.84 hours (3.8 days), with a total of five stalled instances that were automatically relaunched within 2--3 minutes of detection; therefore running on 14 VCPUs at nearly 100\% throughout the duration of the calculation. This is also in comparison to the estimation of a total duration of 112.5 hours (4.6 days) made by extrapolating the 101 point test case out to 9090 points, which is approximately 22 hours longer than the actual calculation lasted. This calculation cost $\sim$\$27.72 ((7 t3.medium @ \$0.0416 per hour + 1 t2.micro @ \$0.0116 per hour) $\times$ 90.84 hours + 5 stalled t3.medium @ \$0.0416 per hour $\times$ 1 hour). Pricing for S3 and ElastiCache are negligible, coming to only \$0.07 for S3 services, and \$0.00 for ElastiCache services, for the entire period of the project not just the duration of this calculation.

Therefore, the original calculation time was reduced by one-sixth by utilizing cloud computing, and the cost was only 0.2\% of the paid price of the original computing resource (0.4\% of today's price for the same computing resource).

\subsubsection{Conclusion}

This project has achieved great success in illustrating the viability and cost-effectiveness of performing materials-science-type calculations in the cloud. I was able to demonstrate a substantial reduction in both the computational time and up-front costs as compared to an identical previously performed calculation. I was also able to fully design and implement a feature-rich computational framework to autonomously perform these calculations. I still plan for this software to be released as a publicly available open source package in the near future via the appropriate LANL software-release process to aid other materials scientists to quickly and cheaply perform complex calculations in the cloud.
\subsection{DOE Climate Models for Risk Assessment on a Cloud Platform}

\subsubsection{Contributors}

\begin{itemize}
    \item Donatella Pasqualini, A-1, dmp@lanl.gov
    \item Derek Aberle, A-4, daberle@lanl.gov
    \item James Wernicke, A-4, wernicke@lanl.gov
    \item Phillip Wolfram, T-3, pwolfram@lanl.gov
    
\end{itemize}

\subsubsection{Motivation}

\begin{figure}[h]
     \begin{center}
      \includegraphics[width=13cm]{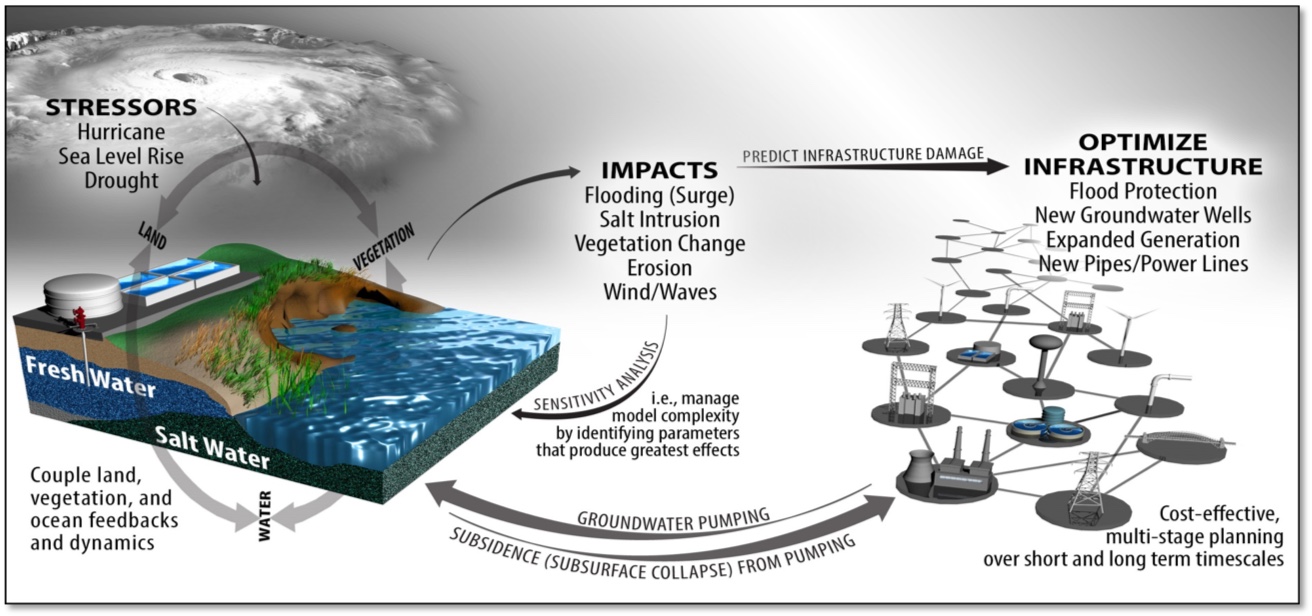} 
     \end{center}
     \caption{Coupled components of a coastal risk and adaptation modeling analysis.}
     \label{fig1}
 \end{figure}

One challenge in climate risk assessment modeling is the need to integrate applications from different disciplines, climate, hydrology, and vegetation models with critical infrastructure systems such as electric power and potable water systems.  The applications coupled within a risk framework may be developed by different institutions, usually run on different environments, and may have different computational requirements.   Science-based risk modeling analyses that guide climate adaptation may necessitate the integration of  high performance computing (HPC) applications (e.g., global climate models) with models that usually need less processing power (e.g., infrastructure networks and fragility models).  The diagram in Figure \ref{fig1} depicts the complex coupled components  included in  a coastal  risk and adaptation modeling analysis of climate change and extreme events.  
Another key aspect of these risk analyses is their probabilistic nature. Climate impact uncertainties can be quantified only through a probabilistic risk analysis that involves running the integrated assessment risk analysis model several times using a large ensemble of different input values. This process requires access to a large computing resource capacity. 

The natural solution to these computational challenges is the deployment of a risk analysis workflow on an HPC system.  An HPC framework may be employed to run single codes within the workflow, the ones that run for a large period of time, and/or to run a large number of integrated workflow instances to estimate uncertainties.   
In this effort we aim to explore the use of a cloud environment for climate risk and adaptation  analysis as an alternative to a more traditional HPC infrastructure. The cloud platform availability to offer users to automatically scale their computing and storage resources on demand, to easily access and configure these resources make the cloud platform a valuable  candidate to run probabilistic climate impacts workflows.  In addition, for decision support applications the cloud can be also beneficial when the models are developed and owned by different institutions providing a neutral infrastructure for integration that can be shared across all developers and offering tools to easily manage the integration, e.g., models synchronization.

In this effort we explored the applicability of cloud computing, investigating the difference between running an HPC  Department of Energy (DOE) global ocean model, MPAS-O \cite{ref1}, on an IC HPC system and running it on the cloud. MPAS-O coupled with critical infrastructure models (e.g., electric power and water distributions network models)  may be used to support strategies for adaptation to climate change and extreme weather events.

\subsubsection{Solution Approach}
As the first step of a larger feasibility study, this work focuses on evaluating the tradeoff of running on the cloud and on an HPC system, comparing the time of running a set of  MPAS-O simulations on the AWS cloud and running it on an IC cluster, {{\it{Grizzly}}}. In this preliminary work we  looked at the runtime and  Input/Output (I/O) write time. 

\begin{figure}[h]
     \begin{center}
      \includegraphics[width=8.75cm]{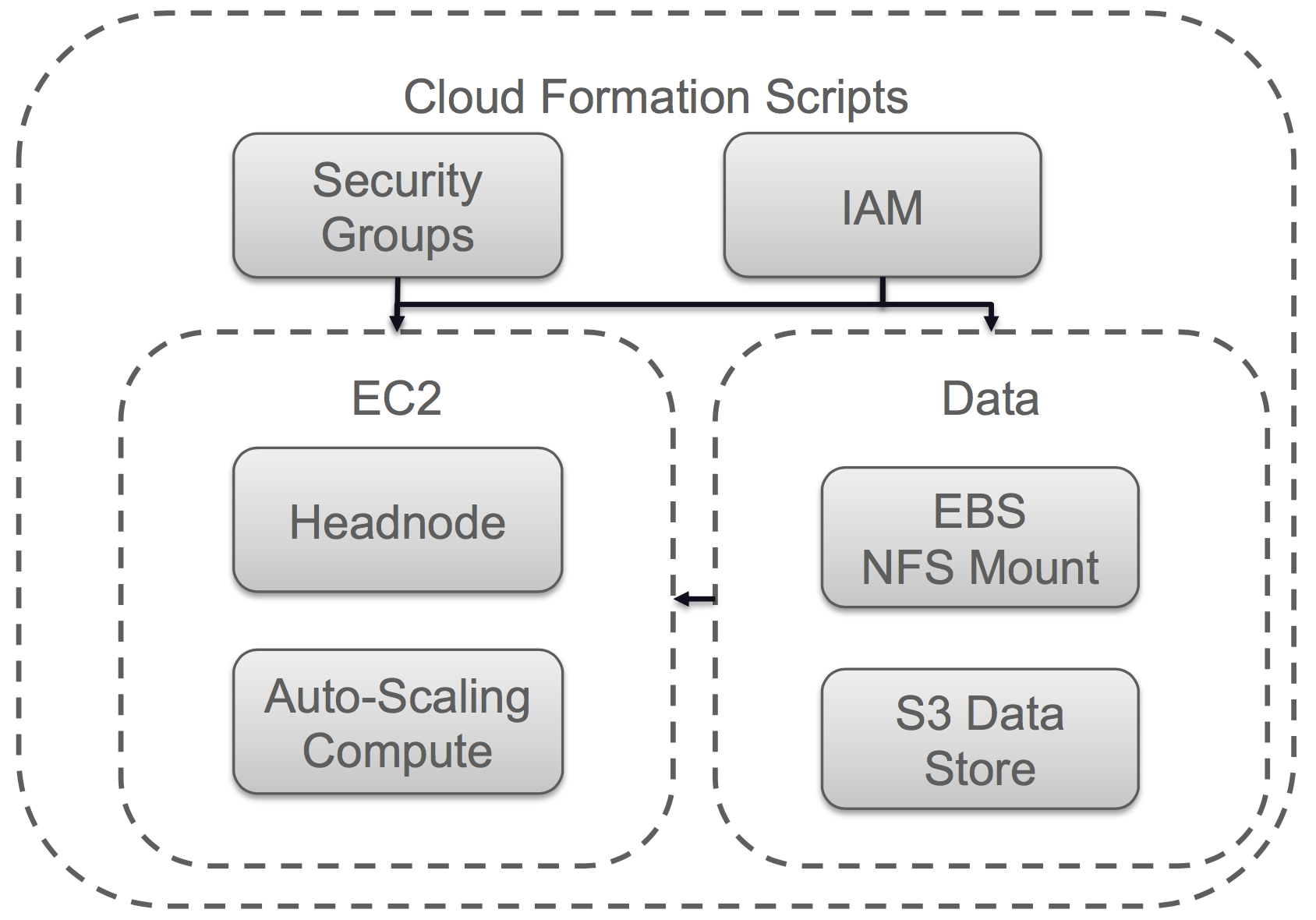} 
     \end{center}
     \caption{A simple HPC architecture within AWS.}
     \label{fig:hpcONaws}
 \end{figure}

In order to deploy a cluster on AWS to run MPAS-O in parallel: 
\begin{enumerate}
\item We locally generated an image of MPAS-O and tested it on a local virtual cluster.
We used an Amazon Linux VM  to build our test environment \cite{ref2} in  a local system. This  enabled us to install dependencies, configure the environment, and verify code execution without using EC2 resources.  We generated an MPAS-O image and locally built a virtual cluster where we tested running the image.  
We developed a set of scripts that automate the process to build the virtual cluster. The scripts take the instance and clone it to create a virtual cluster similar to deploying diskless compute nodes in a physical HPC environment. Those scripts, which helped automate setting up local testing on a virtual cluster environment, can be found on LANL's gitlab \cite{ref3}. 

\item We built an EC2 HPC cluster on AWS where we run our MPAS-O image in parallel.
To build a cluster on AWS, we employed the AWS CloudFormation tool, which can be used to write {\it{recipes}}  for cluster deployment, such as how to create and configure Security Groups, IAM roles, storage resources (i.e., EBS and S3), EC2 instances, and Auto Scaling rules. We defined a Security group that controls network access to EC2 instances and used IAM to define user and machine roles, which can be used to further define Security Groups. The EC2 cluster was formed by (1) creating a head node and compute templates  from which EC2 instances can be created, (2) defining Auto Scaling rules to scale up nodes when cluster CPU usage exceeded 80\% and scale down nodes when they were idle, and (3)  using EBS to create a volume snapshot from which to create virtual disks for new EC2 instances so that they would be configured out of the box for the cluster. We also used an EBS-backed SSD NFS share for hot storage during our HPC runs. Data, such as simulation results, can be stored long-term in S3, which provides its own access control mechanism. Figure \ref{fig:hpcONaws} depicts the HPC architecture within AWS.

Our AWS cluster consisted of 29 { \it{c5.9xlarge}} instances, each providing 36 vCPU cores, 72 GiB memory, EBS storage, and 10 Gbps interconnect \cite{ref4}. We set up a {\it{c5d.9xlarge}} instance as the head node and NFS mount point of home on the EBS storage. The compute mounted the home directory and we ran our simulations from the ec2-user account. 
\end{enumerate}
Once we  built the AWS cluster, we ran MPAS-O code for different scenarios and used a different number of cores on both the AWS cluster and the {{\it{Grizzly}}} cluster. To compare the performance of AWS with the HPC system, we recorded simulation runtime and I/O write time. 

\begin{figure}[h]
    \centering
    \includegraphics[width=14cm]{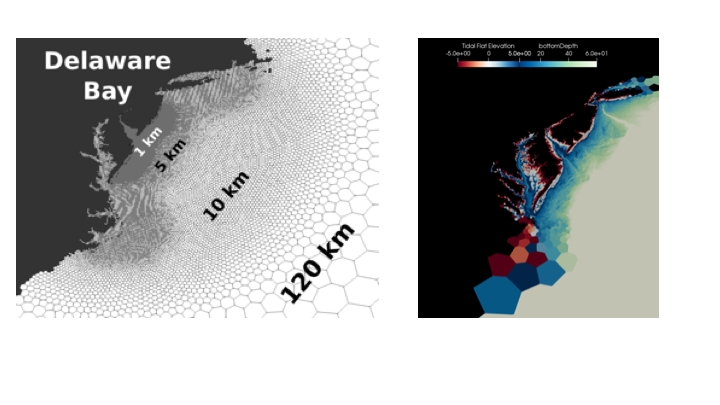} 
    \vspace{-1.1cm}
    \caption{Left, MPAS-O Input: MPAS-O global unstructured mesh zoomed around Delaware Bay, DE. Right, MPAS-O Output:  storm surge-driven coastal flooding.}
    \label{fig:mpasO_mesh}
 \end{figure}

MPAS-O, designed to simulate the ocean system at global scale, is used in this study to model the storm surge generated by a disturbance in the ocean. Using an unstructured global mesh as an input, the model outputs a time series of water depth values.  Figure \ref{fig:mpasO_mesh} shows an example of the MPAS-O unstructured mesh zoomed around Delaware Bay (figure on the left),  and an example of an MPAS-O water depth output---a snapshot in time (figure on the right). Higher is the resolution of the spatial mesh; longer is the time to compute a simulation. 

We considered two different scenarios using two different spatial resolution meshes: 
\begin{itemize}
    \item Scenario {\it{USDEQU300c3}} corresponds to a  global unstructured mesh with 300 km and 3 km as the largest and finest mesh, respectively, and 
    \item Scenario {\it{USDEQU300c0.5}}  uses a global unstructured mesh with 300 km and 0.5 km as the largest and finest mesh, respectively. 
\end{itemize}

In order to compare performance, we ran the two MPAS-O scenarios using a different number of cores:  1, 18, and 36 nodes for the low resolution scenario and 108, 216, and 432 nodes for the high resolution scenario. Unfortunately, due to the limited time available on the IC cluster and the limited scope of this project, we were only able to run the model on {{\it{Grizzly}}} for two configurations: low resolution model {\it{USDEQU300c3}}  with 36 cores and high resolution model {\it{USDEQU300c0.5} } with 432 cores. 

The largest hurdle in our deployment was getting access to the compute resources we wanted.  Our initial intention was to build a system with hundreds of nodes, however we immediately ran into a instance limit within AWS. Each virtual machine counts as a single instance and we were initially only allowed five instances. We were able to submit service tickets and increase the limit up to  30 instances within 24--36 hours. \footnote{It is worth mentioning, we used the Amazon Linux2 image for our operating system, which is a derivative of Red Hat Enterprise Linux. Two differences of note are that UID starts at 1000 instead of 500 and that you will likely need to increase the ulimit.}

\subsubsection{Results}

Table \ref{tab:results} summarizes the results of our preliminary study listing runtime and I/O write time for the simulations we performed. 

\begin{table}[h]
\caption{Simulations Used to Study AWS Feasibility for HPC Applications}
\begin{tabular}{l|l|l|l|l|l|l|l}
\label{tab:results}
                &               &            & AWS                & AWS    & {{\it{Grizzly}}}  & {{\it{Grizzly}}} & AWS \\ 
Scenario        & nCells & Cores & Runtime  & IO     & Runtime  & IO        & Cost  \\ 
                               &            &            &[sec]        & [sec] & [sec]           & [sec]             & [\$/hours] \\ \hline
USDEQU300cr3   & 12350  & 1     & 438.02          & 22.54            &                         &                          & \$3.26                \\
USDEQU300cr3   & 12350  & 18    & 56.43           & 20.4             &                         &                          &\$3.26                \\
USDEQU300cr3   & 12350  & 36    & 56.43           & 20.4             & 166                     & 124                      & \$3.26                \\
USDEQU300cr0.5 & 100171 & 108   & 279.7           & 65.59            &                         &                          & \$6.32                \\
USDEQU300cr0.5 & 100171 & 216   & 214.54          & 68.98            &                         &                          & \$10.91               \\
USDEQU300cr0.5 & 100171 & 432   & 254.32          & 80.83            & 345.57                  & 180.15                   & \$20.09              
\end{tabular}
\end{table}

Examining the runtime on AWS and the one on {{\it{Grizzly}}}, our experiment shows that AWS performs better.  For simulations with the scenario at low resolution (i.e., a mesh of 12,350 cells) using 36 cores, the AWS runtime was 56.43 sec compared to 166 sec obtained running {{\it{Grizzly}}}.  For simulations with the scenario at high resolution, (i.e., a mesh of 100,171 cells) using 432 cores, AWS is still faster: 254.32 sec versus the 345.57 sec of {{\it{Grizzly}}}.  The main reason for this difference is the I/O write time. The I/O write time of AWS is smaller than the  one we recorded for {{\it{Grizzly}}}: 20.4 sec versus 124 sec and 80.83 sec versus 180.15 for the low resolution and high resolution scenario, respectively. MPAS-O code writes a substantial amount of data and as such the storage available has a large effect on the total runtime of the software.  The fact that AWS is faster than {{\it{Grizzly}}} during the I/O process is primarily due to the faster storage AWS-EBS provides.  If we do not consider the I/O write time from the runtime, {{\it{Grizzly}}} has a slightly faster runtime. 

The rightmost column shows the cost per hour to operate the AWS system with the number of cores listed. The AWS cost per hour increases quickly, and the cost of transferring data is also very high. However, for relatively small systems such as the one we are using, the system is very inexpensive. To scale up, if we ran the head node and 29 compute nodes with a total of 1,044 cores it would have cost approximately \$47 per hour.  Defining, and therefore estimating, the costs behind running a simulation on an IC cluster such as {{\it{Grizzly}}} is not a simple task, and the limited time available for this project did not  enable us to complete this task. 

We can summarize the lessons we learned during this work as follows: 
\begin{enumerate}
\item Using Amazon Linux VMs to develop and test locally can eliminate development and testing costs for AWS cluster.
\item AWS imposes service limits \cite{ref4},  however  requests for additional HPC instances were fulfilled within 24--36 hours \cite{ref5}.
\item EC2 pricing grows linearly, however larger instances provide access to better resources including CPU hardware and faster interconnects.
\item Spot instance pricing could reduce costs by only using AWS resources when costs are below a specified threshold.
\item Placement groups can improve performance by localizing HPC resources, and AWS services such as Auto Scaling and CloudFormation can assist in deploying, configuring, running, and shutting down AWS HPC clusters. 
\end{enumerate}
\subsubsection{ Conclusion}
The intention of this project was to explore the feasibility of running distributed software using  a cloud platform.  Our preliminary study shows that the cloud should be considered  as an environment to run HPC applications. We showed that building a virtual cluster to integrate HPC codes into a cloud based workflow is possible, and for some specific needs, the cloud could be also reasonable and probably a more efficient alternative to a traditional HPC cluster.

For decision support the cloud could be a good fit.  Decision makers don’t usually have, built in-house, the needed computational power and storage resources to run risk and adaptation analysis, they have to run workflows that couple models developed on different computational environments and  by different institutions. They need to run the analysis when it is needed not all the time. In this case, AWS could be a good option. It provides computational resources on demand,  managing those resourses, and providing tools to simplify the use of those resources and the couplings of models. Moreover, there is a subset of scientists and engineers with projects that require computing resources beyond their managing capability and are small enough to not require thousands of nodes. These groups fall within a category of users for whom  institutional resources may not be easily available and/or users who may use the resource for a short period of time, and the cloud could be a valuable solution. 

Our effort is only a preliminary work. A more detailed analysis is needed to better identify the use cases where employing a cloud platform is valuable when there is the need to run HPC applications. We are planning to continue this analysis under a current LDRD project.

In addition,  from this AWS study we identified the need to develop a software package capable of fully automating the process of building a cluster within HPC removing the need for the scientist to know anything about how the architecture works. The software could allow a scientist to effectively get ``supercomputing on demand” for only as long as they need it and terminate once they are done removing any long-term costs or commitments to maintain the system.

\subsection{Enterprise Scaling of the Co‐Evolving Attacker and Defender Strategies
for Large Infrastructure Networks Project}

\subsubsection{Contributors}

\begin{itemize}
  \item Adam Gausmann, Missouri University of Science and Technology, ajgq56@mst.edu
  \item Kevin Schoonover, Missouri University of Science and Technology, ksyh3@mst.edu
  \item Clay McGinnis, Missouri University of Science and Technology, cmm4hf@mst.edu
  \item Eric Michalak, A-4, emichalak@lanl.gov
  \item Sean Harris, Missouri University of Science and Technology, snhcn6@mst.edu
  \item Hannah Reinbolt, Missouri University of Science and Technology, hmrvg9@mst.edu
  \item Chris Rawlings, A-4, crawlings@lanl.gov
  \item Aaron Scott Pope, A-4, apope@lanl.gov
  \item Daniel Tauritz, A-4, dtauritz@lanl.gov
\end{itemize}

\subsubsection{Motivation}
Successful attacks on computer networks today do not often owe their victory to
directly overcoming strong security measures set up by the defender. Rather,
most attacks succeed because the number of possible vulnerabilities and entry vectors are too
large for humans to fully protect without making a mistake. Regardless of the
security elsewhere, a skilled attacker can exploit a single vulnerability in a
defensive system and negate the benefits of those security measures. The
Co-Evolving Attacker and Defender Strategies for Large Infrastructure Networks
(CEADS-LIN) project~\cite{Harris2018} uses Galaxy to develop evolved strategies in an emulated
environment.  Galaxy is a high-fidelity virtualized computer network emulation
framework developed by LANL/Missouri University of Science and Technology Cyber
Security Sciences Institute (CSSI) that runs evolutionary evaluations within an
emulated network topology to automatically discover and mitigate network
security flaws.  Initial experiments have been promising but only work in
networks up to 20 nodes with Galaxy due to poor scaling. Emulating enterprise
networks requires the use of a large number of host computers with low
orchestration overhead.

\subsubsection{Solution Approach}
We leveraged AWS to provide Galaxy with the capacity, speed, and elasticity required to scale
building network topologies using the AWS infrastructure. 
\begin{figure}[t]
\centering
\includegraphics[height=2.5in]{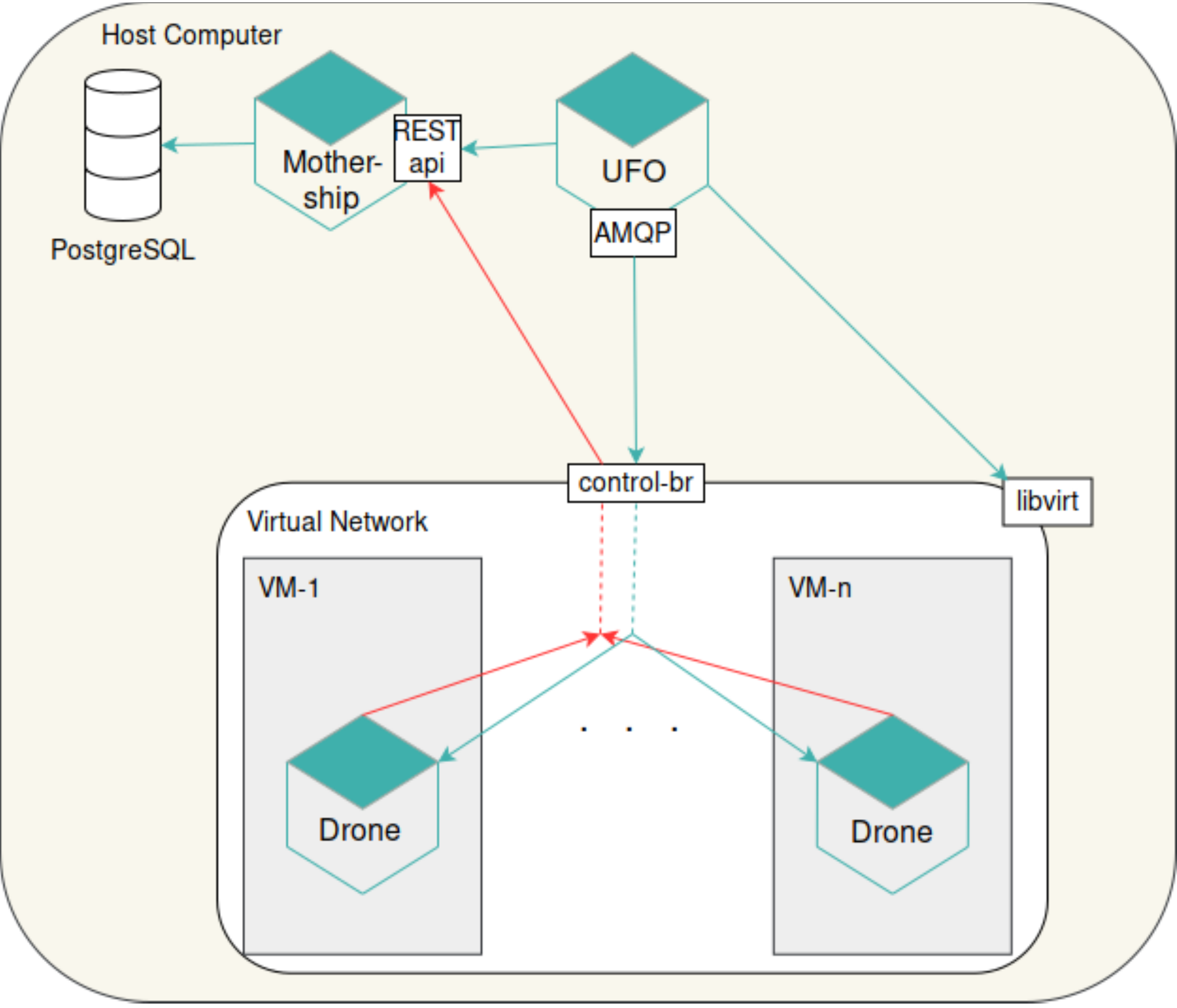}
\caption{Galaxy architecture.}
\label{fig:ceads-architecture}
\end{figure}
Galaxy's legacy design, shown in Figure \ref{fig:ceads-architecture}, consists
of three primary micro-services:
\begin{itemize}
  \item \textbf{Mothership} -- Primary data storage and retrieval REST API for
    researchers and agents. Stores data that persists between each evaluation.
  \item \textbf{UFO} -- Manages and orchestrates each evaluation utilizing the
    libvirt API and the KVM hypervisor.
  \item \textbf{Drone} -- Metric and data collection agent within the emulated
    network.
\end{itemize}
For more information, please see the full publication~\cite{Schoonover2018}.

We refactored the UFO to use the AWS Python SDK Boto3 to test if AWS would scale
better than our legacy system. Libvirt only allows actions on each virtual
machine to be performed serially. Starting a network required each node to be
started one at a time, scaling poorly as the number of nodes increased. In
contrast, EC2 instances can be started psuedo-asynchronously and all at once
through Boto3 which drastically reduced build and overhead time for larger
evaluations.

\subsubsection{Results}
\begin{figure}[t]
\centering
\subfloat[19 Node Network]{
  \includegraphics[width=65mm]{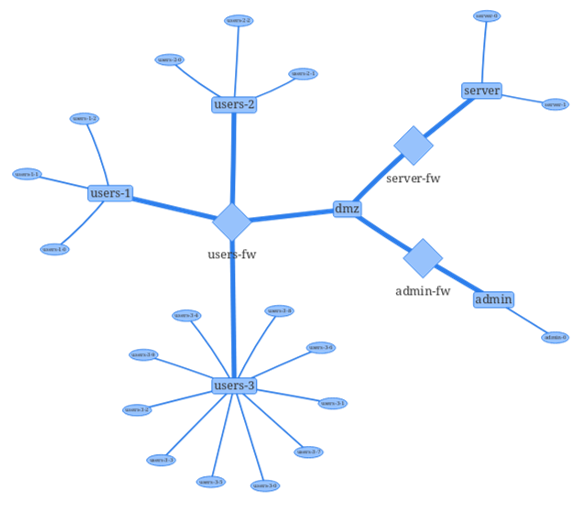}
}
\subfloat[50 Node Network]{
  \includegraphics[width=65mm]{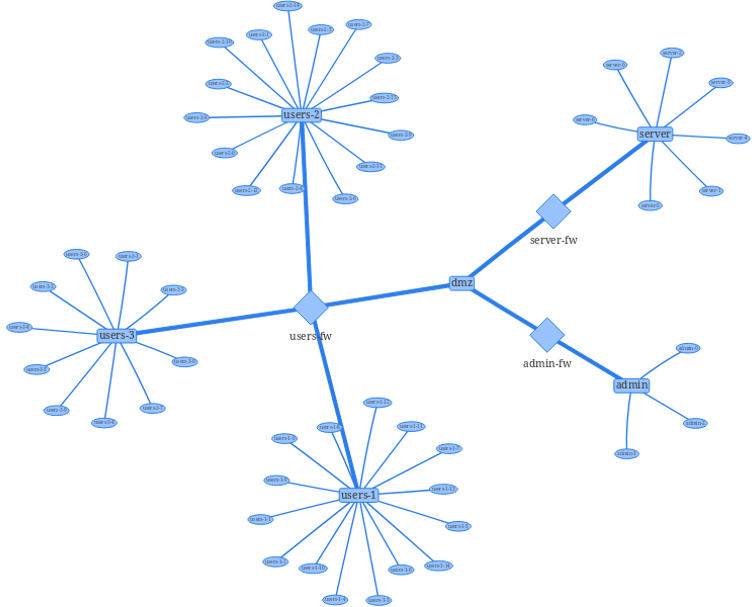}
}
\newline
\subfloat[130 Node Network]{
  \includegraphics[width=65mm]{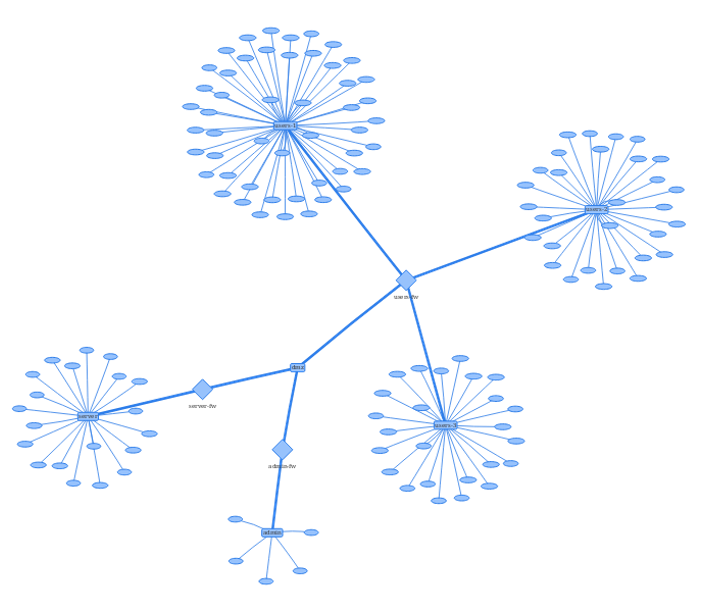}
}
\subfloat[800 Node Network]{
  \includegraphics[width=65mm]{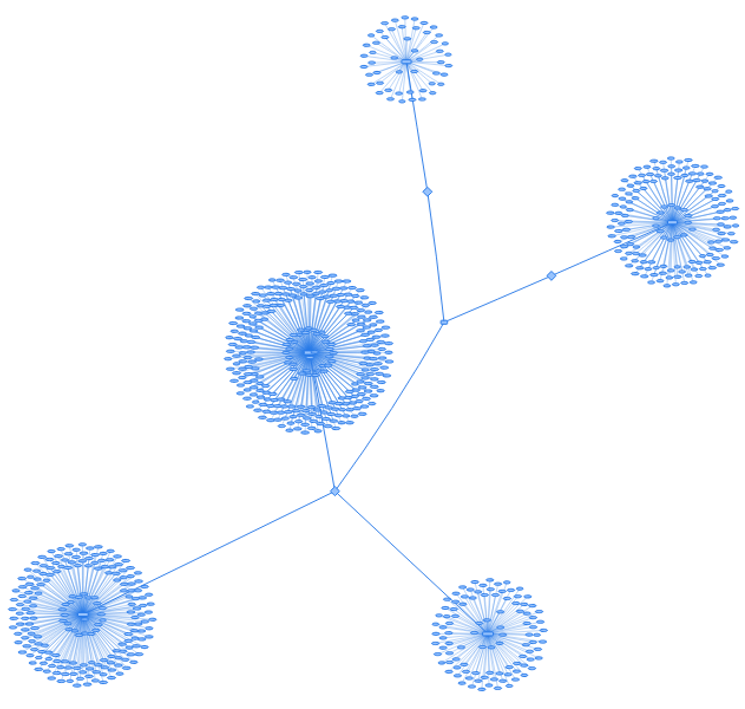}
}
\caption{Networks built during test.}
\label{fig:networks}
\end{figure}

\begin{table}[t]
\centering
\caption{Scaling Results}
\begin{tabular}{l|rr|rr|}
                  &              &   \textbf{AWS}&              &\textbf{Legacy} \\
                  &    Build Time&  8 Evaluations&    Build Time&  8 Evaluations  \\ \hline
\textbf{ 19 nodes}& $\sim$2.3 minutes&   $\sim$35 minutes& $\sim$152 minutes&   $\sim$16 minutes  \\ \hline
\textbf{ 50 nodes}&  3.62 minutes&  47.36 minutes& $\sim$400 minutes& $\sim$53.3 minutes  \\ \hline
\textbf{130 nodes}&  6.09 minutes&  72.75 minutes&$\sim$1040 minutes&  $\sim$138 minutes  \\ \hline
\textbf{500 nodes}&  $\sim$12 minutes&  $\sim$202 minutes&$\sim$4000 minutes& $\sim$533 minutes  \\ \hline
\textbf{800 nodes}& Timeout Error&  Timeout Error&$\sim$6400 minutes& $\sim$853 minutes  \\ \hline
\end{tabular}
\label{tbl:results}
\end{table}
Implementing AWS into Galaxy allowed us to scale to much larger networks with
orders of magnitude less overhead, as seen in Table~\ref{tbl:results}. However,
the overhead shown in the \textit{8 Evaluations} column still scales poorly over many nodes with infeasible wall time for our application. A selection of network topologies
that we tested are shown in Figure~\ref{fig:networks}. 

\subsubsection{Conclusion}
Internal AWS constraints such as rate limiting and total instance limits heavily
affected distributing evaluations across multiple UFOs. Rate limiting prevents
running more than four UFOs concurrently, which means only four networks can be
evaluated at a time. To solve rate limiting, future projects should utilize
Boto3's support for batch requests that allows starting or stopping up to 1000
instances with only one API request. In several cases, AWS ran out of
provisioned nodes of m5.large type within individual regions (at $\sim$ 200 instances) across four UFOs which
prevented scaling even further. The instance limit seems to be correlated with
the number of network interfaces attached to the instance type. Therefore,
minimizing the number of nodes with high interface counts may solve the limit
constraint. Future projects may be expected to encounter similar AWS constraints
when serially requesting \textgreater8 instances within a $\leq$1-second time frame
or requesting a large number of instances of the same type.

Other minor issues we encountered include AWS AMI IDs are not static across regions, AWS instance limits are initially very low for large networks (20 instances per region),  AWS subnets allow a maximum of a 255.255.0.0 (/16) subnet mask forcing a hard cap of machines within the network.
\subsection{Exploring Mesoscale Wind Forecasting on AWS for Integration with Operational Airborne BioAgent Response Models}

\subsubsection{Contributors}

\begin{itemize}
    \item Keeley Costigan, EES-16, krc@lanl.gov
    \item Matt Nelson, A-1, nelsonm@lanl.gov
    \item Patrick Conry, Post-Doc, Univ. of Notre Dame, ptc@lanl.gov
    \item Steve Linger, A-1, spl@lanl.gov
\end{itemize}

\subsubsection{Motivation}

Our goal is to explore how the mesoscale Weather Research and Forecasting (WRF) model could be deployed on AWS for integration with bioevent-driven atmospheric models to predict affected downwind regions. Currently, WRF is run most often in HPC environments. Understanding how WRF could be run in the highly scalable, highly available AWS environment could lead to viable compute options for both research projects and operational capabilities.

The Quick Urban and Industrial Complex (QUIC) and Bio-Event Reconstruction Team (BERT) capabilities at LANL both have models that input data from biodetection events and characterize airborne biological release events. Coupling WRF forecasts with the QUIC and BERT capabilities would expand these tools to include future consequences from biodetection events, providing more situational awareness to emergency responders and decision makers. The scalability of AWS resources would be particularly valuable in operational settings where results need to be calculated quickly and/or periodically, and, depending on model configurations, WRF can benefit from having considerable resources (e.g., processors, memory).

The main participant on this Cloud Computing 2018 project is an atmospheric scientist with experience running the WRF model in LANL’s HPC environment, but no experience with the AWS environment. Additional participants are atmospheric scientists (i.e., staff and Post-Doc) with no experience with AWS or WRF. The project is providing these participants with hands-on experience with the AWS environment and, for the latter participants, experience with the WRF model.

\subsubsection{Solution Approach}

WRF is a public domain atmospheric modeling system that is designed for both atmospheric research and numerical weather prediction. It has tens of thousands of users worldwide. It is run operationally by the National Oceanographic and Atmospheric Administration’s (NOAA’s) National Center for Environmental Prediction (NCEP), as well as other academic and government agencies. It has been applied to research studies on scales from tens of meters to thousands of kilometers. The team’s experience with WRF has been through research studies on LANL’s HPC resources with simulations that can consume tens of thousands of cpu hours and produce terabytes of output. 

The WRF modeling system uses static data (e.g., gridded global topography elevation data and land use data), a preprocessing package, and the model code. It is typically run with limited area domains and supports nested grid domains. Initial, boundary, and nudging conditions are prepared in the WRF Preprocessing System (WPS), which requires large scale gridded weather data for real (non-idealized) case simulations. The analysis fields of global weather prediction models are often used for this input data. Compiling the WPS and WRF codes is dependent on a number of libraries (e.g., netcdf) that need to be installed on the compute platform.

Given the large number of WRF users, our approach began with a search of previous efforts to run WRF on AWS. A couple of papers about WRF on AWS were found. Hacker et al.~\cite{hacker2017containerized} mostly looked at reproducibility of results on different AWS instances. However, it only used a single compute node. Goga et al.~\cite{GogaSpringerBook2018}  compared performance and speed-up with multiple nodes/CPUs on AWS. It used CfnCluster to access multiple nodes. 

Two demonstration tutorials were identified and used for the project. The first was an instructional session at the 2018 Joint WRF/MPAS Users’ Workshop, June 11--14, 2018, in Boulder, CO. The first tutorial included a presentation by Kevin Jorissen (Amazon Web Services) at~\cite{JorissenAWS} and a tutorial~\cite{NCARAWStutorial}. The second tutorial employs Docker containers and was available from the Developmental Testbed Center~\cite{DTCenter1}~\cite{DTCenter2}. 

Containers could potentially be useful for WRF applications. WRF code installation can be time consuming and containers could reduce spin-up time to build the libraries and code components. Containers are highly portable, which is useful in cloud computing, and make it easy to replicate procedures for preprocessing input data, running the code, running analysis, and plotting codes on the model results. They could simplify running the model on multiple instances (e.g., for ensemble simulations or multiple cases), taking the same Docker container to multiple instances and knowing that the computing environment and model configuration is the same for all runs.

\subsubsection{Results}

The first tutorial from the WRF/MPAS Workshop was straightforward to follow and employed a public AMI in the AWS West Region of Northern California. The AMI included the compiled code for real cases and some input data. However, it did not include code compiled for idealized cases, and the input namelist files for WPS and WRF appeared to be the default files that are included in the model package and not set up as a demonstration real case. Perhaps the workshop provided a real case to participants that was not available to the web community. However, a NCAR Command Language (NCL) script was run to produce a plot of the model grids defined in the namelist.wps file and the plot was successfully transferred to the user’s LANL desktop, using \texttt{scp}. A private AMI was created from the public AMI, for future reference.

It seems possible to use \texttt{scp} to transfer a WRF case study that we have previously run on HPC resources to our AWS instance using the AMI and run it, provided the appropriate static data for the domains in the real case is available and provided our namelists are compatible with the version of WRF that they have in the AMI. There were some challenges attaining the larger (e.g., c4.8xlarge) instance types (recommended in the tutorial for NWP),  a smaller one had to be utilized, which would not be sufficient, for instance, for running WRF-Chem simulations which typically use ~200 processors on the HPC clusters. We would also need to make use of some storage capability to save our results for larger simulations, maybe Amazon EFS. 

For the second tutorial from the DTC website, project participants learned about Docker containers and how to use them on AWS. The AWS tutorial on deploying Docker Containers was difficult to follow because the tutorial needs to be updated or steps are left out. It also was not particularly useful for NWP purposes. However, it did go through using Amazon ECS and used an Elastic Load Balancing (ELB) load balancer, which may be needed to run WRF on a cluster. 

Other documentation~\cite{dockerbasics} appeared to be instructive for installing Docker on an instance, however it failed on the fourth step, which was using the command ``sudo yum install -y docker”.  It could not find docker, and the failure might be related to the type instance (Red Hat Linux). Eventually, the steps in~\cite{hackernoondocker} proved to be most useful for installing Docker when using an Amazon Linux instance. Note that after the command ``sudo usermod -aG docker ec2-user”, one needs to logout of the instance and then SSH back in.

Once Docker was installed on an instance, project participants were able to follow the DTC tutorial. The first attempts of both participants who followed this tutorial failed because the instance storage space was not increased. In addition to Docker, ``sudo yum install git” could be used to install git or the steps in DTC tutorial could be used to install from Github using the difference method.

\begin{figure}[htp]
\begin{center}
\includegraphics[width=0.9\textwidth]{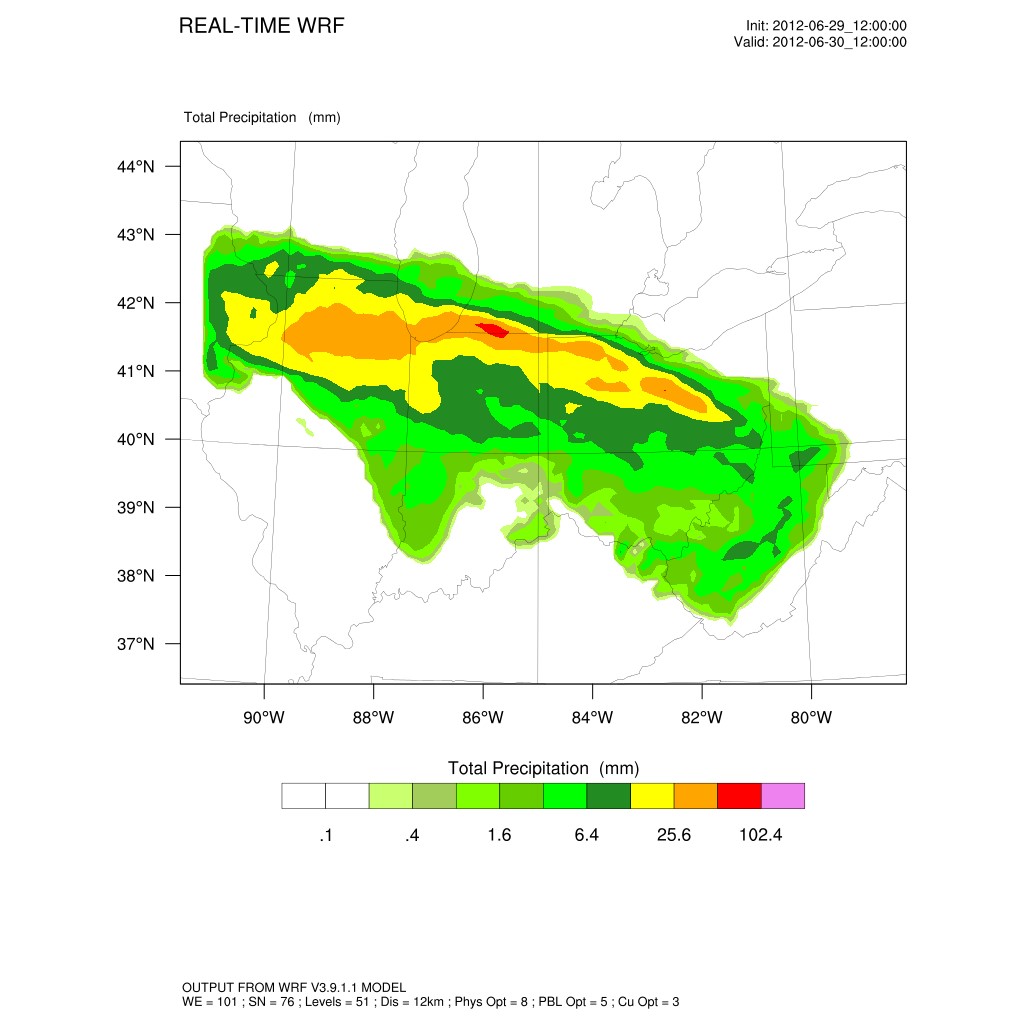}
\caption{Precipitation results for Derecho test case.}
\label{precip-results-derecho-case}
\end{center}
\end{figure}

Keeley Costigan ran the derecho case, very strong, high-impact derecho (i.e., straight-line, high winds from meso-scale storms) on June 29, 2012. It ran on four processors. A second terminal window on Mac Pro was used to SSH into the instance and verify that it was writing output files. Keeley was able to make plots with NCL, put them in a TAR file, and use \texttt{scp} to bring them to desktop (Figure  \ref{precip-results-derecho-case}). The Model Evaluation Tools (MET) container’s output, however, could not be visualized because the command ``docker-compose up -d” gave the error ``command not found.” 

\begin{figure}[htp]
\begin{center}
\includegraphics[width=0.9\textwidth]{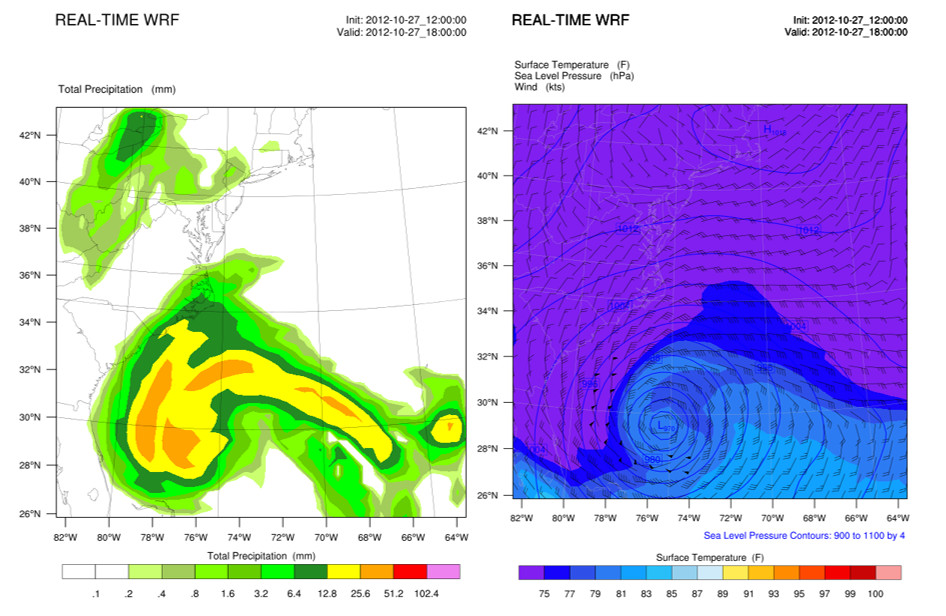}
\caption{Hurricane Sandy precipitation (left) and surface winds (right).}
\label{hurricane-sandy-precip-and-surf-winds}
\end{center}
\end{figure}

Both Keeley and Patrick Conry ran the Hurricane Sandy test case. This run only simulated 6 hours, so it went much faster on instances using a small number of processors. Keeley was able to make plots with NCL, put them in a TAR file, and use \texttt{scp} to transfer them to desktop as in the Derecho test case (Figure  \ref{hurricane-sandy-precip-and-surf-winds}). Patrick was unable to make plots as the NCL command instructed by DTC tutorial failed with an error. As before with Derecho MET container failed to visualize results with error ``command not found.” While Docker containers succeeded in running WRF on AWS instances, the tutorial visualization containers were lacking, and experienced WRF users would need to export results to their own desktops or adapt the container for easier and more flexible visualization.

Other disadvantages or challenges with using AWS include: (1) Many services are available, which caused confusion in knowing what is needed and what is best to use and when. (2) Much time was spent trying to figure out what to use, which led to time wasted going down some paths before determining that they were not needed. (3) Keeley spent almost an entire day trying to figure out why SSH was not connecting to an instance and repeating what had been previously done in the first tutorial. She tried launching new instances, key pairs, etc. Finally, at the end of the day, she tried working in another region, which worked, and the new instance could be accessed with ssh. (4) Patrick and Matt Nelson were unable to connect to their first instance for a few hours, and after an experienced AWS user in A-1, Daniel Frank, tried for one hour, the problem was solved by simply rebooting the instance. There was also an experience of an instance unexpectedly rebooting. This inconsistent behavior with accessing instance via SSH is problematic for users with limited time frame and budget to run instances (i.e., users are charged for the entire period that the instance is running). (5) It would be helpful to have someone to call when experiencing problems to figure out what caused them and how to fix them.

\subsubsection{Conclusion}

Going forward, we will probably want to create our own containers tailored to our needs. The last section of the DTC tutorial website talks about customizing the runs. The WRF/MPAS Workshop website also had some suggestions. We would need to look into adding storage for saving simulation results and ways to increase the number of processors that can run a given simulation.

The cloud computing efforts allowed project participants to directly experience the advantages and disadvantages of using AWS for NWP applications. WRF simulations were successfully run by both experienced and novice WRF users, highlighting a key advantage of using Docker containers which greatly simplify the complexity of running WRF. AWS instances were sufficient in computing problem to run simple tutorial cases, but project participants expressed concern about AWS’s applicability and costs for running more complex NWP cases.



\subsection{Evaluating a C++ Big Data Framework for HPC and Cloud Platform}

\subsubsection{Contributors}

\begin{itemize}
    \item Li-Ta Lo, CCS-7, ollie@lanl.gov
\end{itemize}

\subsubsection{Motivation}

Recently, distributed big data analytics frameworks like Apache Spark \cite{zaharia2016apache} and Apache Flink \cite{carbone2015apache} are gaining popularity among data science practitioners. Those frameworks provide general purpose, high level, application programming interfaces (APIs) that hide many details of distributed parallel computing. Abstractions over distributed data, such as Resilient Distributed Data (RDD) in Spark, relieve users of the need to explicitly distribute their data and computation among compute resources. Users express their distributed algorithms through a set of scalable parallel primitives such as \texttt{Map}, \texttt{Reduce}, \texttt{GroupBy}, etc. The frameworks then convert the user programs into computational tasks and their dependency graph and schedule and submit tasks to the runtime system for execution.

There are several hurdles for the HPC community, especially  DOE labs, to adopt those frameworks. The frameworks are written in programming languages based on the Java Virtual Machine, which is not available on HPC platforms for security reasons. They use TCP/IP for communication thus do not take advantage of the high speed interconnect available on HPC platforms. Users also need to reformulate their distributed algorithms in terms of the the parallel primitives provided by the frameworks.

This leads to some interesting questions: (1) Is there some big data framework that is written in a programming language that is more palatable to the HPC environment and can take advantage of the high speed interconnect? (2) How can we expression some common data analysis operations that are commonly provided by big data frameworks?

\subsubsection{Solution Approach}

We explored using the Thrill \cite{bingmann2016thrill} library to track the most energetic particles for the results of VPIC plasma physics simulation \cite{guo2015particle}. Thrill is a research project that aims to provide a bridge between big data analytics and HPC platforms. It is written in C++, thus providing performance of native code and avoiding the security concerns of the Java Virtual Machine. It supports using both TCP/IP and MPI as a communication back-end providing a way to take advantage of high speed interconnect. It also provides a high-level application programming interface similar to Spark.

Thrill uses a hybrid parallel execution model. On each node, there is one MPI process (rank) with a configurable number of threads. One of the threads is the \emph{host} thread that performs MPI communication and the others are \emph{worker} threads that actually perform computations. By default, the number of worker threads on each node is the same as the number of cores available. Data is automatically distributed among worker threads by the runtime system, and application programmers normally do not have to worry about it.

We used EC2 and EFS to reproduce an environment that mimics a traditional HPC system. We used eight instances of m4.4xlarge as compute nodes. The HDF5 particle data is stored in the EFS and shared by all the compute nodes. The data file contains 100 time steps. Each time step has around 1 million particles, and each particle has three floating point values for $x, y, z$ position, three floating point values for $x, y, z$ components of momentum, and an integer particle ID.

Our analysis program contains two parts. In the first part, we load the last time step from the data file on the ``root" serially. We compute kinetic energy of each particle based on momentum and sort particles according the their kinetic energy. This gives us the IDs of the most energetic particles. In the second part, we use Thrill's \texttt{Generate} primitive to generate a sequence of integers $[0, N)$ where $N$ is the number of time steps. The generation of the sequence of integers is evenly distributed among MPI ranks and workers. For example, when using only one node with 16 worker threads, each worker thread on average generates six consecutive integers. When using all eight nodes with 128 worker threads, 28 of them remain idle. We then use the \texttt{Map} primitive with a user-defined function to load particle data from the data file in parallel. The worker threads call the user-defined function with the time step number, the user-defined function loads the particles of the corresponding time step by calling the HDF5 library routine. It then selects the most energetic particles based on the particle ID computed in part 1. After all time steps are processed by the workers, the particle positions are then saved as a CSV file using the \texttt{WriteLines} primitive. Since multiple worker threads are calling the HDF5 library concurrently, we have to compile the HDF5 library using ``thread safe" mode.

\subsubsection{Results}

\begin{figure}[t]
     \begin{center}
     \includegraphics[width=8.75cm]{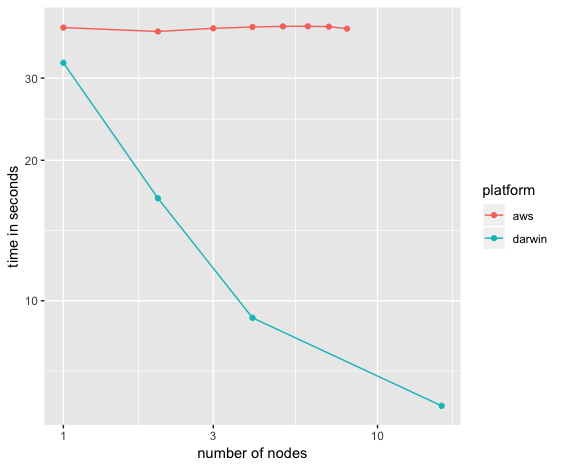}
     \end{center}
     \caption{VPIC Trajectory Strong Scaling.}
     \label{fig:vpic_strong_scaling}
\end{figure}

We performed a strong scalability study on both AWS instances and the \emph{scaling} partition of the Darwin cluster at CCS-7. The timing result is shown in Figure \ref{fig:vpic_strong_scaling}. As we can see from the figure, our solution scales well on Darwin but not on AWS. Due to limited duration of the project, we were not able to do a detailed profiling of our solution to find out the reason behind the discrepancy. However, we think it may be due to certain characteristic of EFS, such that concurrent reads by multiple instances are not scalable.

\subsubsection{Conclusion}

We implemented an analysis operation for the VPIC simulation and conducted a scalability study on both a traditional HPC cluster and AWS. Our experiment is scalable on the HPC platform but did not show scalability on AWS. Further investigation is required to determine the true cause of this discrepancy.

\subsection{Enabling Wildland Fire Managers to Explore the Fire ``Response Space"}

\subsubsection{Contributors}

\begin{itemize}
    \item Marlin Holmes, EES-16, mjholmes@lanl.gov
    \item Rodman Linn, EES-16, rrl@lanl.gov
    \item Sara Brambilla, A-1, sbrambilla@lanl.gov
    \item Michael Brown, A-1, mbrown@lanl.gov
\end{itemize}

\subsubsection{Motivation}

\begin{figure}[ht]
     \begin{center}
    \includegraphics[width=8.75cm]{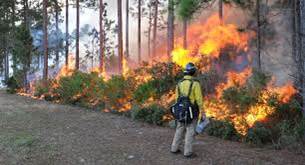}
    \end{center}
     \caption{A U.S. Forest Service fire detailer watches the flames during a prescribed burn Nov. 21 at the Silver Flag Exercise Site, Tyndall Air Force Base, FL. The burns clear undergrowth and allow for healthier forests while lowering the possibility of wildfires. (U.S. Air Force photo by Airman 1st Class Alex Echols)}
    \label{fig:firec}
 \end{figure}

Land and fire managers face challenges not only responding to wildfires but also in preparing for wildfire season, optimizing tactics for prescribed fires to meet their objectives, and evaluating fuels management options. Figure \ref{fig:firec} shows what is at stake. These managers are not only responsible for protecting ecosystems and property but, most importantly, human lives. Increased access to process-based computational coupled fire/atmosphere models could provide valuable information concerning the interaction between fires and their environment and could be used to explore the response of fires to a range of conditions or even ignition tactics in the case of prescribed fires. Furtermore, an increase of access to these computational tools could lead to better understanding of risk in wildland fire operations. Unfortunately, sampling the parameter space of fire behaviour comes with appreciable computational cost and time delays. In this proposal we aim to display how cloud computing combined with fast running models can be leveraged to increase access to computational tools for land and fire managers. 

\subsubsection{Solution Approach}

The computational tool that was used in this proposal was QUICFire, a reduced order fire/atmosphere code developed at LANL. QUICFire utilizes the knowledge gained from the high fidelity physics based model FIRETEC to simulate a limited set of parameters such as fire spread, fuel consumption, and plume/wake dynamics. QUICFire can simulate an approximately 400 square meter domain on a single processor at nearly real time with the ability to add additional processors for larger domains. This tool would allow fire/land managers to explore the uncertainties or sensitivities of fire behavior to a variety of conditions at low cost. 

\begin{figure}[ht]
     \begin{center}
    \includegraphics[width=8.75cm]{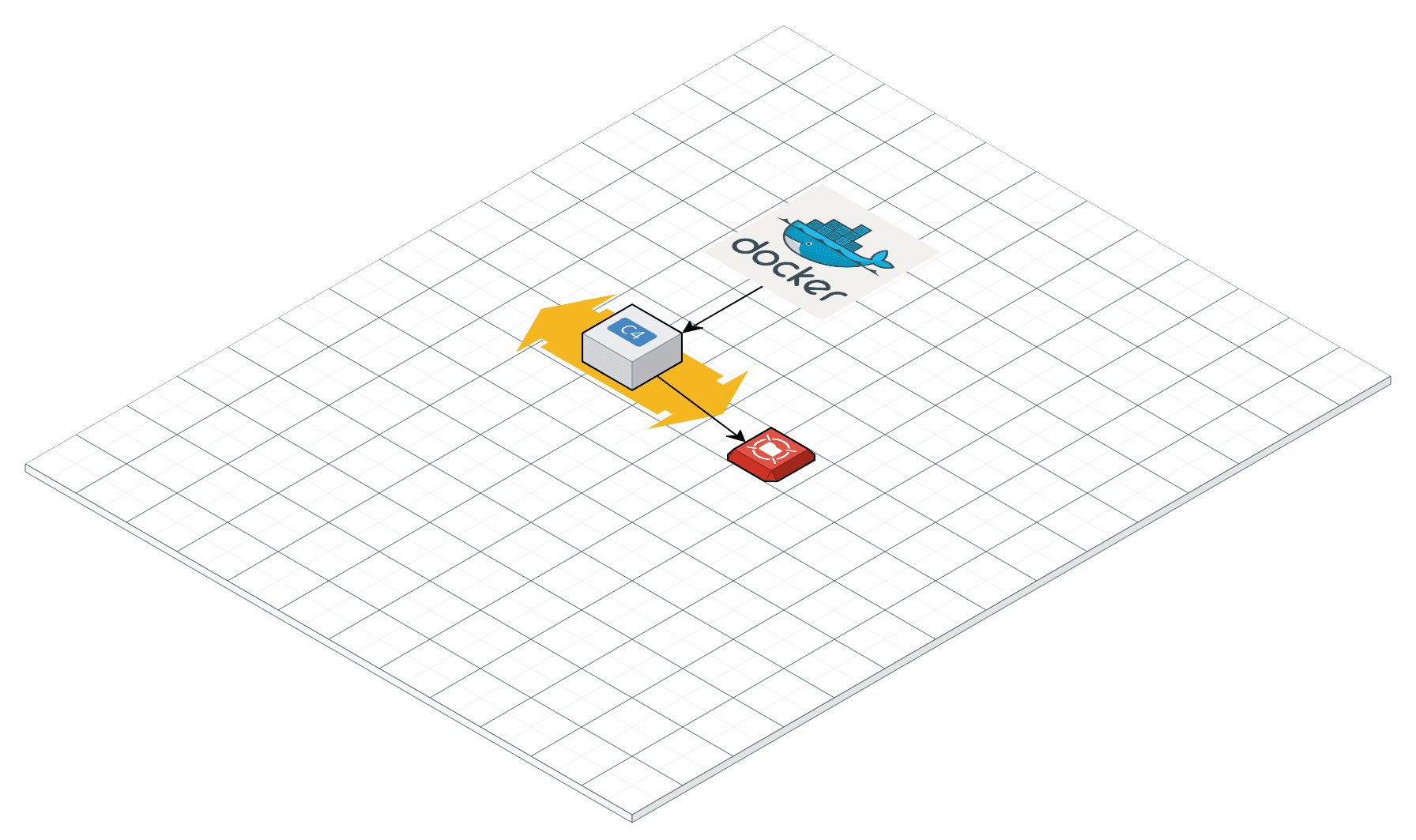}.
    \end{center}
     \caption{Diagram of cloud architecture for QUICFire deployment.}
    \label{fig:cloudc}
 \end{figure}

A diagram of of our planned solution approach can be seen in Figure \ref{fig:cloudc}. QUICFire will be packaged in a Docker container on an AWS EC2 instance with EFS providing persistent storage across multiple instances. Docker will allow the user to utilize the application without becoming overly concerned with software dependencies and machine architecture as long as the Docker program is installed. EC2 leveraged with or without spot pricing will provide the end user with sufficient and specific computation power for the problem/task at hand. EFS once mounted to an instance will provide persistent storage of simulation runs which can be leveraged across an organization, for example the U.S. Forest Service.  Nearly all of these processed can be set up through either the AWS console, or using a command line interface of the users choice. 

\subsubsection{Results}
\begin{figure}[ht]
     \begin{center}
    \includegraphics[width=8.75cm]{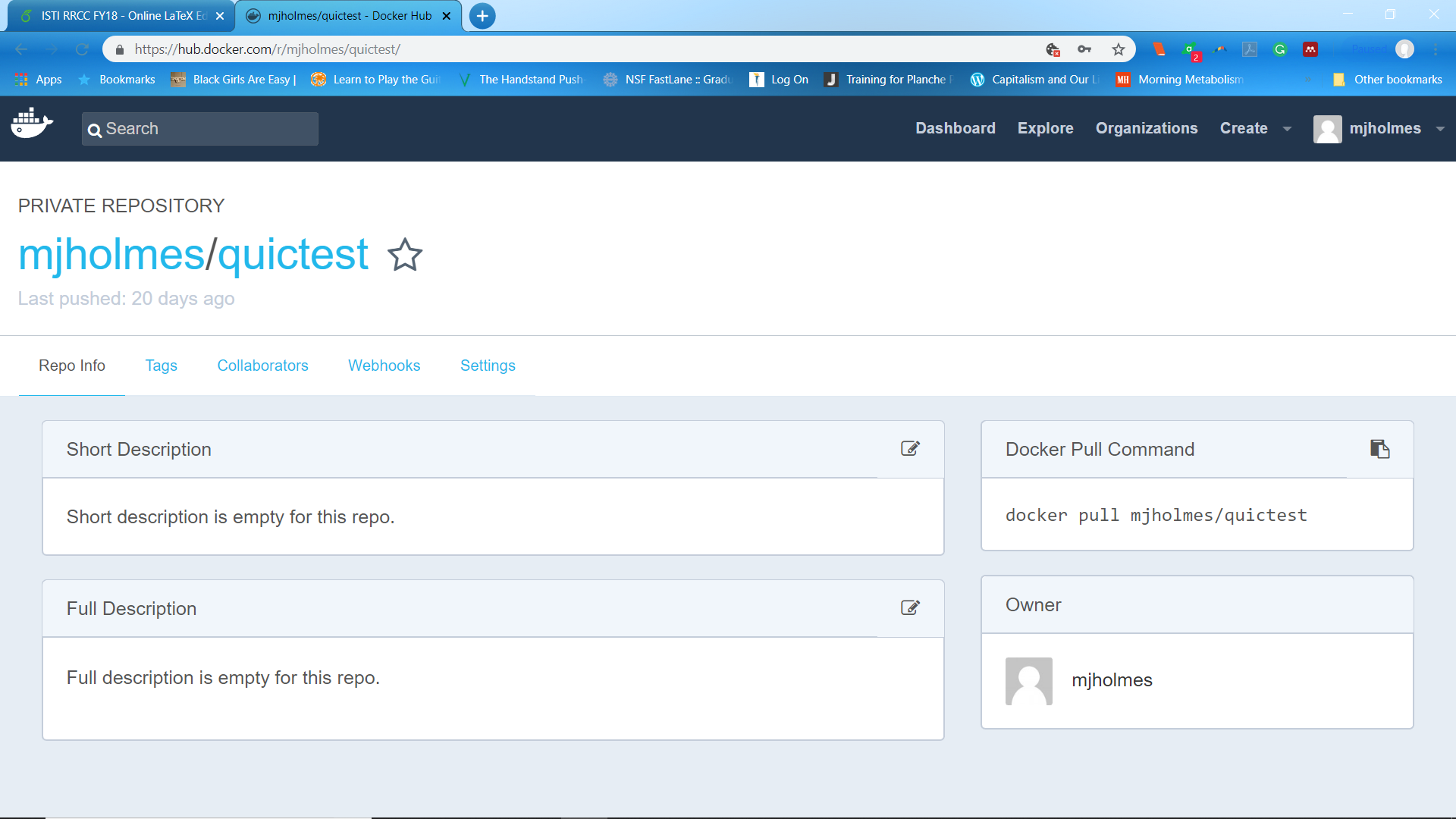}.
    \end{center}
     \caption{Screen shot of QUICFire container on Docker hub.}
    \label{fig:quichub}
 \end{figure}

The first step to providing platform independence for the AWS implementation on EC2 was to ``dockerize" our application. This part of our solution actually proved to be the most time consuming, further emphasizing two points: AWS is relatively straightforward to deploy as a machine/architecture on LANL systems, however effectively moving applications in and out of the firewall at LANL can still prove a challenge to the inexperienced. The hurdles with deploying our applications as a container largely had to do with two issues tied to how Docker, as a program/tool, is set up. First, Docker requires the ability to pull the requisite composite pieces into each container, which means appropriately setting up the Docker application and the container to work with the LANL proxy. Second, Docker in it's current iteration requires super user privileges to execute every single command. This last portion can be quite onerous, however LANL has developed an application called CharlieCloud that can remove the necessity for super user privileges to run Docker but still leave the requirement for super use to build Docker images. This application was build by LANL staff with the specific intent of decreasing the security risk of utilizing Docker but still only for image execution. For this project to mitigate security risks, we decided it was simpler to create a Linux virtual machine where all Docker building and configuration was performed which was then pushed to Docker hub when completed. Figure \ref{fig:quichub} shows the private repository we created to host our image of QUICFire for AWS deployment. 

After setting up a container for our application almost every additional step can performed in the AWS GUI/console.  An EC2 instance of the Amazon Linux was reserved, as well as an Amazon EFS file system. Connecting these two systems together is fairly straightforward with nearly exact commands listed on the console which can be copied and pasted into your command line interface of choice once the user has logged into the EC2 instance (usually accomplished through ssh). After this it is also fairly straightforward to set up Docker on the EC2 instance. Once the ``docker pull" command is used, EFS can be mounted as a run directory for the container, and QUICFire can run as one would locally. At this point the user simply needs to decide whether to move the data back locally or perform their analysis in the cloud first. 

\subsubsection{Conclusion}

To conclude, a reduced order coupled fire-atmosphere code was successfully ported and run on AWS with all dependencies included. This application which runs at nearly real time speeds combined with cloud computing is just one example how we can provide increased access to physics-informed tools for our comrades out in the field. However, with this success duly noted, we would be remiss if we did not note the limitations in this approach. First, cloud computing in its current form is best suited for applications that do not require consistent and fast inter node communication. This will likely prove to be a limitation on the complexity and size of tools deployed on AWS depending on their underlying algorithms. Second, while it is of nearly no cost to move date on AWS systems and even to keep it there, a user can begin to incur significant cost when data is removed \emph{off AWS}. As such, we argue that currently it is not enough to simply ``cloud compute," we must also begin to consider ``cloud data reduce/analyze/visualize" so that only the essential and most relevant information is actually brought back to local machines. This will only prove more and more critical as cloud computing, and thus cloud computing tools, grow and the output of these tools swells with time.

\clearpage
\section{Machine Learning}
\label{sec:ml}

\subsection{Accelerating Hydrodynamics and Turbulence Modeling with Efficient Deep Learning on Amazon
AWS}

\subsubsection{Contributors}

\begin{itemize}
    \item Arvind Mohan, T-4/CNLS, arvindm@lanl.gov
    \item Dima Tretiak, T-4, dtretiak@lanl.gov
\end{itemize}

\subsubsection{Motivation}
Turbulence in fluids plays an important role in a wide variety of science and engineering problems, with its inherently chaotic multi-scale nature making modeling extremely challenging. At LANL, hydrodynamics and turbulence modeling efforts are critical to the success of campaigns in inertial confinement fusion (ICF), astrophysics, earth sciences, and applied research. Although computational fluid dynamics (CFD) has significantly improved our understanding of turbulence physics, high-fidelity CFD like direct numerical simulation (DNS) requires powerful clusters, which are prohibitively expensive to deploy for problems of practical interest. Over the years, machine learning-based approaches have made significant strides in
modeling complex nonlinear phenomena in several scientific domains, with
a computational cost orders of magnitude lower. Our focus is on developing deep learning (DL)-based reduced order models (ROMs) to learn key high-resolution dynamics from DNSs and model them with desktop-scale computing power. A major challenge in
ML for turbulence is the chaotic, high-dimensional, and spatiotemporal
nature of the data, which can make the learning process ineffective and/or expensive. Past work by the author~\cite{mohan2018deep} in modeling turbulence with long short term memory (LSTM) neural networks demonstrated its
capability to capture temporal nonlinearities. In this work, we extend this
capability to modeling spatiotemporal features of turbulence using the
Convolutional LSTM (ConvLSTM)~\cite{xingjian2015convolutional} neural network. ConvLSTM augments the
traditional architecture of an LSTM cell with a convolutional layer to capture
spatial correlations in multidimensional datasets. We demonstrate the
potential of ConvLSTM in learning and predicting the dynamics of a DNS
homogeneous isotropic turbulence dataset. We perform statistical tests on
the predicted turbulence to quantify the quality of the learned physics. We will upload useful codes/scripts to a LANL internal Gitlab repository at \url{https://gitlab.lanl.gov/arvindm/AWSCC_ISTI_codes.git}.

\subsubsection{Solution Approach}

\begin{figure}[htp]
    \begin{center}
    \includegraphics[width=10.75cm]{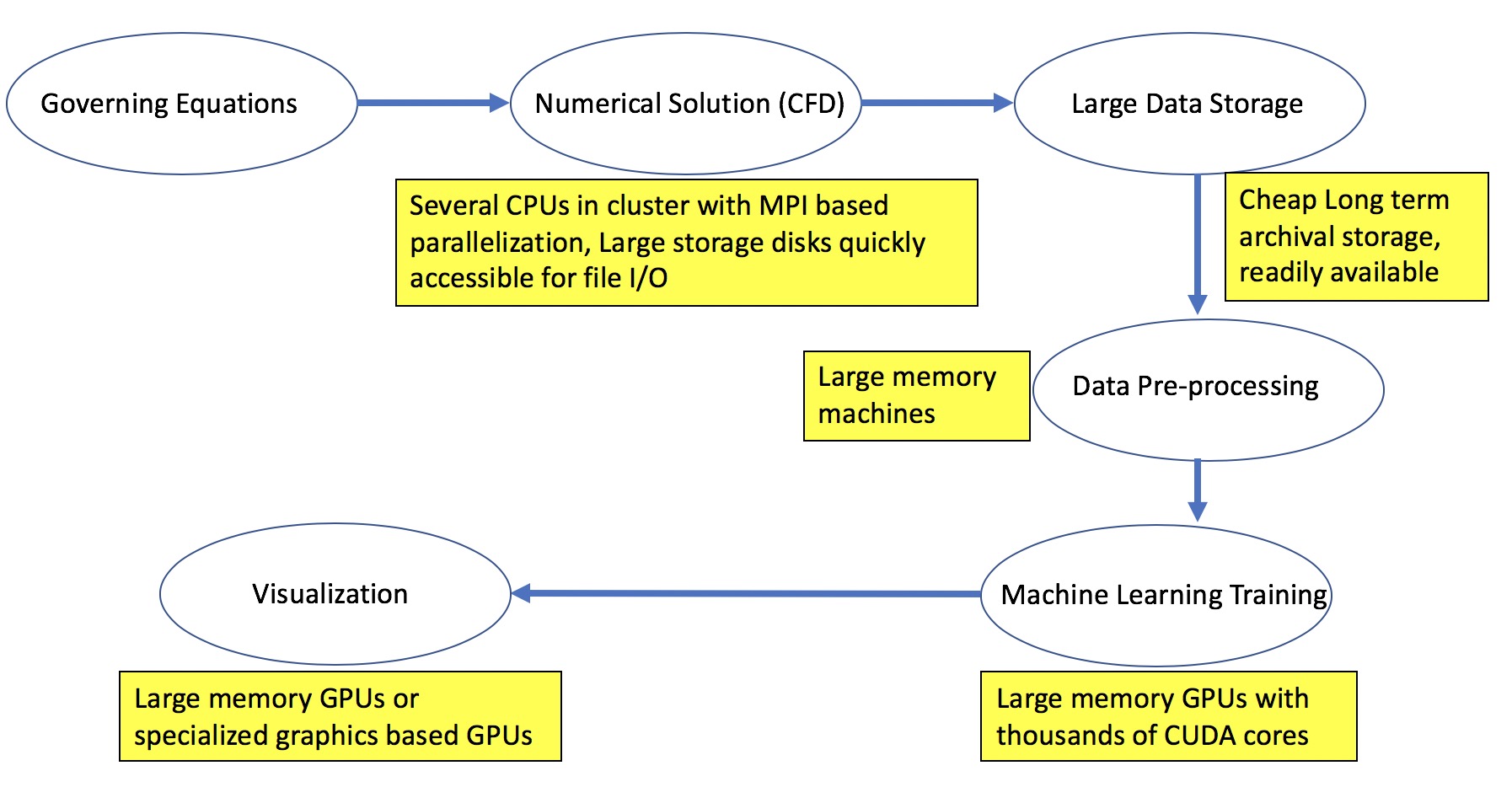}
    \end{center}
    \caption{The Machine Learning for Turbulence workflow, from simulation to modeling.}
    \label{fig:workflow}
\end{figure}

The overall workflow of the effort is shown in Figure \ref{fig:workflow}. The workflow not only includes the machine learning process, but also data generation with CFD simulations, data transfer, and preprocessing.

It can be seen that at every stage of the process, the preferred compute architecture is different. CFD typically prefers large MPI based clusters to perform high fidelity simulations, while machine learning operations are extremely fast on GPUs. Furthermore, RAM-intensive operations like data munging and visualization require powerful high-memory desktops. It is expensive to acquire such hardware for each research workflow, and the workflows also change several times during the course of a project. As a result, cloud computing can be a boon to such projects where the various compute architectures needed can be invoked and collapsed on demand---at a much lower cost and effort. We use the following AWS services for our workflow.

\begin{itemize}
    \item \textbf{CFD} -- EC2 Compute Optimized c5.18x large nodes, AWS EBS filesystem for low latency hard disk storage
    \item \textbf{Readily accessible archival storage} -- AWS S3, medium latency
    \item \textbf{Data-preprocessing/Analysis} -- EC2 Memory Optimized r5d.4x large instance
    \item \textbf{Neural Network training} -- EC2 GPU compute instances p3.8x and p3.16x with Nvidia V100 GPUs
\end{itemize}

The ML code was written in Keras using the TensorFlow backend. For problems of this scale, the network training can get tedious and slow. Neural network training can be greatly expedited through the use of more than one GPU in a distributed fashion. Doing so in TensorFlow can be complicated and tedious, therefore Horovod, an open source framework created by Uber intended to easily parallelize TensorFlow, was chosen to accelerate our task. Horovod’s Github (\url{https://github.com/uber/horovod}) provides more details. AWS offers Deep Learning Amazon Machine Images (AMIs) with numerous pre-built ML frameworks like TensorFlow, PyTorch, and Horovod, which makes creating the software environments for EC2 instances smooth and quick without resorting to Docker.

In AWS, users typically launch instances on-demand. However, an alternative approach is to take advantage of the machines not currently in use by other users by requesting a spot instance. Spot instances cost far less than on-demand instances and could yield up to a $90 \%$ decrease in hourly rate. Instead of launching an instance for immediate access, one requests a spot instance with configurations to run the selected job if it becomes available. We used a shell script, which ran on instance launch, to configure and start/resume training. To run the shell script on launch, upload it to ``User data” under ``Advanced Details” in the ``Configure Instance” settings. Subsequently, every time a new spot instance launches, the shell script creates a new directory, uploads the code and data to that directory, activates the TensorFlow environment, and then runs the code. 

The caveat with spot instances is that they may---at any time---be terminated when demand rises and the machine is no longer available. Therefore, neural nets that are trained on spot instances must automatically create checkpoints in order to avoid losing progress. Fortunately, TensorFlow’s \textit{tf.train.MonitoredTrainingSession} handles creating checkpoints and restoring from checkpoints locally. We used the Python Boto3 library for AWS in order to save these checkpoints to an S3 bucket. This way, whenever the spot instances are terminated, the next spot instance can load the checkpoint from S3 and continue training from where it left off. Shell scripting combined with checkpointing allows the spot instance to be fully autonomous after the initial spot request is made. 

\subsubsection{Results}
We train our ConvLSTM network on DNS datasets of 3D homogeneous, isotropic turbulence~\cite{mortensen2016high}. As a proof of concept, from this 3D dataset, we collect several 2D slices to use as a training dataset. The goal is to generate turbulent fields at future timesteps, given turbulent fields from the previous timesteps as input. We use the DNS data as our training and ground truth since the HIT problem in turbulence has been well studied and extremely high-fidelity numerical simulations are possible for this case. However in general, CFD simulations, especially for engineering flows, are not always representative of ground truth conditions. The ML models were built using Keras with TensorFlow as a backend. The networks were trained mostly on AWS p3.16xlarge GPUs which consist of eight Nvidia V100 GPUs which scaled very well with distributed TensorFlow. The network took $1:35$ hours on a single V100 p3.2x GPU, while the distributed version took us $00:16$ hours on the p3.16x, which is an impressive speedup. In comparison, LANL's Kodiak cluster has P100 GPUs and the same training on a single P100 GPU took 3:10 hours.

\begin{figure}[htp]
    \begin{center}
    \includegraphics[width=13.75cm]{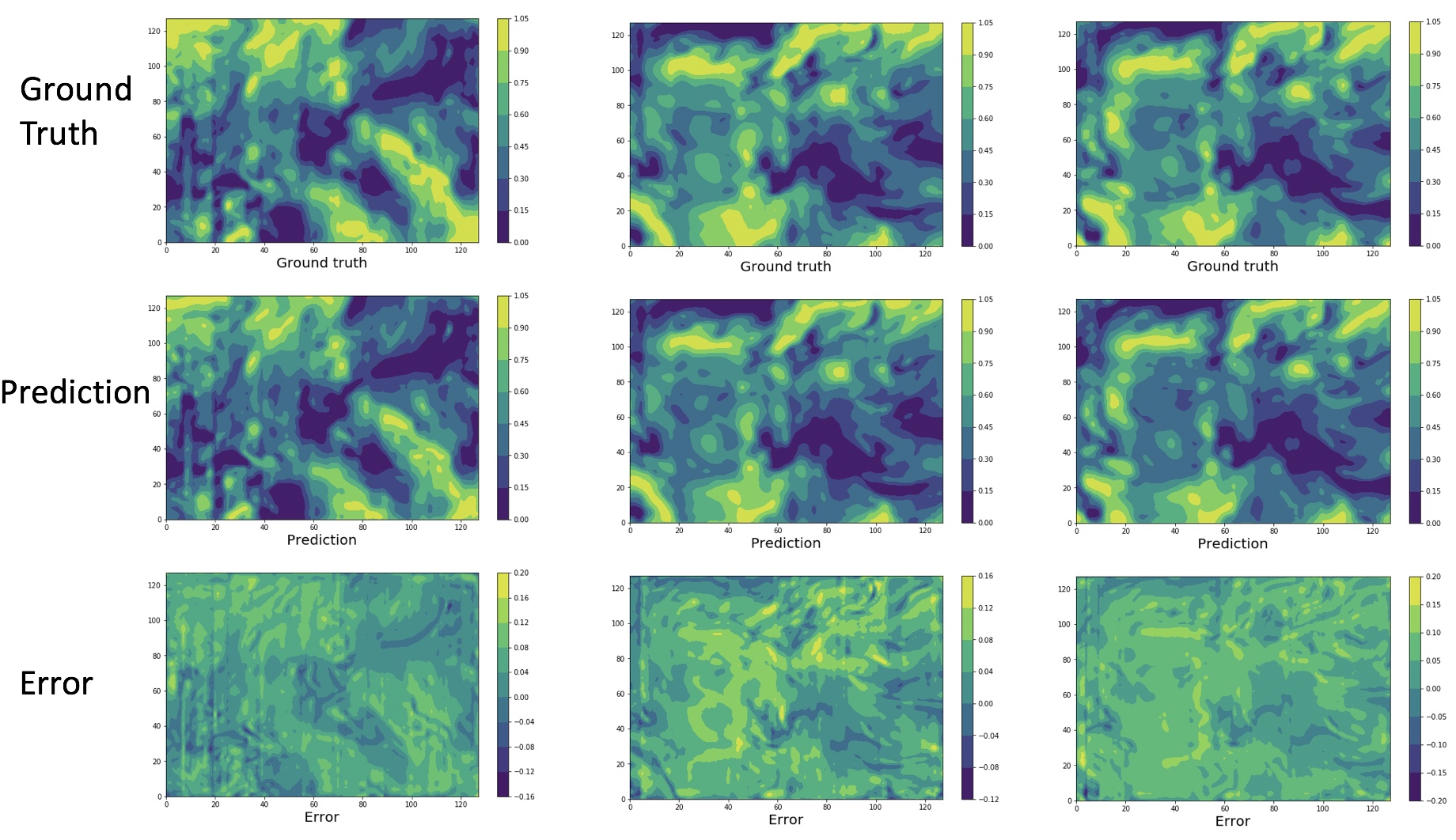}
    \end{center}
    \caption{DNS turbulence vs. ML generated turbulence.}
    \label{fig:results}
\end{figure}

In Figure \ref{fig:results}, we show a few samples of our true DNS dataset (which we call our ``ground truth") and the predicted turbulence from ConvLSTM at various timesteps. The data is scaled between $0$ and $1$, while the actual turbulence data has negative values as well. We found that this helped in faster training of the LSTM cells in the network, and the data can also be scaled back to true turbulence ranges.

\begin{figure}[htp]
    \begin{center}
    \includegraphics[width=10.75cm]{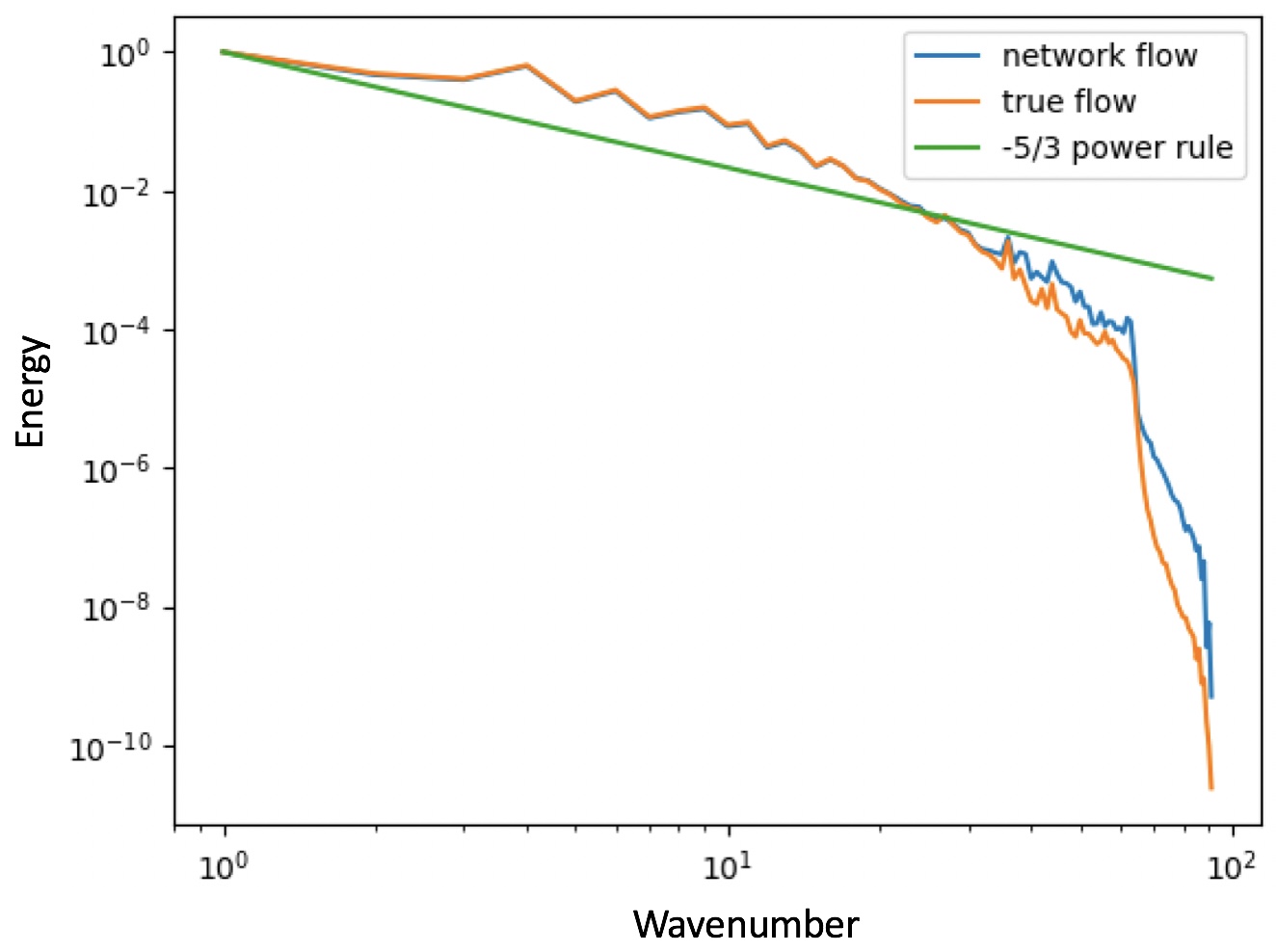}
    \end{center}
    \caption{DNS turbulence vs. ML-generated turbulence.}
    \label{fig:Espectra}
\end{figure}

The error field is computed as the difference between the Truth and Predicted datasets at all points in the domain, and it can be seen that ML model generates very realistic looking turbulence. In order to physically quantify this, the energy spectra for both the DNS and predicted flow is compared, along with the theoretical spectra from Kolmogorov theory. The plot is shown in Figure \ref{fig:Espectra}, and it shows that the predictions exceeded our initial expectations for an ML approach, especially since the flow is fully turbulent. In particular, we can see that low wavenumbers (which correspond to larger scale features) are accurately resolved, while there are some discrepancies in the high wavenumbers, which are representative of smaller scales. 

It is important to note that these are preliminary results obtained with minimal hyper-parameter tuning. The quality of these results and the powerful on-demand GPU resources on AWS demonstrate considerable promise in further exploring this area.
Finally, we also attempted to use AWS SageMaker for automated hyper-parameter tuning. While SageMaker's capabilities seemed very promising, our experience is that it falls short in the range of applicable problems. For instance, SageMaker has its own Python API which has to be used to train machine learning models. Furthermore, SageMaker currently only supports Python 2.7, while we use Python 3.6. This requires a complete rewrite of our existing codes, which is a lot of overhead compared to coding our own hyper-parameter tuning modules. Since SageMaker is a fairly new product, we hope it will mature with time and be more flexible for all types of neural networks.

\subsubsection{Conclusion}
In the course of our work, we gained valuable experience with different capabilties of AWS, the idea of cloud computing in general, and the author's past experience with AWS EC2. A few tips and efficiencies we learned along the way that might be helpful to new users include:
\begin{itemize}
    \item Utilize EBS for large datasets on which we frequently operate. They can be attached and detached
    to various EC2 instances and have low latency. However, it is important to launch an EC2 instance \textbf{in the same subnet} where the EBS volume lives to be able to use it. The default EC2 launch subnet is often different.
    \item Data transfer costs---though cheap and seemingly benign---can quickly eat into budgets for HPC/large simulations.
    \item Spot instances offer heavy discounts (especially for expensive onDemand P3 GPU instances) and are a good choice for non time-critical workloads.
    \item Use Horovod instead of TensorFlow's native distributed computing module. The former is quick to implement and works in a distributed computation with minimum modifications to the serial code.
    \item Deep Learning Ubuntu AMI is extremely useful as the conda environment comes with all useful packages by default.
    \item Explore automation with Python Boto3, which, in our experience, is more seamless with our python code than AWS CLI.
\end{itemize}

Our experience with AWS has been quite positive, mainly because it provides extreme flexibility in choosing our computational resources and architecture based on the task at hand. This may help reduce capital expenditure on hardware for research problems, where the workflow (and hence compute requirements) are constantly changing. We will continue exploring machine learning for hydrodynamics in the LANL ``Machine Learning for Turbulence" LDRD project, and cloud resources have potential to rapidly accelerate the research progress.
\subsection{RF Modulation Classification}

\subsubsection{Contributors}

\begin{itemize}
    \item Bill Junor, ISR-2, bjunor@lanl.gov
    \item Alexei Skurikhin, ISR-3, alexei@lanl.gov
\end{itemize}

\subsubsection{Motivation}

The use of deep neural networks to recognize modulation protocols in radio frequency (RF) time series is a new and rapidly evolving field.  These techniques have been shown to out-perform expert feature searches using multiple decades of experience \cite{RFModulationClassification}. However, training a complex deep neural network on a large dataset can take a significant amount of time. Therefore, accelerated training of neural network models is very important. Cloud computing is ideal for these kinds of scenarios.

Our goal was to investigate the scalability of AWS cloud computing resources for machine learning and getting experience which can be transitioned to develop RF data analysis capabilities in mission-related projects. 

\subsubsection{Solution Approach}

\begin{figure}[t]
     \begin{center}
     \includegraphics[width=6.75cm]{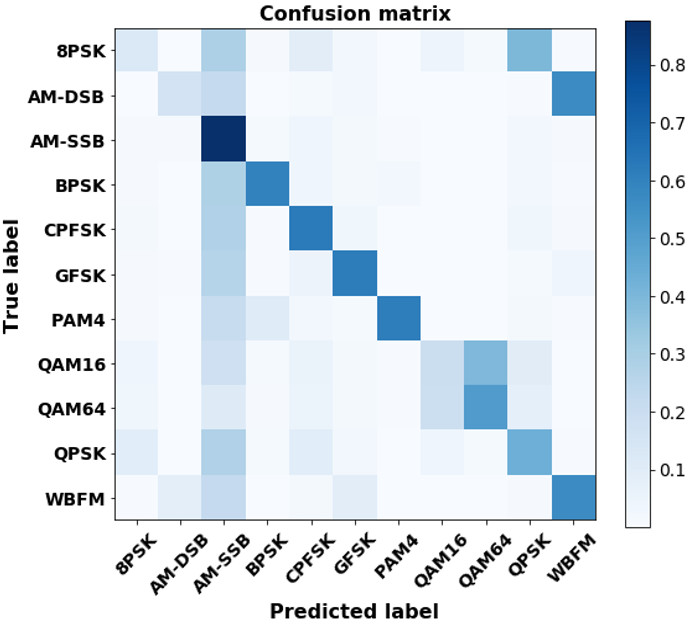}
     \end{center}
     \caption{Results of evaluating CNN classification performance: 11-modulation confusion matrix. }
     \label{fig:Fig1}
\end{figure}

\begin{figure}[h]
     \begin{center}
     \includegraphics[width=8.75cm]{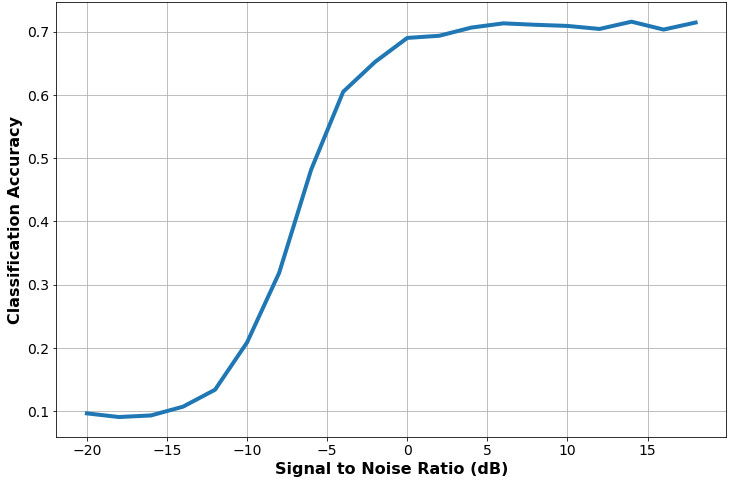}
     \end{center}
     \caption{Results of evaluating CNN classification performance: Classification accuracy vs. SNR levels. }
     \label{fig:Fig2}
\end{figure}

We concentrated on the Amazon SageMaker machine learning platform for RF modulation classification based on a convolutional neural network (CNN). CNN was designed to recognize modulation types in radio signals. We used prototype software written in Python and a dataset prepared by O'Shea et al.  \cite{RFModulationClassification} that is publicly available \cite{RFDataLocation}. The dataset contains complex-valued temporal radio signals corresponding to 11 different modulation schemes (eight digital and three analog modulations), which are often used in wireless communications systems. The 11-modulation dataset has 220,000 examples of temporal radio signals with SNR levels ranging from --20 dB to +18 dB. Each example consists of 128 samples of a complex-valued temporal radio signal. The total dataset is approximately 600 MB. Results of the performance evaluation of the trained CNN are shown in Figures \ref{fig:Fig1} and \ref{fig:Fig2}, including how classification accuracy breaks down across SNRs.

The goal of our experiments was to measure the time-efficiency of scaling neural network training using the SageMaker service which provides access to different AWS Sagemaker notebook instances. We have chosen SageMaker notebook instances because they allow us to quickly build, train, and evaluate machine learning algorithms, including neural networks, using Python and Jupiter notebook.
We performed benchmarks testing of several implementations of CNN classification model with Keras+TensorFlow, Keras+Theano, and Keras+MXNet. Keras is a high-level API to quickly construct and test neural networks using TensorFlow, Theano, or MXNet libraries as back-end. Our experiments were focused on the RF 11-modulation dataset, comparing different non-GPU- and GPU-enabled AWS cloud infrastructures. We built and evaluated the CNN model that consisted of six layers, including input and output layers, and two convolutional layers of 256 and 80 nodes. Total number of CNN trainable parameters were 22,982,747.

\begin{figure}[h!]
     \begin{center}
     \includegraphics[width=10.75cm]{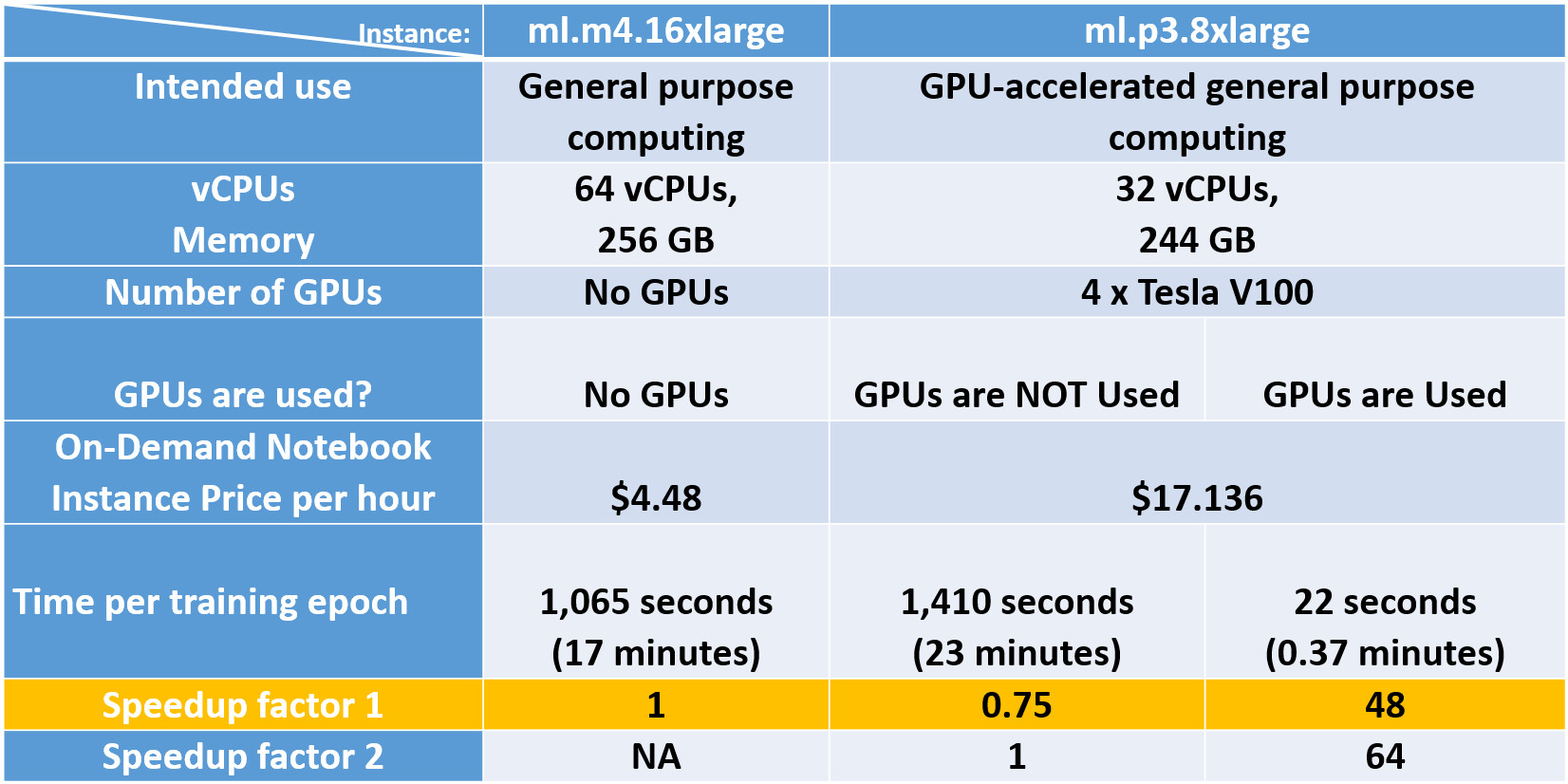}
     \end{center}
     \caption{Comparison of time consumed by one training epoch with Keras+Tensorflow on different AWS instances.}
     \label{fig:Fig3}
\end{figure}

\begin{figure}[h!]
     \begin{center}
     \includegraphics[width=8.75cm]{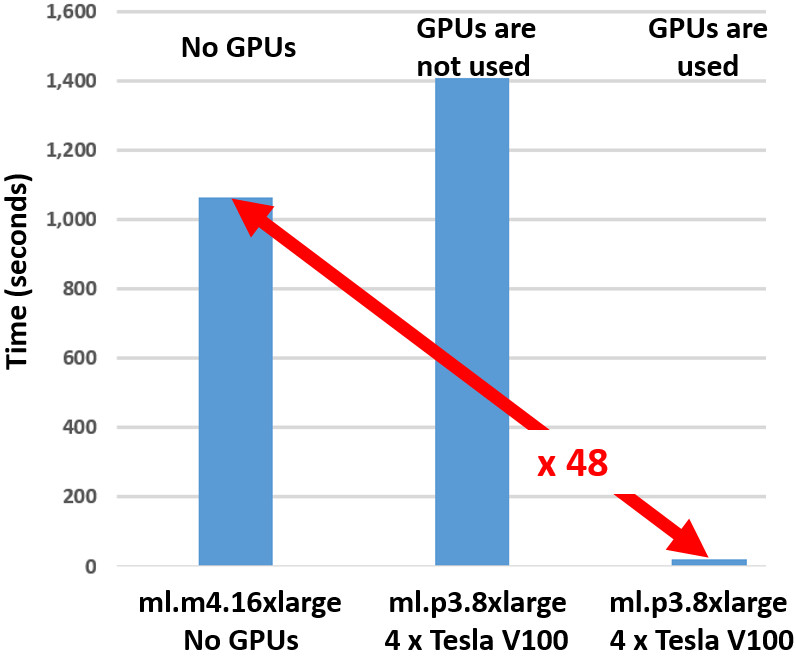}
     \end{center}
     \caption{Neural network training speed-up achieved for one training epoch with Keras+TensorFlow on p3.8xlarge SageMaker notebook instance compared to non-GPU enabled m4.16xlarge instance.}
     \label{fig:Fig4}
\end{figure}

\subsubsection{Results}

Results of our experiments are summarized in Figures \ref{fig:Fig3} and \ref{fig:Fig4}. We have found the use of Keras+TensorFlow for implementation of neural network to be the fastest. Figure \ref{fig:Fig4} illustrates training speedup factors achieved with AWS GPU-enabled SageMaker notebook instance. It is interesting to note that running GPU-enabled instance without the use of GPUs results in slowing down the processing by roughly 25\%.

\subsubsection{Conclusion}

The use of GPU compute SageMaker notebook instances, in particular p3.8xlarge with 4 Tesla V100 GPUs, resulted in neural network training speed-up by a factor of 48 against the use of general compute CPU-based m4.16xlarge instance. In our experiments, the use of GPUs reduced overall neural network training time from several hours to 15 minutes. Such speed up is critical for future applications of RF modulation data analysis that will have large-scale and very dynamic data loads with needs for near-real-time solutions. The use of AWS GPU compute SageMaker instances allows us to quickly optimize and evaluate our machine learning models, as well as process incoming data streams using pre-trained neural network. Our next steps include the use of larger-scale RF datasets and larger (deeper) neural network models of various architectures.

\subsection{Exploring Cloud Computing for Collection and Analysis of Image Data }

\subsubsection{Contributors}

\begin{itemize}
    \item Diane Oyen, CCS-3, doyen@lanl.gov
\end{itemize}

\subsubsection{Motivation}

The world wide web hosts a wealth of information, however finding data on the web often requires massive bandwidth and processing capabilities. We have an ongoing project in which we search for technical information in images that are openly available on the internet. As a demonstration of using cloud computing for image analysis, we analyze images with the goal of identifying images of bicycles and bicycle parts.

Cloud computing offers an ideal solution for this project because massive bandwidth and processing is required on an intermittent basis. Furthermore, because images are being harvested from the internet, there is a risk of inadvertently downloading objectionable or malicious content. The cloud computing platform provides virtual machines that can be spun up and taken down easily, thereby providing a buffer layer, between the open internet and the LANL network, where we can scan images quickly before deciding whether to bring them onto the LANL network.

\subsubsection{Solution Approach}

\begin{figure}[h]
\begin{center}
\includegraphics[width=8.75cm]{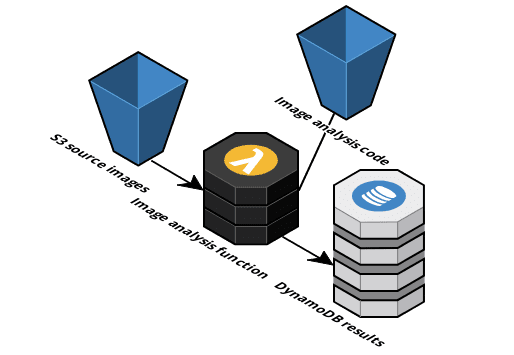}
\end{center}
\caption{Architecture of cloud computing using Lambda serverless functions.}
\label{fig:lambda_arch}
\end{figure}

To spin up processing on-demand and for ease of parallelization of independent processes for analyzing images, the core of our AWS solution is the Lambda serverless function. The image analysis algorithm is defined in a Lambda function. This Lambda function is triggered whenever a new image is copied into a specific S3 storage bucket. The Lambda function performs the function defined by code in another S3 bucket on the new image, then writes out the results to a DynamoDB. Our results are simply labels and confidence scores for each image. A simple spreadsheet or text file would suffice for the results, but because each Lambda operates independently, we cannot write out a centralized result file except through the use of a database. Each Lambda invocation enters an item into the DynamoDB. Figure~\ref{fig:lambda_arch} shows the overall architecture of the system.

For this project, our image analysis code is an ML model that assigns a label to an image. We used the Inception model \cite{szegedy2016rethinking} available on GitHub.com which uses Keras and TensorFlow. TensorFlow is a Python-based deep learning library that facilitates  fast computation of machine learning algorithms \cite{tensorflow2015-whitepaper}. Keras is a Python library that provides wrapper functions and pre-trained models for TensorFlow \cite{chollet2015keras}. For the code to be executed by Lambda, all of the source code---including dependencies---must be packaged in a zip file and uploaded to an S3 bucket. We used the Anaconda virtual environment manager to package source code appropriately. One limitation of the Lambda service is that there is a storage capacity limit on the deployment code of 250 MB. An additional 512 MB of non-persistent disk space can be used for loading additional resources like dependency libraries or data sets during function initialization. This storage capacity is adequate for the Inception model, but some deep learning models approach 1 GB in size.

\subsubsection{Results}

For a comparison against the Inception model image labels, we used the AWS Rekognition service which labels images. Rekognition gives several labels that it finds relevant, along with a confidence score for each label. Inception, on the other hand, gives a ranked list of labels with confidence scores that must sum to 1. For the image we looked at, the top label from Inception has high confidence, and so we only look at the top label.

\begin{figure}[h]
\begin{center}
\includegraphics[width=\textwidth]{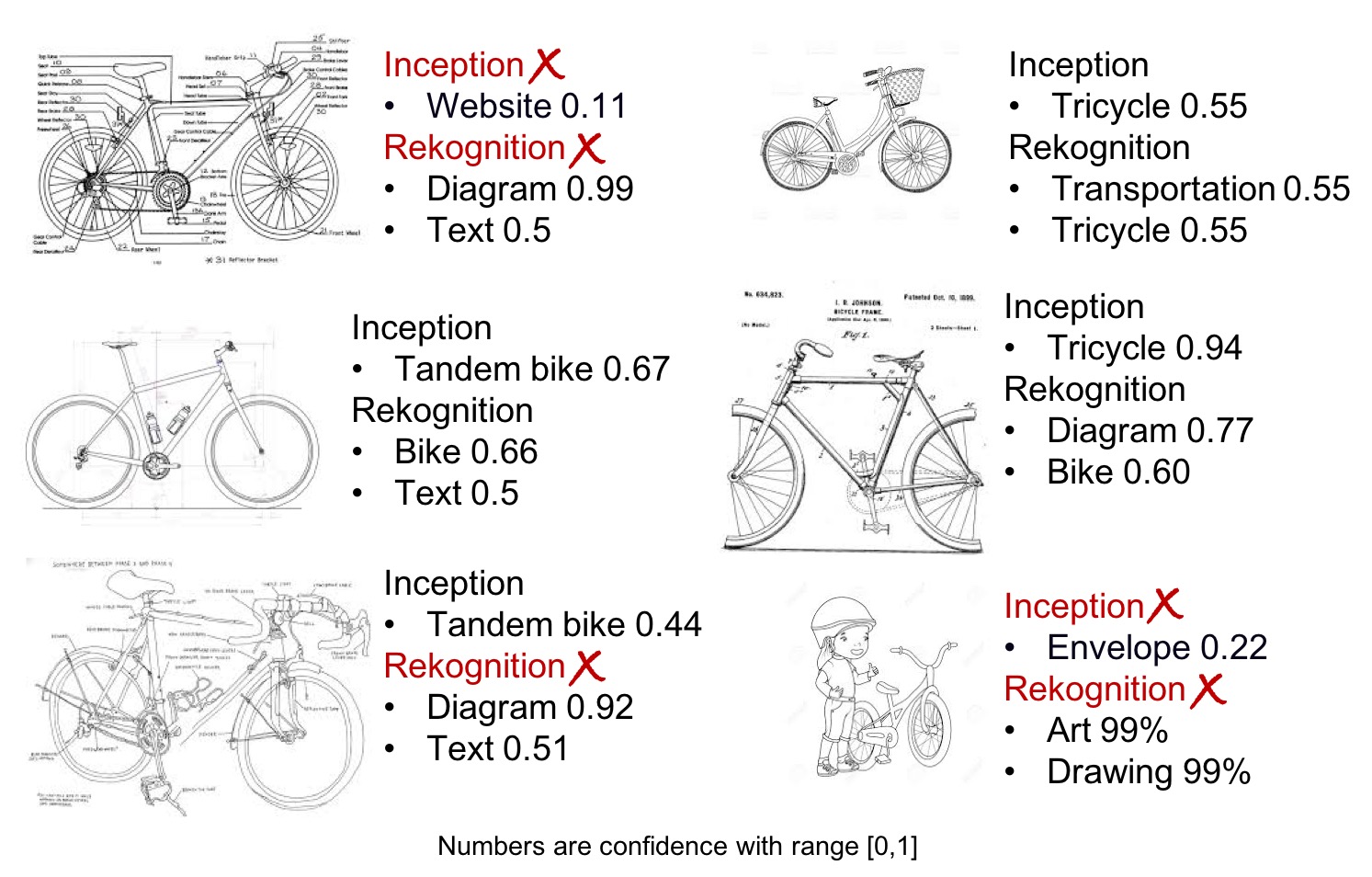}
\end{center}
\caption{Sample results from labeling drawings of bicycles. The machine learning model name is highlighted in red and marked with a red "X" if the label prediction is incorrect. Rekognition is highlighted in red if none of the Rekognition labels are bicycle-related. }
\label{fig:image_label_results}
\end{figure}

Figure~\ref{fig:image_label_results} shows that the image labels often do not identify drawings of bicycles as bicycles. The Inception model was not trained to identify ``bicycle," but is trained to recognize ``mountain bike," ``tricycle," ``tandem bike," and ``unicycle." Therefore, we consider any of those bike-related labels to be correct for this exercise. We consider Rekognition labels to be wrong if none of the labels include a bicycle-like object.

The accuracy of both image labelers are much better when the images are photographs with primarily a single object contained in the photograph.

\subsubsection{Conclusion}

Cloud computing provided us with not only a cost-effective and scalable solution for processing large quantities of images quickly, but also a security buffer that will allow us to prescreen images before bringing them onto the LANL network. Image analysis algorithms that are currently available need to be fine-tuned for our purposes. One problem that we care about is finding diagrams, and so a screening algorithm would filter out all images that are not diagrams.

\subsection{An Autonomous Security System with AWS DeepLens}

\subsubsection{Contributors}

\begin{itemize}
    \item Csaba Kiss, A-2, csakis@lanl.gov
\end{itemize}
\subsubsection{Motivation}

\begin{figure}
\centering
\includegraphics[width=5cm]{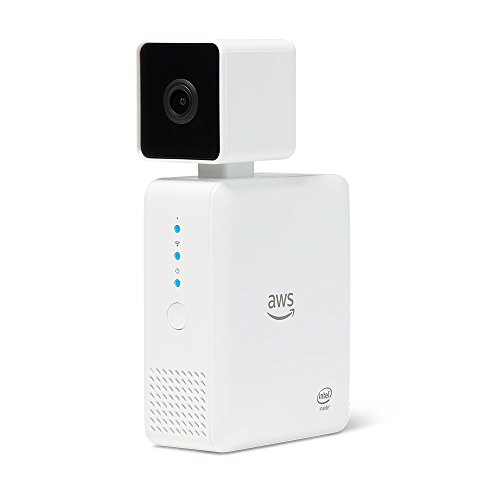}
 \caption{AWS DeepLens device.}
 \label{fig:deeplens}
\end{figure}

My personal motivation for this project was to get started with ML and familiarize myself with the processes, vocabulary, and common language scientists use when they talk about ML. The aim of this proposal was to create an autonomous wildlife monitoring/warning system using Amazon’s deep learning-enabled video camera called DeepLens with the help of multiple AWS services such as S3, EC2, SageMaker, Rekognition, DeepLens, Lambda, and Greengrass. This brief project demonstrates the power of ML for image recognition applying to a practical security problem of recognizing bears and mountain lions. 

 \subsubsection{Solution Approach}
 For this project, I used the ML-enabled computer named AWS DeepLens, illustrated in Figure \ref{fig:deeplens}.
 This device and the DeepLens AWS service are a very good start for scientists and developers on their ML journey.

 At first, I studied how to deploy pre-made models on the DeepLens device. These models are: Artistic Style Transfer, Object Recognition, Face Detection and Recognition, Hot Dog Recognition, Cat and Dog Recognition, Action Recognition, and Head Pose Detection. These pre-built models all use different frameworks, and the source codes are available on github. They are a great resources for learning about machine learning.

\begin{figure}[h]
     \begin{center}
    \includegraphics[width=6cm]{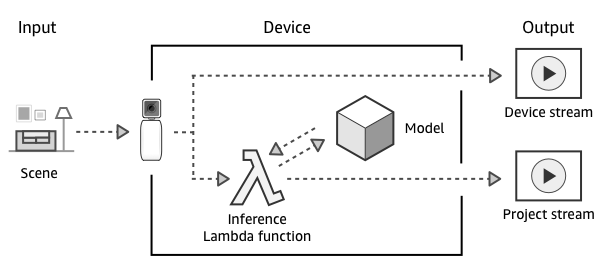}
    \end{center}
 \caption{DeepLens deployment scheme.}
     \label{fig:deeplens_deployment}
 \end{figure}

 My strategy was to use AWS SageMaker to train a model that could distinguish between bears and cougars.  SageMaker also provides plenty of tutorials and sample Jupyter notebooks to get started with machine learning. I chose a \href{https://aws.amazon.com/blogs/machine-learning/classify-your-own-images-using-amazon-sagemaker/} {tutorial} which focused on distinguishing 10 different clothing articles. 
 After training the model on SageMaker, the model needs to be deployed to the DeepLens device. In order to launch the model, a special Lambda function had to be written that publishes the inference result to AWS IoT using the Greengrass service, illustrated in Figure \ref{fig:deeplens_deployment}.
 
\subsubsection{Results}
\begin{figure}
\centering
\includegraphics[width=4.5cm]{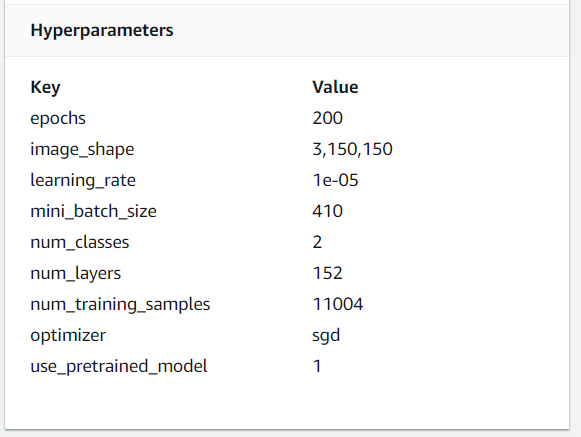}
 \caption{SageMaker training hyperparameters.}
     \label{fig:training}
\end{figure}
 
\begin{figure}[h]
     \begin{center}
     \includegraphics[width=15cm]{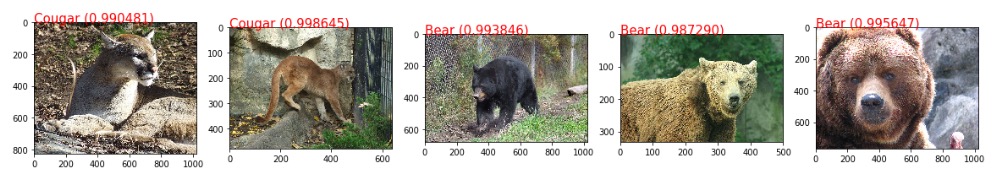}
     \end{center}
     \vspace{-0.5cm}
     \caption{Model testing with API endpoint.}
     \label{fig:testing}
 \end{figure}

\begin{figure}[ht]
     \begin{center}
     \includegraphics[width=6cm]{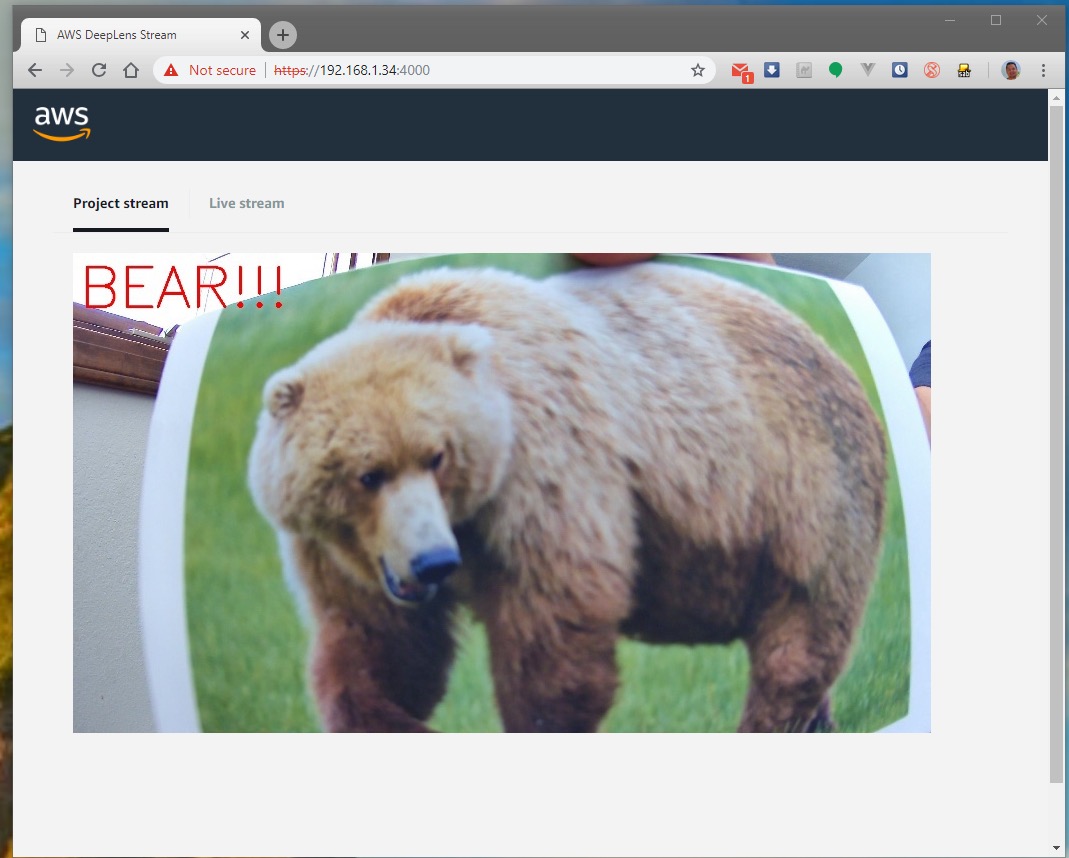}
     \end{center}
     \caption{Inference stream monitored in a browser window.}
     \label{fig:stream}
 \end{figure}

\paragraph{Data Gathering}
In order to perform the training, I needed to gather lots of animal pictures from the internet. I used search engines and photo-hosting websites to collect approximately 8,000 bear and cougar images. The images were hand-filtered for accuracy. 
\paragraph{Model Training}
The SageMaker tutorial used a ResNet-152 model. The images needed to be resized and reformatted to RecordIO format. The dataset was randomly divided into a training (0.8)/testing set (0.2). Training was performed with the hyperparameters on Figure \ref{fig:training}.
\paragraph{Model Deployment and Testing}
First the model was deployed as an API endpoint using SageMaker. SageMaker makes training and deployment as easy as clicking a few buttons. The deployed model was tested with five random images. The result of the test can be seen on Figure \ref{fig:testing}.
\paragraph{Model Deployment on DeepLens}
DeepLens is able to host the model and the inference code as a standalone computer. I used AWS Greengrass and a Lambda function to deploy the model and the code to the device. Once deployment is complete, inference is automatically started as an infinite loop. The result of the inference streamed from the device can be monitored in a browser window, as seen in Figure \ref{fig:stream}.
 
\subsubsection{Conclusion}
At the beginning of this project, I had no experience with ML and was reluctant to even think about projects implementing ML. By the end of this short project, I have become much more comfortable understanding the common language used by scientists and developers when they discuss ML. I also learned how to gather and reformat the data necessary for deep neural network models and learned how to train and deploy custom models on SageMaker and DeepLens. I successfully trained a ResNet-152 model to distinguish bears and cougars and deployed it onto a standalone AWS DeepLens device.

\subsection{Smart Machine-Learning Workflows in the Cloud}

\subsubsection{Contributors}

\begin{itemize}
    \item John Ambrosiano, A-1, ambro@lanl.gov
    \item Nidhi Parikh, A-1, nidhip@lanl.gov
\end{itemize}

\subsubsection{Motivation}

In today's world, an unprecedented amount of data is being collected every second, but extracting useful information out of it is challenging. Machine learning approaches can be used to extract quantitative and meaningful information out of this data. However, machine-learning (ML) analyses can be very complex. There are typically several steps that must be applied to process raw data before any learning methods can be applied, and several learning methods may be used at different stages in an ML workflow, or in parallel, to arrive at a consensus. Constructing ad hoc workflows for machine learning increases the overall time to apply such methods to any given problem and can make it awkward to extend or reuse successful analysis workflows in new problems.

The smart workflow concept is one where the semantics of the workflow are understood by the system through semantic models. The system is aware of each available learning component, its type, input requirements, expected output, computational cost, and so on. The semantics of any constructed workflow are also known, including initial state, data dependencies, and process sequence. Smart workflows can, in principle, be compared, extended, and reused in other problems, as well as within larger workflows.

Our goal was to use AWS to implement a test workflow application for nowcasting disease using internet data sources and evaluate which data source leads to better nowcasting.


\subsubsection{Solution Approach}

\begin{figure}[htp]
\begin{center}
\includegraphics[width=0.9\textwidth]{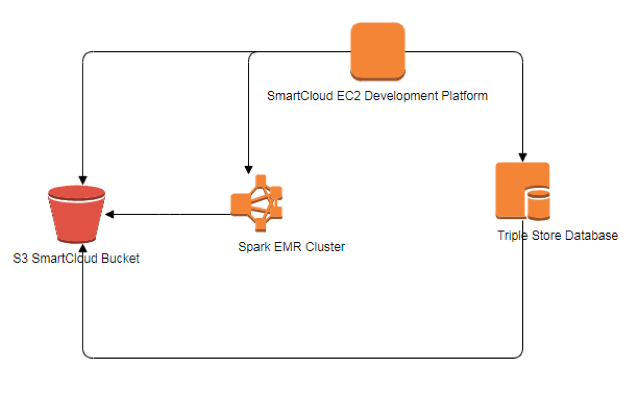}
\caption{Cloud architecture.}
\label{fig-cloud-architecture}
\end{center}
\end{figure}

Figure \ref{fig-cloud-architecture} shows the cloud architecture for our project. We used a t2.large EC2 instance for development, an S3 bucket for data storage (including 745 GB of Twitter data), a Spark EMR cluster with variable size (from 2 to 16 m4.large instances as needed) for the ML pipeline, and a GraphDB standalone instance on the t2.large for the semantic model.

\paragraph{Machine Learning Workflow and Pipeline}

\begin{figure}[htp]
\begin{center}
\includegraphics[width=1\textwidth]{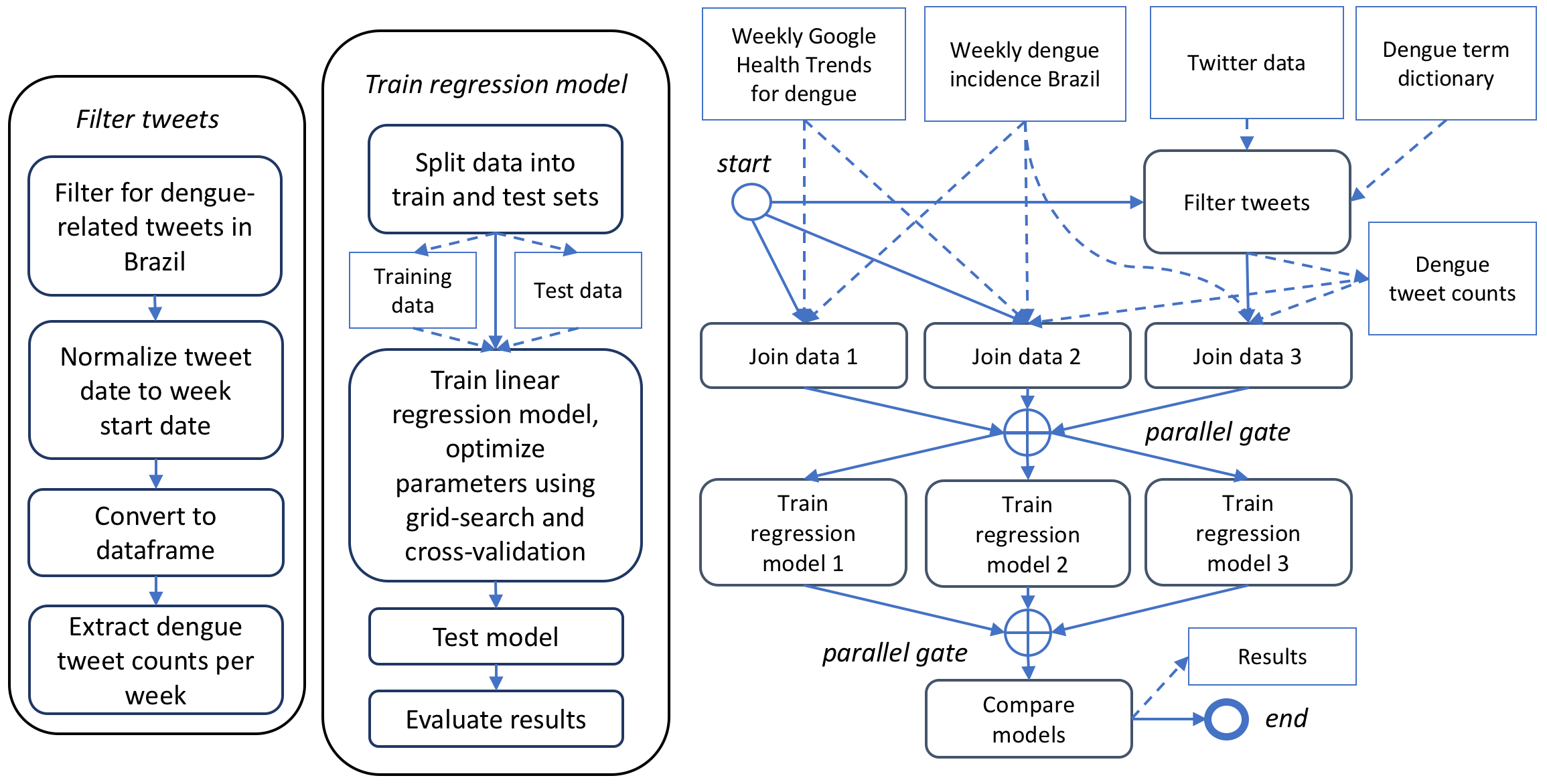}
\caption{Machine learning pipeline and workflow.}
\label{fig-ml-workflow}
\end{center}
\end{figure}

We developed a machine learning pipeline and workflow for nowcasting dengue incidence for Brazil using Google Health Trends (GHT) and Twitter data and evaluate which data source (or their combination) leads to better nowcasting as shown in Figure~\ref{fig-ml-workflow}. 

The data about dengue incidence in Brazil at weekly interval was obtained from the Ministry of Health in Brazil. GHT provides a measure of how much the given term is searched in the given geographic region. We used various dengue-related terms identified from an earlier study and downloaded GHT data for these terms at the country-level and weekly interval for Brazil using GHT API~\cite{GoogleHealthTrends}. Twitter data was obtained using Twitter search API~\cite{TwitterSearchAPI} which provides 1\% of all tweets. We used tweets collected for 2014 and 2015. 

Since the Twitter data contains 1\% of all tweets (including non-dengue-related tweets around the world), the first step in the workflow was to filter this data. We used dengue-related terms to filter tweets in Brazil. Consistent with the standard practice of forecasting/nowcasting at the weekly interval in epidemiology, we counted dengue-related tweets in Brazil per week. Since this was a big data problem (with 745 GB of Twitter data), we increased the size of the Spark EMR cluster to 16 instances, and it took about 4.7 hours to process.

We used linear regression with L1 and L2 regularization to nowcast dengue. We built and compared three models using different internet data: (1) using Twitter data (2) using GHT data (3) using Twitter and GHT data. First, for each model, the dataset is split into training and test datasets. The training dataset is used to train the regression model and find optimal parameters using grid search and cross-validation. The model with the optimal parameters is then tested on the test dataset. Finally, all regression models are compared.

\paragraph{Machine Learning Workflow Ontology}

One of the main ideas behind this project was to explore the concept of a ``smart workflow" for machine learning. ``Smart” means mapping an ML pipeline or workflow implemented in the cloud to a semantic model or ontology. The semantic model would serve two main roles:

\begin{enumerate}
	\item Facilitate or automate the construction of a workflow. Since the semantic model describes the workflow in detail in a machine-interpretable form, it should be possible for the system to instantiate the workflow in the cloud from the semantic representation. A demonstration of this capability was beyond the scope of the project.
    \item Preserve workflow details. The idea here is to capture the specific details of the workflow as metadata, so that the model can be queried after the workflow has been executed. These details would include specification of computational components and their relationships, as well as the origin and disposition of the data consumed and produced by the workflow. The model can also specify which cloud resources were used to execute the workflow and could record performance.
\end{enumerate}

\subsubsection{Results}

In this section, we summarize our results. First we describe results from the machine-learning workflow instantiated and executed on the EMR cluster. Then we describe the semantic model of that workflow and suggest how it could be integrated with an executable ML workflow to capture details of the workflow in relation to portions of the architecture on which it was executed, and later queried for these details.

\begin{figure}[htp]
\begin{center}
\includegraphics[width=0.8\textwidth]{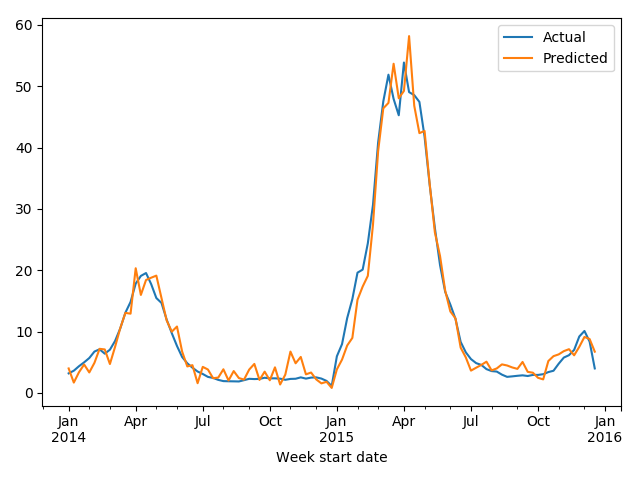}
\caption{Actual and predicted dengue incidence for the best regression model (using Twitter and GHT data).}
\label{fig-ml-results}
\end{center}
\end{figure}

\paragraph{Machine Learning Results}
\begin{table}
\caption{Machine Learning Results}
\begin{center}
\begin{tabular}{c c c c}
\hline
Measure & Twitter & GHT & Twitter + GHT \\
\hline
RMSE train & 10.3443 & 2.5788 & 2.2824 \\
RMSE test & 18.2959 & 3.2906 & 2.3355 \\
$R^2$ train & 0.3181 &  0.9630 & 0.9722 \\
$R^2$ test & -0.0080 & 0.9478 & 0.9673 \\
\hline
\label{table-ml-results}
\end{tabular}
\end{center}
\end{table}
Table~\ref{table-ml-results} shows performance for all regression models. The model trained only on Twitter data had the worst performance. We believe, this is because of the small number (approximately 4600 tweets for 2014 and 2015) of dengue-related tweets in Brazil collected. This happened because the data was collected for another project that did not specify search criteria and could be fixed by specifying dengue-related terms as keywords and the geographic region as Brazil to Twitter search API. The model trained on GHT data performed much better with $R^2$ value of $0.9478$ on the test dataset, but the model trained on both Twitter and GHT data had even better performance with $R^2$ value of $0.9673$ on the test dataset. The predicted dengue incidence for the best model is quite close to the actual incidence as shown in Figure~\ref{fig-ml-results}.

\paragraph{Semantic Model}

\begin{figure}[tp!]
\begin{center}
\includegraphics[width=0.7\textwidth]{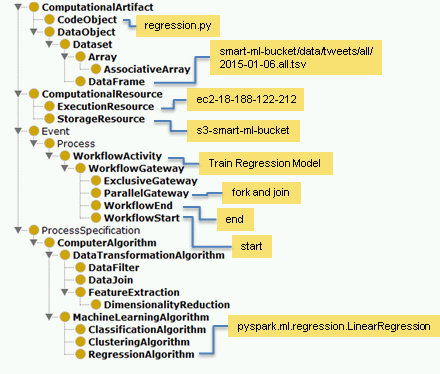}
\caption{A notional machine-learning workflow ontology.}
\label{fig-ml-wf-onto}
\end{center}
\end{figure}

For the purposes of this project, we created a notional workflow ontology based on prior work at LANL to describe workflows in high-performance computing. The class hierarchy for the workflow is shown in Figure~\ref{fig-ml-wf-onto}.
\begin{figure}[tp!]
\begin{center}
\includegraphics[width=0.7\textwidth]{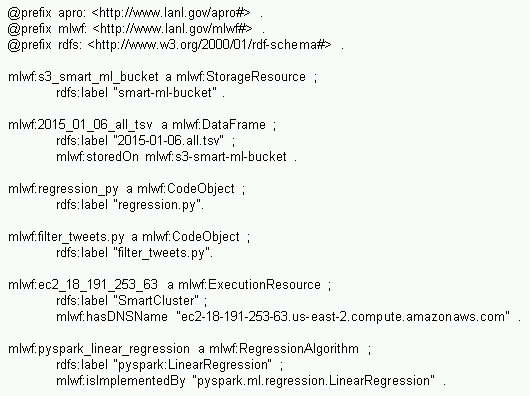}
\caption{RDF triples for artifacts of the workflow written in the Terse Triple Language (TTL) called ``Turtle."}
\label{fig-art-triples}
\end{center}
\end{figure}
At an abstract level, the ontology describes the flow of processes, which are a type of event, and process specifications, which describe what happens when a process occurs. For workflows, a WorkflowActivity is a kind of process. To control the flow, we also defined special control-related activities, often called ``gateways" which were designated by the class WorkflowGateway. Common gateways include those for start and end, as well as parallel gateways. Parallel gateways like forks and joins are usually considered to be inclusive---that is, the collection of flows all executed in parallel. There are also exclusive gateways in which only one flow is selected based on logical conditions.

Machine learning (ML) algorithms appear as process specifications. Here we see familiar categories such as classification, clustering, and regression. There are other process specifications that are common in data preprocessing steps such as filters and joins.
The activities or tasks in the workflow generally consume and produce data in the context of computing and are enabled by storage and execution resources. In-cloud computing storage resources are things like databases and S3 buckets, while execution resources are elastic computing nodes and clusters. These resources perform their functions on data objects and code objects respectively.
In this way the ontology is able to capture details about how a particular ML workflow was performed on a particular cloud architecture.

For the purposes of illustration, the figure has been annotated with details, using examples from the ML workflow previously described. These details can be captured as RDF (Resource Description Framework) triples and stored in a special database called a ``triple store." To illustrate this, we captured triples related to the ML workflow performed in this project and stored them in a triple store. The triples include specifications on selected artifacts of the workflow. These are shown in Figure~\ref{fig-art-triples}.

\begin{figure}
\begin{center}
\includegraphics[width=0.6\textwidth]{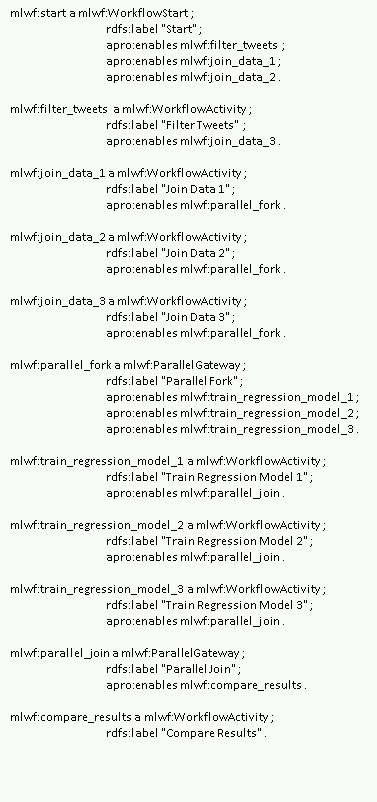}
\caption{RDF triples describing the workflow activities.}
\label{fig-flow-triples}
\end{center}
\end{figure}

The actual flow of the workflow and its pipeline are described by other triples shown in Figure~\ref{fig-flow-triples}. Once the triples have been stored, they can be queried using a standard query language called ``SPARQL." In Figure~\ref{fig-flow-triples}, we have performed a query for all workflow activities based on the relationship that binds them---an object property called ``enables." The resulting graph is shown in Figure~\ref{fig-flow-triples}, which reproduces the flow described in the previous section.

\begin{figure}
\begin{center}
\includegraphics[width=0.8\textwidth]{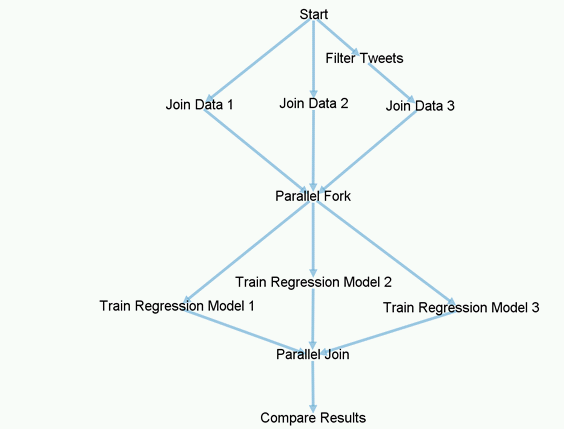}
\caption{The flow of activities reconstructed from a SPARQL query of the semantic model.}
\label{fig-flow-graph}
\end{center}
\end{figure}

\subsubsection{Conclusion}

The project described in this section was intended as a challenge to motivate learning how to build an AWS architecture for big data. In particular, we learned how to build and work with EMR instances to solve a real-world machine-learning problem. We found the ability to easily instantiate clusters of varying sizes and performance characteristics, already fully-equipped with the Spark ML environment, extremely helpful.

A further aim was to explore the idea of a ``smart" workflow based on using semantic technology to capture and query metadata for complex ML pipelines and workflows. We created a notional ML workflow ontology and used it to model such a workflow. We were further able to import the model to an instance of the open-source database GraphDB, which installed on a t2.large EC2 instance. However, integrating the semantic model in some automated fashion with the ML workflow fell out of scope given the resources available. We welcome the opportunity to continue these explorations and would especially like to test the AWS Neptune platform for managing semantic data in this context.

\vspace{2.0cm}
\subsection{Deep Learning Image Segmentation on Cloud Computing}

\subsubsection{Contributors}

\begin{itemize}
    \item Phil Romero, HPC-ENV, prr@lanl.gov
\end{itemize}

\subsubsection{Motivation}

Nuclear forensics applies morphological analysis for materials \cite{mama} in an attempt to resolve particle properties from images obtained from microscopes.  The properties to resolve include the pixel area of particles, their circularity, their convexity (area/perimeter), their ellipse aspect ratio, and their orientation axis.  An expert can take 2,000 hours to annotate a single image produced by a microscope.  The motivation is to capture the expertise of an individual expert in producing these annotations to produced annotations of new images.

\subsubsection{Solution Approach}

\begin{figure}[t]
     \begin{center}
     \includegraphics[width=12.75cm]{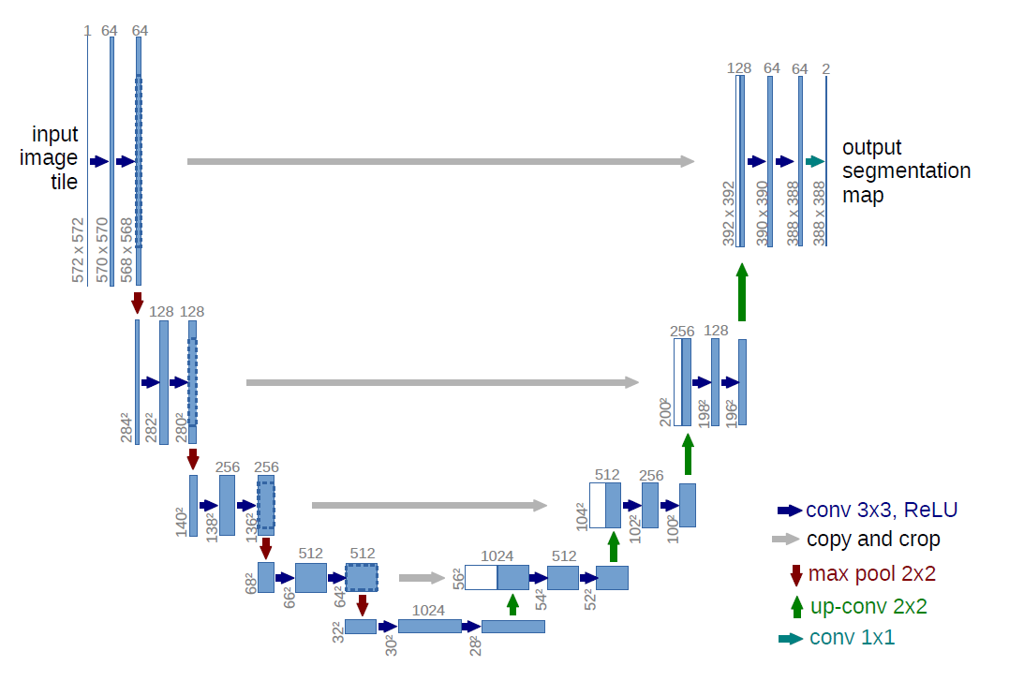}
     \end{center}
     \caption{U-Net is a fully convolutional neural network developed for use in biomedical image segmentation.  It includes both contracting and expanding paths to produce an output segmentation from an input image.}
     \label{fig:Unet.png}
\end{figure}

 More than 40 annotations of images have been collected at this time.  These images can then be used as ground truth image segmentations to aid in producing image segmentations from new images.  This will drastically decrease the time necessary to produce the valuable image segmentations.  It will also attempt to capture the expertise of the individual producing the ground truth image segmentations by applying deep learning to the problem.  A specific type of deep neural network that is applied to this problem is called U-Net \cite{unet}.  U-Net is a fully convolutional neural network that includes both contracting and expanding paths to produce an output segmentation image from an input image, this is shown in Figure \ref{fig:Unet.png}.  In short, deep learning is used to find a pattern that an expert applies to an image to produce an image segmentation.  
 
Cloud computing was leveraged to train the deep neural network from input images and subsequently used to provide image segmentations from many more images.  Amazon SageMaker \cite{sagemaker} was utilized as a platform to run TensorFlow \cite{tensorflow} on source code written in Python and process more than 50 GB of data.  TensorFlow is a widely used open source software library that provides support for machine learning and deep learning.  

The data was uploaded with the AWS command line application to an S3 bucket.  The code was input into a Jupyter notebook and run on an instance of ml.p3.2xlarge.  This is an NVIDIA V100 \cite{v100} based general purpose graphics processing unit (GPGPU).  Processing on a GPGPU is much more efficient than processing on standard CPU-only based hardware.

\begin{figure}[th!]
     \begin{center}
     \includegraphics[width=12.75cm]{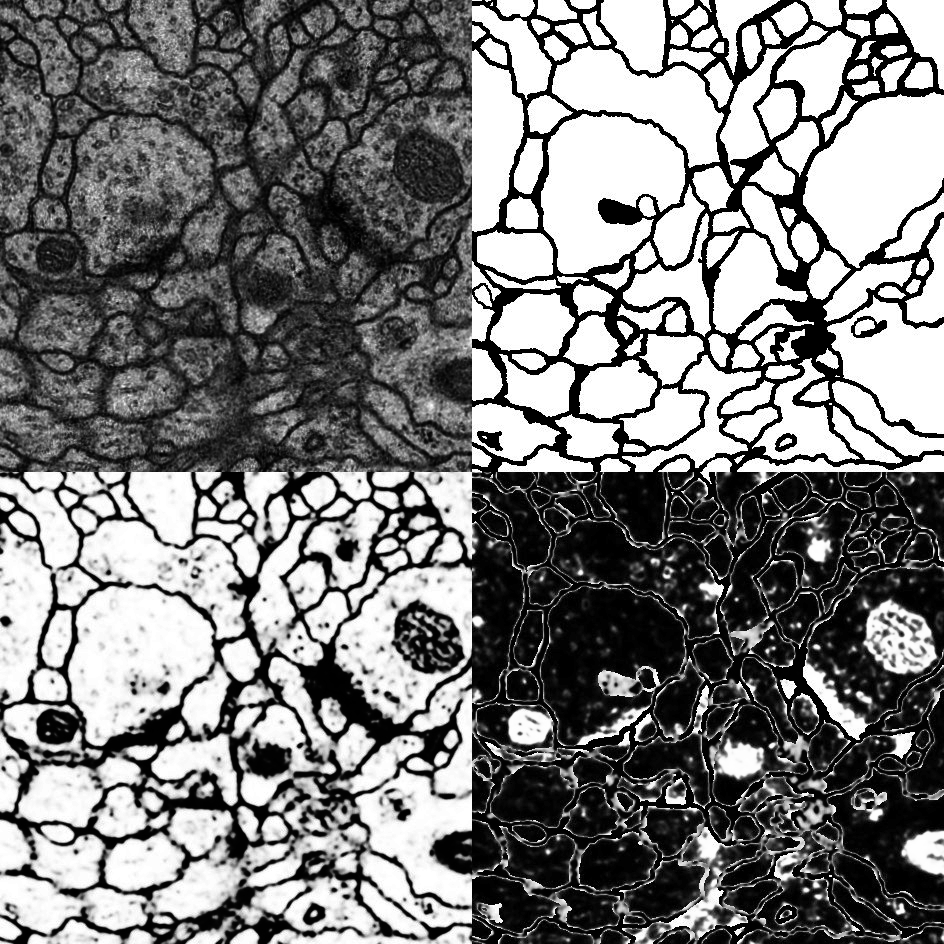}
     \end{center}
     \caption{There are four images above, the upper left is the raw input image from a microscope, the upper right is the ground truth image segmentation done by an expert from the raw image, the lower left is a U-Net produced image segmentation, and the lower right subtracts the U-Net produced image segmentation from the image segmentation produced by an expert.}
     \label{fig:CombinedImage.png}
\end{figure}

\subsubsection{Results}

A major challenge was finding a way to get more than 50 GB moved into an S3 bucket.  The web interface would not successfully transfer the data, so the AWS command line interface utility was installed on a Linux computer, a Mac OS X computer, and a Windows 10 computer.  The Linux version of the AWS command line interface was installed through a package manager and would not work due to it's inability to recognize the us-east-2 region in Ohio.  Upgrading probably would have fixed this; it is only mentioned here because it is odd for a syncing utility to have the limitation of only contacting certain destinations that are provided within the software.  Both the Mac OSX version and the Windows 10 versions worked as designed.    

Once the data was moved, the standard SageMaker limitations on instances would not allow for the completion of training the deep neural network.  It took several days to approve and implement the V100-based instance in part due to a confusion between instances available on EC2 and SageMaker.  It is important to note that there are two differing hardware limitations listings, otherwise the change would have been implemented in a matter of a day or two at most.  

Since SageMaker is a relatively new service, I expect its limitations will be decreased as the service matures.  One particular limitation I expect to be alleviated is that certain versions of TensorFlow would not work with the S3 storage system.  This severely limits SageMaker as TensorFlow tends to be upgraded frequently and code tends to be associated with certain versions, it is not atypical that code will not work if attempting to use other versions of TensorFlow.  

A sample of an image segmentation is shown in Figure \ref{fig:CombinedImage.png}.  The upper left image is the raw image obtained from a microscope, the upper right image is the ground truth image segmentation produced by an expert, the lower right image is the U-Net produced image segmentation, and the lower right image is the ground truth image segmentation minus the U-Net-produced image segmentation.  Notice the differences where the ground truth segmentation differs from the U-Net segmentation.  One reason for the difference is that the U-Net image segmentation includes probabilities in the form of utilizing a full range of pixel values ranging from white to black. It should be noted that the image on the lower right would be completely black if there were no differences in the image segmentations.  Part of the discrepancy exists due to the grey scale inclusion of pixel values (as representations of probabilities), and some of the other errors are minor boundary differences that occur around the lines of the image segmentations.  Other discrepancies arise from how an expert would produce the segmentation.  As a whole, this is a great improvement of the process as an expert could start with a pre-segmented image and have less than 2\% of the pixels to adjust/verify in order to produce a new ground truth image segmentation.

\subsubsection{Conclusion}

Amazon SageMaker services were utilized to produce image segmentations of microscope images depicting particles by utilizing deep learning.  The process drastically speeds up the process necessary as input to a quantification process.  Although SageMaker is a relatively new service, it shows promise in providing a platform for deep learning applications.  As the service matures, I would expect it can be utilized with lesser concern for the details of the problems including what versions of software are needed and complexity of the code that needs to be modified/restructured.

\subsection{Cloud-based Image Analysis}

\subsubsection{Contributors}

\begin{itemize}
    \item Brad Wolfe, P-25, bwolfe@lanl.gov
    \item Zhehui Wang, P-25, zwang@lanl.gov
    \item Pinghan Chu, P-23, pchu@lanl.gov
\end{itemize}

\subsubsection{Motivation}
Image data can capture complex physical phenomena, which makes imaging a viable tool for diagnostics and analysis. One issue with image data is that conventional tools to analyze image data require manual intervention. This leads to large amounts of image data not being analyzed. Unsupervised machine learning allows for physically relevant features of an image to be extracted automatically. However, training these methods typically requires parameter tuning and can be computationally expensive. Effective training of algorithms requires GPUs which can be quite expensive. HPC systems such as LANL's Kodiak cluster can accommodate these needs, although access to resources is competitive. Cloud computing allows for fast and inexpensive access to multi-GPU architectures, which means training of machine learning algorithms for physical image analysis can be easily scaled.

\subsubsection{Solution Approach}
The main frameworks outside of AWS used were TensorFlow, Keras, and Docker. TensorFlow is a Python library for building ML models and performing general mathematical computing. The library provides implementation of commonly used operations such as gradient descent, automatic differentiation, and convolution. TensorFlow supports both the usage of multi-CPU and multi-GPU architectures and provides visualization tools such as TensorBoard. Keras is a Python library for building neural networks and can be built on top of TensorFlow. Docker is a tool used to make containerized applications. Containers are similar to virtual machines since they are both methods of virtualization. However, containers do not require their own operating system, instead they use the resources located on the host machine. 

\begin{figure}[t]
    \centering
	\includegraphics{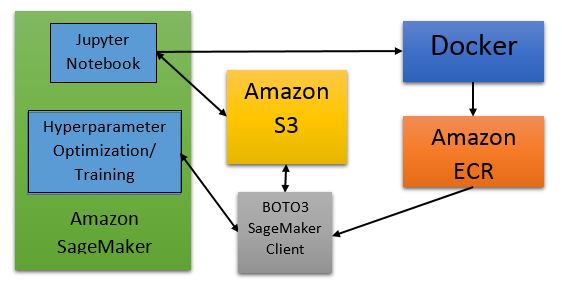}
	\caption{Workflow for designing and using new models in SageMaker on AWS.}
	\label{fig:cloud_chart_sagemaker}
\end{figure}

Our cloud computing structure is mostly set up around SageMaker (see Figure  \ref{fig:cloud_chart_sagemaker}). Though SageMaker provides multiple out-of-the-box methods, it does not provide image-to-image models such as representative and generative models. Thus, models that are not specified in SageMaker require extra services to implement. First, SageMaker's Jupyter notebook instances allow for easy design and simple testing of ML code. Then the model code can be built into a Docker container. This allows for the custom model to be easily implemented in SageMaker. An image of the docker container is then made and pushed to Amazon's Elastic Container Repository (ECR). After an image of the model is available, the parameters of the model are tuned by initializing a SageMaker hyperparmeter tuning job. Finally, SageMaker can start a training job to effectively train the model. Data for this training process is stored using an S3 bucket, which means that the implementation of the model and the actual data can be kept separate. This has the benefit of being able to easily swap data or models. 

\subsubsection{Results}
\begin{figure}[t]
    \centering
	\includegraphics[width=.5\textwidth]{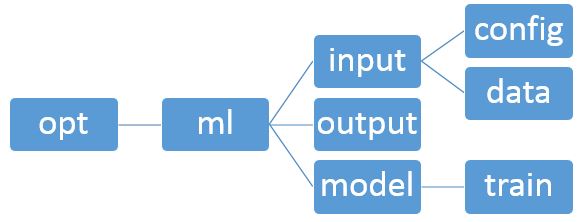}
	\caption{File system structure inside Docker container that can run in SageMaker.}
	\label{fig:file_tree_structure}
\end{figure}

\begin{figure}[t]
    \centering
	\includegraphics[width=.5\textwidth]{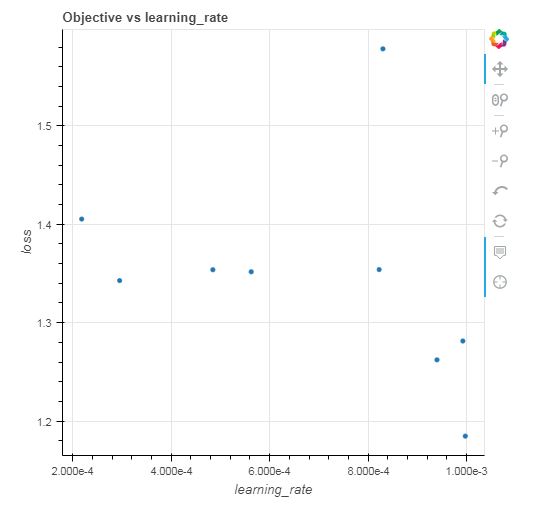}
	\caption{Hyper-parameter optimization for a custom model.}
	\label{fig:hopt}
\end{figure}

The main challenge of using SageMaker was that the platform is designed around the stock ML algorithms. Though SageMaker's API does provide support for common frameworks, such as TensorFlow, this support is only available for older Python 2 versions. This is the main reason why the use of Docker containers was necessary. One issue with this is that the file structure of the Docker container must follow a certain structure (Figure \ref{fig:file_tree_structure}). This is due to the fact that certain directories in SageMaker are reserved, for example training data is always stored in the /opt/ml/input directory. This means moving models into SageMaker may require parts of code to be adapted. Another issue is that the size of the Docker container affects the start-up time of SageMaker. This can make projects with large amounts of code/executables a bit prohibitive, however in our case a base image for Keras with a TensorFlow backend is around 2 GB, which does not add much to the start time. After the image is produced, it was able to be pushed to ECR. There is one extra step besides a push command. Docker required credentials to move the image to the ECR repository. These credentials were easy to obtain using the AWS-CLI tool. Once this was completed, starting the hyperparameter tuning job on the custom model was simple. Boto3's SageMaker client provides an object for the hyperparameter tuning job. All it requires is credentials for the account, the location of docker image, and specifications for the tuning job. The specifications include settings such as which optimization method to use, number of instances to use, and ranges for the hyperparameters. These should be given in a JSON format. Figure \ref{fig:hopt} shows usage of SageMaker's hyperparameter tuning tool.

The second challenge for this project was obtaining access to GPU instances. When using deep neural networks and convolutional models, GPUs are essential for speeding up computation. However, initially there were no available GPU instances. This was true for training/hyperparameter optimization, Jupyter Notebook, and EC2 instances. This was solved by requesting a limit increase, however this has to be done for each type of instance. This caused some slight difficulties at the start but was not an issue once the limit increase occurred. For scalability this provides an issue for short term use, though for long term use this should likely be less of an issue.

\subsubsection{Conclusion}
Using Jupyter Notebook instances and ECR, we were able to design and build custom image-to-image ML models that are not provided by SageMaker. We were able to utilize hyperparameter tuning instances to choose model parameters without performing grid searches or manually guessing values. This shows that AWS can serve as a platform for training ML algorithms. Due to the resource limits placed on the service, it seems unlikely that this will be a replacement for HPC resources such as the Kodiak cluster. However, the affordability of instances provides easy access to a limited number of GPUs. This is more economically viable than groups purchasing their own equipment which can run into the range of tens of thousands of dollars.

\clearpage
\section{Software Deployment}
\label{sec:deploy}

\subsection{Virtual Reality (VR) and Augmented Reality (AR) Applications}

\subsubsection{Contributors}

\begin{itemize}
    \item Donovan Heimer, NEN-3, dheimer@lanl.gov
\end{itemize}

\subsubsection{Motivation}

Conventional 3D application deliverables have large file sizes making delivery by email impossible. Other delivery methods, such as large file transfer, only allow access to the deliverable for a limited period of time. If the user loses the original deliverable, the user must contact our team and wait for re-delivery. Another major challenge involves hardware expenses and limitations. 3D applications that run from local executables consume sufficient hardware resources. Before the application is built, assets need to be developed, and these assets consume significant storage space. Additionally, for VR applications, different clients may have already invested in different VR hardware. Another technical challenge involves verbal communication of information. Text-to-speech and voiceover are both options for resolving this challenge, but these solutions are respectively clunky and time consuming.

\subsubsection{Solution Approach}

\begin{figure}[htp]
\begin{center}
\includegraphics[width=0.9\textwidth]{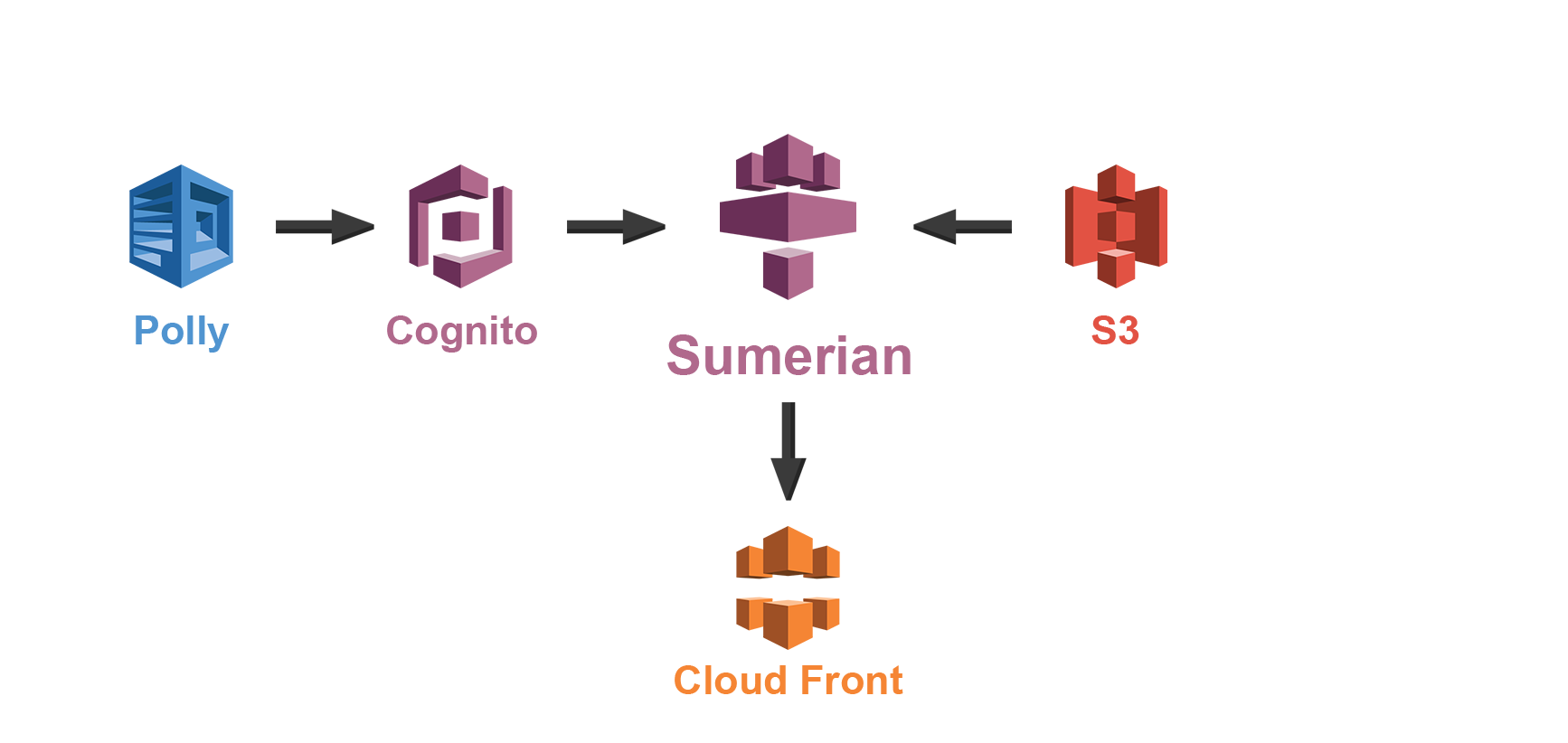}
\caption{AWS tools.}
\label{VRAWSToolsDiagram}
\end{center}
\end{figure}

The solution involved developing an application that generates a virtual hands-on training environment, where the user interacts with the environment via web browser. The application and all of its data were stored on the AWS Sumerian Binary Storage and users can access the training through a URL link run on AWS CloudFront. Amazon Polly was also used for text-to-speech features in the training application. AWS service such as Cognito and S3 were used to connect the aforementioned AWS services. See Figure \ref{VRAWSToolsDiagram} for a diagram of AWS services used for this project. Non-AWS platforms were used for asset generation, such as the Foundry’s Modo for 3D models and animation and Adobe Photoshop for 2D texture development.

\subsubsection{Results}

\begin{figure}[htp]
\begin{center}
\includegraphics[width=0.9\textwidth]{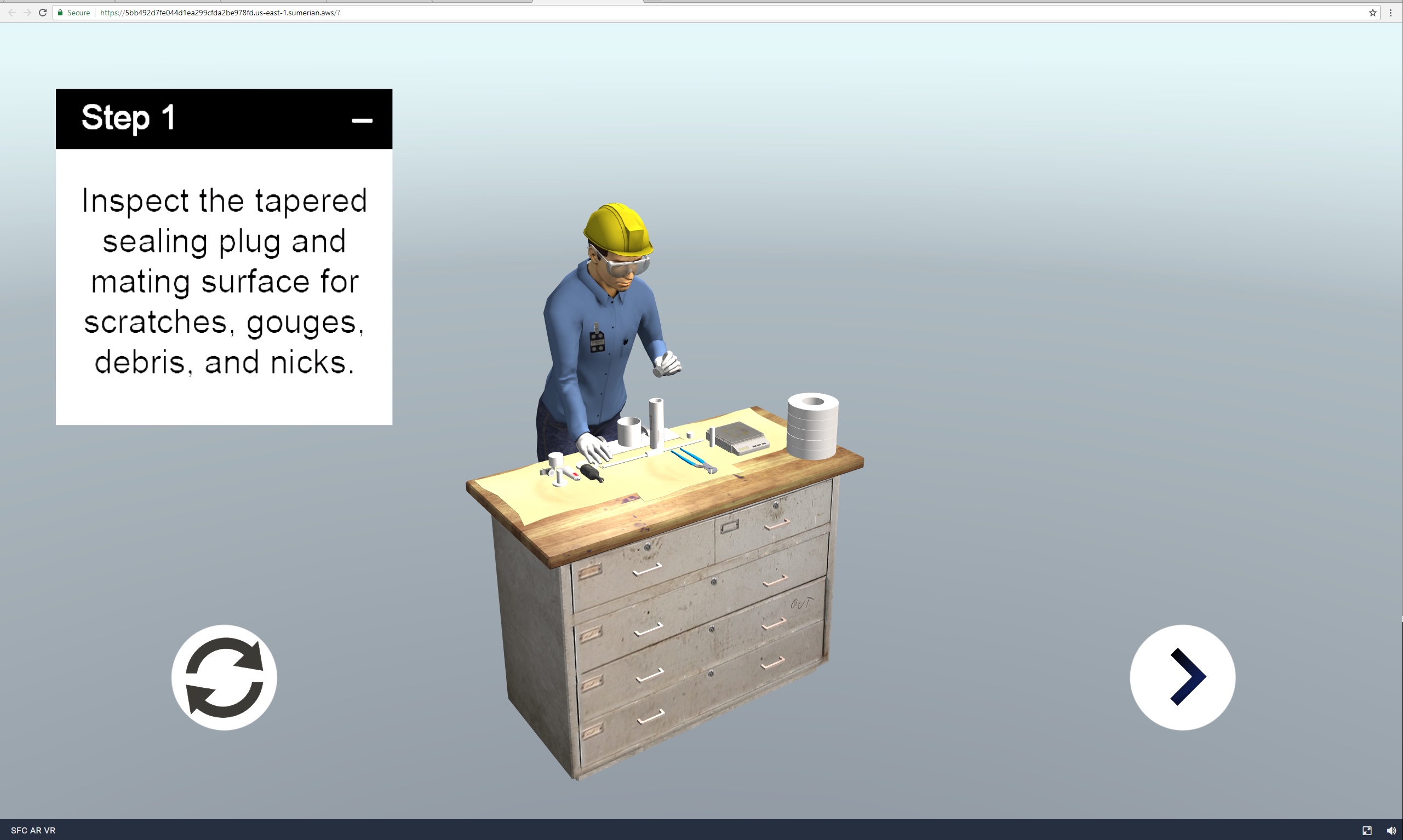}
\caption{VR application in action.}
\label{VRApplicationScreenshot}
\end{center}
\end{figure}

\begin{figure}[htp]
\begin{center}
\includegraphics[width=0.5\textwidth]{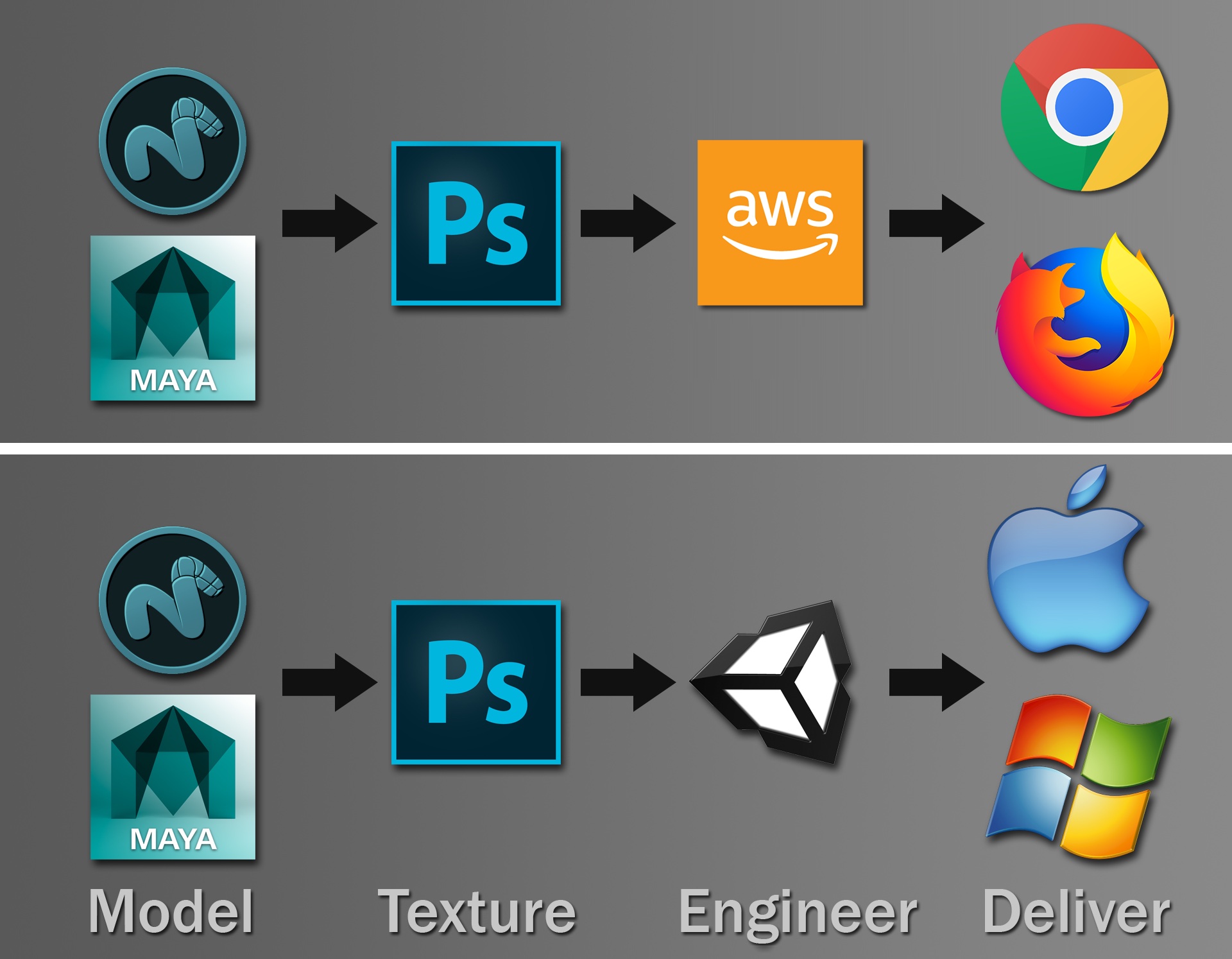}
\caption{New (top) and old (bottom) development pipeline.}
\label{softwarePipelineAWS}
\end{center}
\end{figure}

There were several challenges involved with implementing the solution. Sumerian’s web browser interface would sometimes cause loading errors with FilmBox (FBX) files.  FBX is a cross platform file type that contains 3D geometry data and animations.  One such error involves vertex weight problems, where a 3D model's geometry, improperly, stretches or contorts. Also, Sumerian's FBX load times can be slow with large files, and FBX import progress is uncertain until after the FBX upload completes. Sumerian auto-saving would sometimes interrupt workflow, but this issue was a minor inconvenience. Another challenge with this solution was pipeline adjustments. In a conventional 3D pipeline, if a developer wanted to animate an object with no articulating parts, such as a pencil, the developer would simply animate the object's pivot point. However, in Sumerian, these animations are ignored, unless the object has a mesh deformation associated with it. Manual asset reload and manual adjustment of asset parameters (especially shading materials) after reload were additional challenges. One last challenge was web browser version requirements where only certain versions of Firefox or Chrome support Sumerian and WebVR.

Despite some challenges, there were many successes. Sumerian was compatible with our FBX models and animations to a degree, and it was compatible with shading materials and textures in the FBXs. This compatibility freed us from needing to reapply animation, materials, and textures that already existed in the model. The deployment of the application via URL was also a success. Once set up the text-to-speech functions were simple to implement. The application used 123,344 characters for text-to-speech using Polly. The project and application uses 26 GB of storage space on the cloud, which would normal take up space locally. The cost for using the cloud for this project for one month was \$3.30, which is cost effective. See Figure \ref{VRApplicationScreenshot} for the results of this solution.

There are many similarities between our established approach to developing 3D applications and a cloud-based approach. Both approaches still use assets developed on local machines. These assets still use FBX files with models and animations. However, the cloud-based solution required animations be split out into separate actions, whereas the established method allows the animations to be split out after FBX import. As mentioned before the AWS solution requires all animated objects to have mesh deforms, even if they only simple move in space without deformation. The conventional methods allow an object to move in 3D space via animating the object’s position. See Figure \ref{softwarePipelineAWS} for a visual comparison of workflow differences.

\subsubsection{Conclusion}

AWS’s compatibility with our team's current workflow and assets demonstrates that cloud computing has potential to develop functional applications for our team in the immediate future. Having access to an intuitive text-to-speech component can aid in communicating information during training. Beyond this current solution AWS Lex could also be added to the application and allow users to ask questions during training and get specific answers. Although this project still needed a local machine to run modelling applications and store the original assets, using AWS’s virtual machine and server services could result in a completely cloud-based workflow in the future. Otherwise our team could implement this solution with minimal changes to our already existing pipeline.

\subsection{Cloud-based Scientific Knowledge Retention}

\subsubsection{Contributors}

\begin{itemize}
    \item K. Nolan Carter, NEN-3, knc@lanl.gov
    \item Alexander McQuarters, NEN-3, amcquarters@lanl.gov
\end{itemize}

\subsubsection{Motivation}
A Knowledge Retention System (KRS) is used to capture and make available organizational memory. An organization is a network of inter-subjectively shared meanings that are sustained through the development and use of a common language and everyday social interactions \cite {doi:10.1177/003803858001400219}. As shared knowledge evolves, how does an organization remain constant even after key individuals have departed? In general, an organization may exist independent of particular individuals, but it should be recognized that individuals acquire information in problem-solving and decision-making activities. The thread of coherence that characterizes organizations interpretations is made possible by the sharing of those interpretations. Thus, through this process of sharing, the organizational interpretation system in part transcends the individual level. This is why an organization may preserve knowledge of the past even when key organizational members leave \cite{doi:10.1037/h0030461}. The construct of organizational memory is composed of the structure of its retention facility, the information contained in it, the process of information acquisition and retrieval, and its consequential effects \cite{doi:10.2307/258607}. In an organization such as LANL, this concept extends beyond an individual and refers to the organization's ability to effectively collect, store, and retrieve knowledge and information. 
The primary goals of this project were to: (1) determine the distribution and cost effectiveness of AWS cloud services and create a foundation for a multimedia knowledge retention system; (2) gain an understanding of AWS media services' ability to convert and transcode, store, backup, and package; and (3) determine its capabilities as a delivery mechanism to multiple device types and end users regardless of location.

\subsubsection{Solution Approach}

S3 buckets were created for audio, video, text, and image files, and public ``read" access was established for publication on the web. S3 buckets can be viewed as folders or directories.  Raw data files (primarily video) were uploaded to the previously created buckets. To convert the raw video files we used AWS MediaConvert to create web compatible media files. A similar method was used through AWS ElasticTranscoder for audio files. An additional S3 bucket was created for non-transcoded files (i.e., documents, text, and images). 

\begin{figure}[h]
\begin{center}
\includegraphics[width=0.82\textwidth]{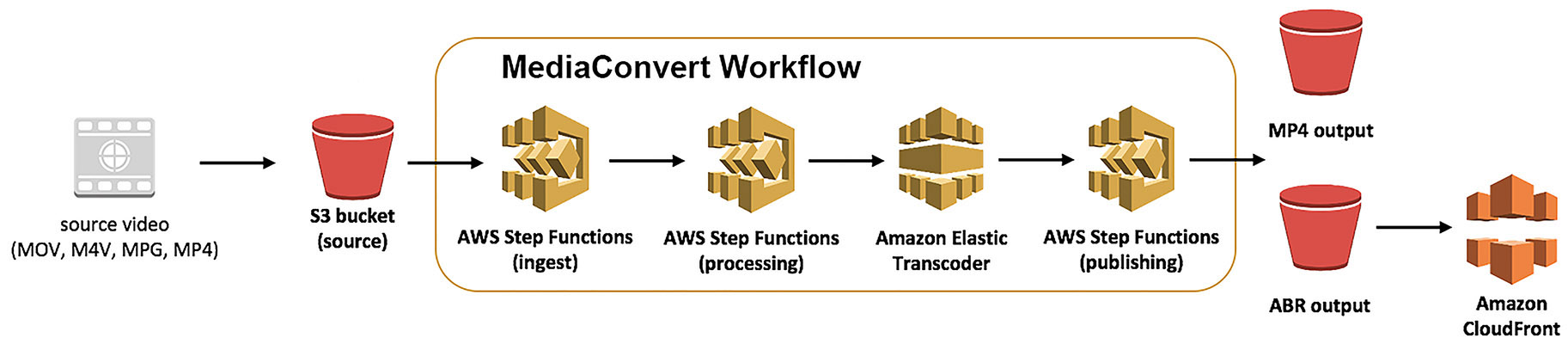}		
\end{center}
\caption{AWS MediaConvert workflow.}
\label{aws-framework}
\end{figure}

An initial expectation of AWS MediaConvert was that it supplied templates for file distribution. This turned out not to be the case thus requiring the creation of an EC2 LightSail instance to serve up a simple interface for testing. As a result, AWS Route 53 was required for creating a test domain and modification of DNS tables.
This simple site, based on Hypertext Markup Language (HTML) and Cascading Style Sheets (CSS), established links to converted files that resided in S3 buckets distributed through CloudFront. Each section was intended to signify a different media type: a ``Video" section to demonstrate video files captured at LANL in multiple media formats; a ``Presentation" section to demonstrate common presentation types seen at LANL; an ``Audio/Spoken" section created to demonstrate audio recordings of talks or audio books; an ``Online Books" section to demonstrate text and pdf delivery; a ``Notes" section to demonstrate captured notes or drawings; and an ``Offsite Video" section to demonstrate the ability to link to other videos located at an off-site location such as a sister laboratory, a field location, or on an external website.     
\begin{figure}[ht]
\begin{center}
\includegraphics[width=0.82\textwidth]{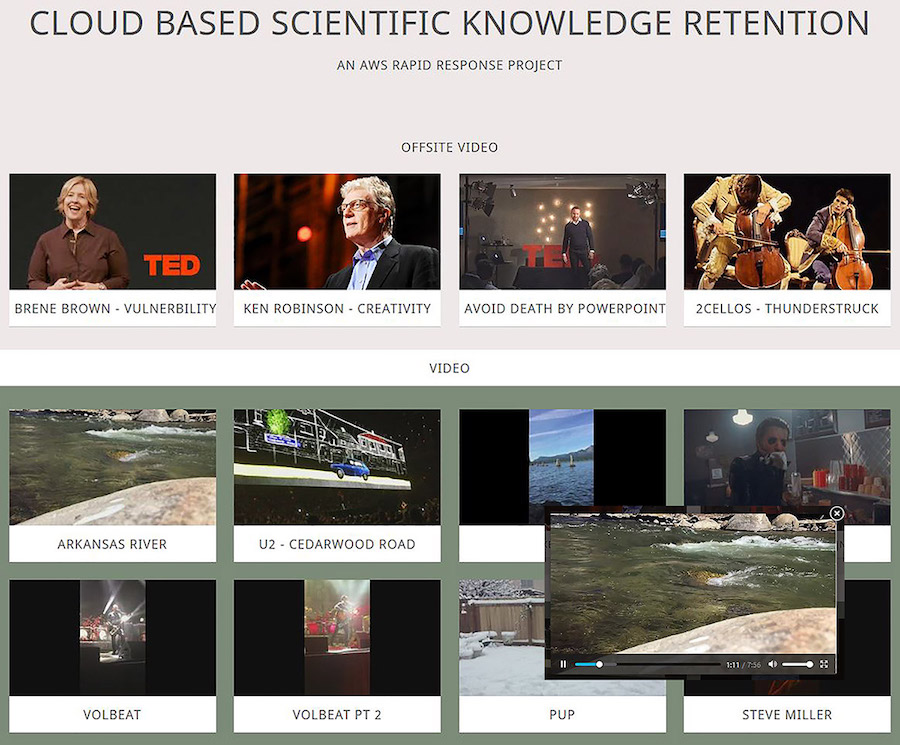}		
\end{center}
\caption{Simplified knowledge retention website. Inset, active video screen. (nentestdomain.com) }
\label{aws-framework}
\end{figure}

The AWS services used in this implementation include, S3 (cloud storage), MediaConvert (video converter), ElasticTranscoder (audio converter), EC2 LightSail (virtual private cloud compute node/server), CloudFront (distribution), and Route 53 (domain and routing). The primary goal was to establish a Netflix-style framework for multi-media file retention and distribution.

\subsubsection{Results}
The AWS interface is unparalleled in cost effectiveness and quick start. The price of this project for two months was just over \$30 USD; a bargain for services that were utilized including just over 14 GB of storage, a virtual machine, media conversion services, and a registered test domain. Due to the fact that MediaConvert did not supply output templates, the project was scaled back to not include a database backend that would feed videos on demand. However we were able to meet the goals desired by this project to determine if AWS could handle multiple file types and store, backup, and distribute these files in a cost effective manner. All-in-all we did find out that not only was AWS affordable but easy to startup, build, and deploy.

\paragraph{Lessons Learned}
The initial expectation that MediaConvert provided delivery templates caused an early setback and required a reduced scope as the project time would not allow time for the creation and implementation of a searchable video framework similar to Netflix\textsuperscript{TM}. With more time this option should definitely be investigated in the future. 
It was also found that, although MediaConvert can accept files from any S3 bucket, ElasticTranscoder requires an established ``pipeline."

\begin{figure}[h]
\begin{center}
\includegraphics[width=0.80\textwidth]{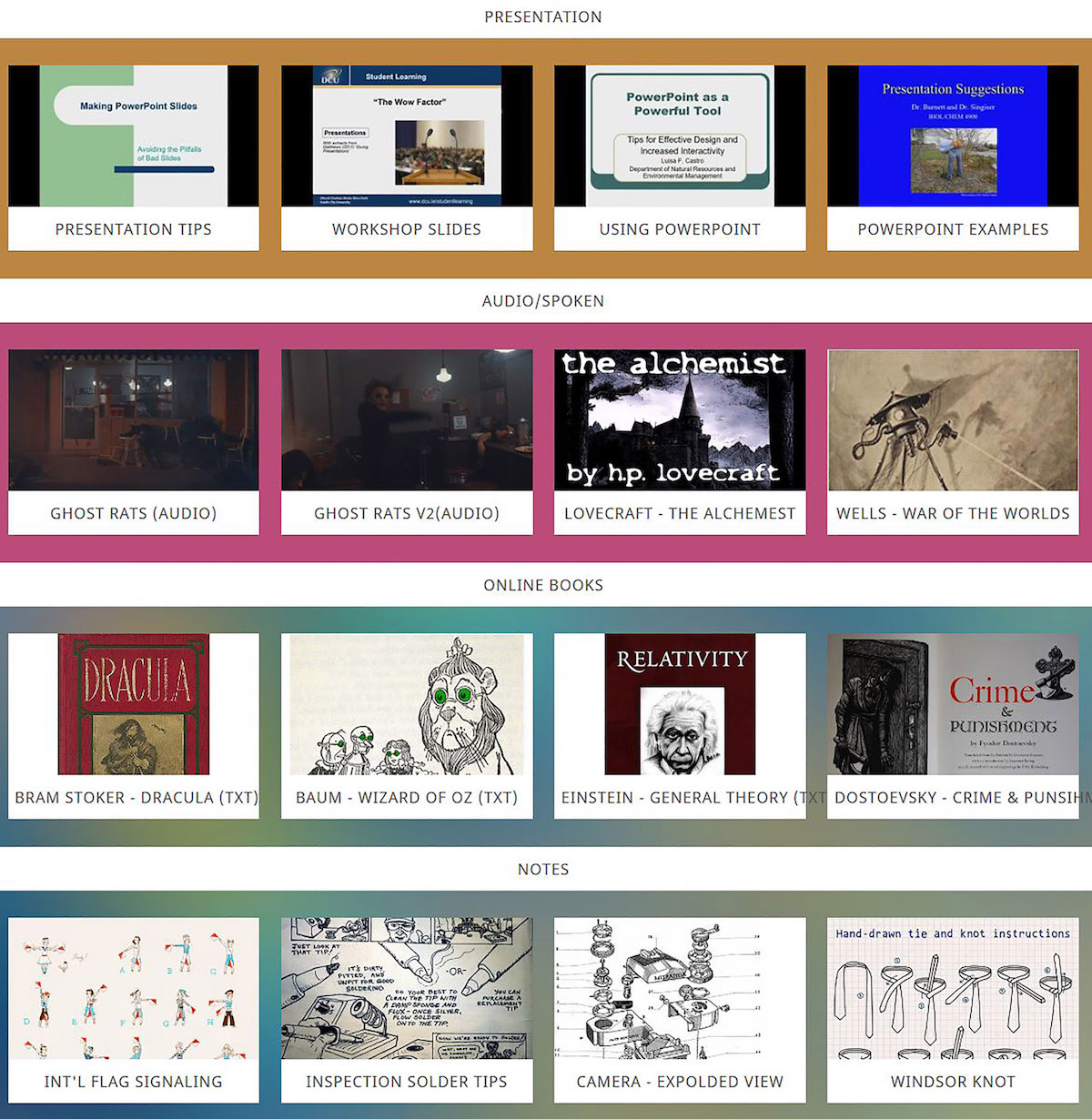}		
\end{center}
\caption{Other media categories. (nentestdomain.com) }
\label{aws-framework}
\end{figure}

\subsubsection{Conclusion}
AWS is certainly a robust and viable solution for storing and creating data storage through cloud computing with abundant features. However, time did not allow the exploration of all features thoroughly. Media Services' documentation was lacking in terms of creating working links between various services. In relation to video conversion, basic presets that can be utilized by users who are not familiar with video quality settings, content delivery networks, etc. Besides the presentation setbacks and insufficient amount of documentation for the use of the services, AWS offers a complete solution for most IT cloud operations. In our case, we were able to store, convert, deliver, and present various types of multimedia using only AWS.

\subsection{SASSY Cloud: Scalable Architectures for Serious Signal Analysis}

\subsubsection{Contributors}

\begin{itemize}
    \item Reid Porter, CCS-3, rporter@lanl.gov
    \item Reid Rivenburgh, CCS-3, reid@lanl.gov
\item Ariane Eberhardt, ISR-2, ariane@lanl.gov 
\end{itemize}

\subsubsection{Motivation}

The purpose of this project was to gain familiarity with AWS services and demonstrate how LANL signal analysis capabilities developed on LANL internal networks can be deployed and demonstrated on AWS. 

LANL has developed a number of signal, image, and video analysis tools over the years that provide unique analysis capabilities. For example, the Los Alamos Maximum Likelihood Pipeline (LAMP) is currently the only quantitative method for detecting EM signatures that are statistically defensible and able to be completely automated. Unfortunately, the technical advantages and the programmatic potential of these capabilities are often lost in the challenge to deploy these tools outside LANL and the specialized communities in which the tools were developed. At the same time, the quantity and types of data collected for LANL applications have rapidly increased and analysis is often limited by the challenges associated with processing large volume data in desktop computing environments.  

The main motivation for the SASSY (Scalable Architectures for Serious Signal Analysis) project was to address the data volume and deployment challenges by integrating and demonstrating LANL signal and image analysis tools with a modern web-centric platform for big data. In addition to increased data throughput and more timely analysis, the SASSY project could enhance the interpretation of multiple datasets and multiple sensors. The primary motivation for the SASSY-Cloud project was to demonstrate that the SASSY solution could be easily deployed to AWS, and thereby demonstrated to a wider community of potential users and customers. 

The LANL signal analysis capabilities that were considered include GENIE (Genetic Image Exploitation), ACD (Anomalous Change Detection), MAMA (Morphological Analysis of Materials), and LAMP. These tools have been previously deployed as Windows desktop applications and were written primarily in Python and C/C++. Interactive user interfaces have also been a key component of most of these tools. 

\subsubsection{Solution Approach}

\begin{figure}[t]
\begin{center}
\includegraphics[width=10.75cm]{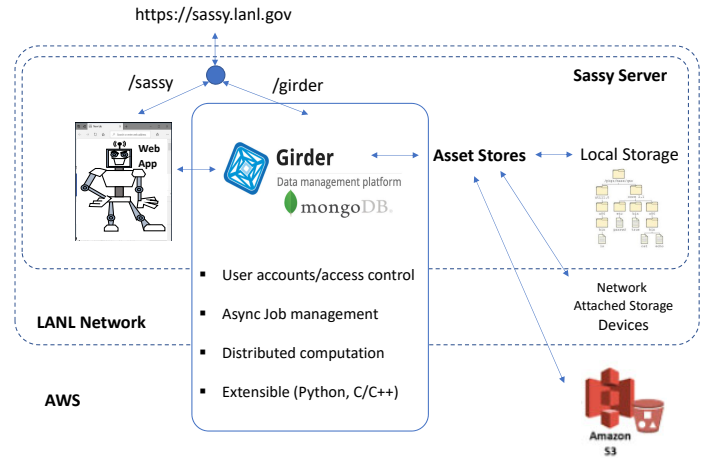}
\end{center}
\caption{The SASSY architecture aims to satisfy our LANL internal network needs today and also provide a path to deployment and demonstration on AWS.}
\label{fig:sassy}
\end{figure}

Figure \ref{fig:sassy} illustrates the main SASSY components. Central to the approach is the open source Girder platform (\url{http://girder.readthedocs.io/en/latest/}).  Girder provides a unified REST API for accessing and organizing a variety of file-centric data collections. Specifically, the Girder Asset Store provides abstraction between the web applications and analysis code and the files themselves. This is useful for many large data applications at LANL where data has been collected over many years and is currently stored in a variety of file systems, formats, and locations often with different levels of access and control. Girder also provides SASSY with a variety of other components required to migrate traditional desktop tools to a server environment including user account management and control. 

Girder makes it straightforward to add and customize Rest API endpoints in Python, and this is the main mechanism we used to wrap the LANL analysis tools---typically C++ tools---for server-side processing. Girder also provides a light weight job management API that enables custom code, such as ours, to report job status and progress in a standard format so that jobs can be monitored through a Girder administrator interface. 

A prerequisite to reliable deployment of LANL analysis tools as Girder endpoints was the integration of the analysis tools into an automated build and continuous integration system. LANL analysis tools are typically developed on internal networks, and these tools have many dependencies that often include a number of specialized and legacy libraries (C++/Fortran). Continuous integration uses standard version control (git), build management (make/cmake), and test and deployment frameworks (Jenkins) to document, optimize, and update these dependencies during development and in deployment. 
With LANL analysis tools wrapped as Girder endpoints and our continuous integration system producing nightly builds and deployments to a LANL network server, deployment to AWS proved to be relatively straightforward. We launched an EC2 instance with a configuration similar to our LANL server, copied the relevant binaries, and launched a mirror of our internal server at very low cost.  

\subsubsection{Results}

The advantages of our straightforward---albeit low-level---AWS deployment approach included simplicity (with little additional knowledge of other tools and services) and complete control and support for the tailored security configuration for the multi-service application as implemented on our internal server.
The disadvantages of our low-level approach are mostly a function of scale and deployment update frequency. These disadvantages will not likely be a factor for our short-term objective to use AWS as a demonstration platform but will become a factor if and when customers request that LANL provide operational services through Docker or AWS. Providing Docker containers for the independent services used in SASSY would be relatively straight forward and enable greater scaling of compute and storage as required. We note that Docker images for Girder are made available through Docker-hub. 
\subsubsection{Conclusion}

AWS provides an incredible variety of services targeting customers with various legacy systems and needs. In our case, the main motivation was to provide a way to deploy and demonstrate legacy analysis codes on AWS as a surrogate for deployment and demonstration on government and secure cloud services. For this use case we found that our upfront effort in continuous integration made it very easy to deploy using the most basic AWS services associated with instance management.

\subsection{Migration from a Public Cloud to a Secure Cloud}

\subsubsection{Contributors}

\begin{itemize}
\item Zachary Baker, CCS-7, zbaker@lanl.gov
\item Jonathan Woodring, CCS-7, woodring@lanl.gov
\end{itemize}

\subsubsection{Motivation}

One of the main software tools for running applications in the cloud is
\textit{Docker} \cite{docker}, a middleware package for the Linux kernel to
Kernel-based Virtual Machine (KVM) \cite{kvm} services. Through it and
\textit{Docker Hub} \cite{dockerhub}, it allows users to quickly deploy
\textit{user-defined software stacks}: self-contained software environments
that only contain a minimal set of software tailored to user-specific needs.
This allows users to install software without requiring \textit{root}
(superuser) access on the cloud hardware nodes via Docker \textit{images}.
Per-user images get around versioning conflicts between user and system
requirements, allowing users to pick specific software for their needs.

Docker has resulted in a large open-source community, allowing users to easily
create custom software environments and deploy them into the cloud.  This ease
comes with the risk of running malicious untrusted code via Docker images
hosted in Docker Hub. Even though software running in a Docker environment does
not have root access on a cloud node, there are plenty of things that a
malicious Docker image could do if left unchecked. This is particularly
sensitive when migrating code and data from unsecure environments (Docker Hub
and unsecure Amazon) to secure (Amazon C2S), where an untrusted image could
attempt to report the results of its computations or try to perform denial-of-service (DoS) attacks on the cloud environment.

In the following, we discuss \textit{spock}, a prototype for a front-end tool
to \textit{Docker} (the command line implementation of Docker), that reduces
the likelihood of running malicious Docker images and increases the trust,
security, reliability, and provenance of Docker application code running in any  cloud environment (not just a Amazon C2S).

\subsubsection{Solution Approach}

\textit{Spock} is a software front-end for \textit{docker}, to securely control
image access and increase code trust within a Dockerized environment, such as a
set of cloud nodes or cluster.  \textit{Spock} is a portmanteau of
\textit{``signing ports for docker.''} In this context, \textit{signing} stands
for cryptographic signing and security.  \textit{Ports} refers to BSD
ports or Gentoo Portage \cite{portage}, a software packaging
paradigm built around the idea of distributing build \textit{recipes} rather
than binary distributions, like Redhat \textit{yum} or Debian \textit{apt}.
At the core \textit{spock} cryptographically signs and verifies build recipes for generating trusted
Docker images.

In more detail, \textit{spock} keeps a record of \textit{Dockerfiles}, the
build instructions for generating a Docker image. These Dockerfiles must be
cryptographically signed and checksummed by a trusted authority, such as an
administrator or peer-user in a trusted user group. The Docker images, which
are built from the set of signed Dockerfiles, are cryptographically signed as
well, which generates a pool of trusted Docker images. With \textit{spock} in
place, users are only allowed to run valid, trusted Docker images.  Additional
capabilities of \textit{spock} include: provenance, trust revocation, and
auditing, which controls the set of valid images and allows administrators
to go back in time and to digital forensics on untrusted images.

\begin{figure}[htb]
\begin{center}
\includegraphics[width=8.75cm]{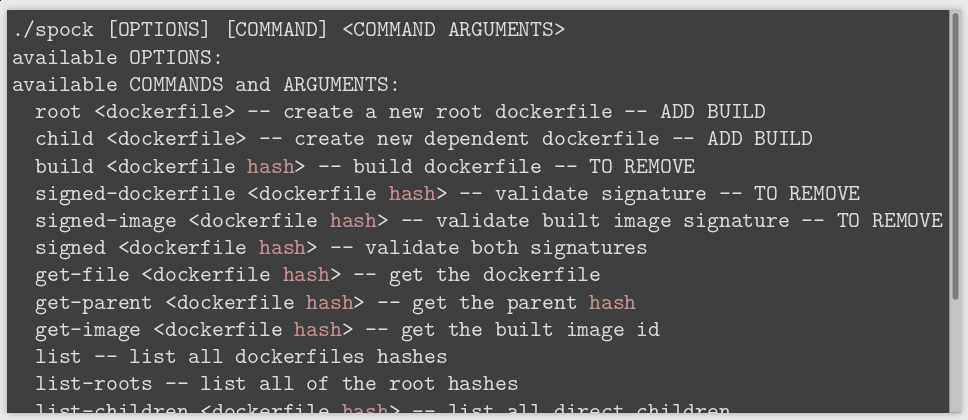}
\end{center}
\caption{Example commands in the \textit{spock} prototype.}
\label{fig:example-commands}
\end{figure}

\textit{Spock} has been prototyped as a \textit{bash} application, allowing for
deployment to nearly all Linux and cloud environments. It utilizes other common
components, found on most major installations, as well: \textit{openssl} for
signing and checksum verification, \textit{sqlite3} for provenance tracking and
data storage, \textit{base64} for data conversion for cryptographic signing and
storage, and \textit{docker,} itself. Figure \ref{fig:example-commands}
shows a sampling of the current state of the \textit{spock} prototype.

\subsubsection{Results}

In the following, we show results from demonstrating our \textit{spock}
prototype in an example workflow, including creating an initial Docker image, creating dependent Docker images, querying the \textit{spock} database, and invalidating an image and removing it and its children from the image pool.

\begin{figure}[htb]
\begin{center}
\includegraphics[width=8.75cm]{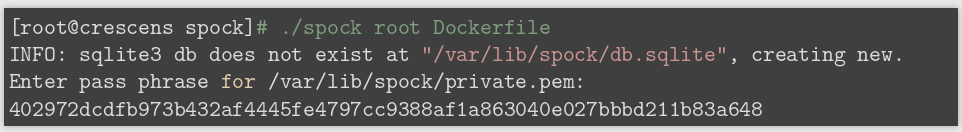}
\end{center}
\caption{Introducing a \textit{root} Dockerfile into the system.}
\label{fig:root-dockerfile}
\end{figure}

\begin{figure}[htb]
\begin{center}
\includegraphics[width=8.75cm]{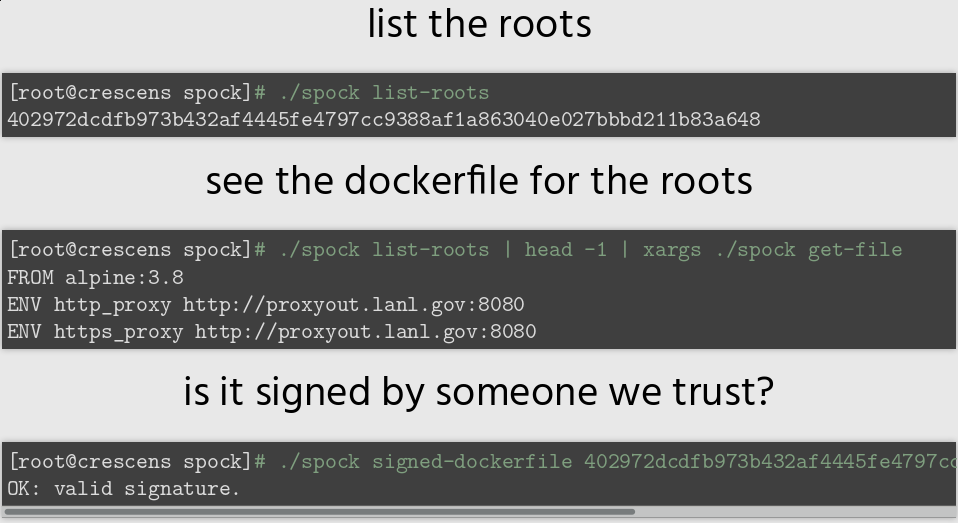}
\end{center}
\caption{Querying different aspects of a \textit{root} Dockerfile.}
\label{fig:root-info}
\end{figure}

\textbf{Creating a root Dockerfile:} A \textit{root} Dockerfile in
\textit{spock} is a build recipe that references an outside Docker image
source (i.e., Docker Hub) that subsequent trusted Docker images are built from.
In Figure \ref{fig:root-dockerfile}, a new Dockerfile (the build recipe for
making a Docker image) is introduced to \textit{spock}. This file is signed by
our private key using \textit{openssl}, identifying ourselves as a trusted user
to the system; \textit{spock} then records this information in its database.  In
Figure~\ref{fig:root-info}, we query and display various information about the
build, such as the contents of the Dockerfile and if it is signed by a trusted
entity. Trusted entities are recorded by \textit{spock} in its data store
through public keys and verifying the signatures of signed data~\cite{signing}.

As seen in both Figures \ref{fig:root-dockerfile} and \ref{fig:root-info}, the
Dockerfile is identified by a cryptographic hash, in this case SHA-256; signing
is performed and verified through \textit{openssl} with an AES-256
public-private key pair (an NIST standard encryption method)\cite{aes256}.
Future versions will incorporate the use of Personal Identity Verification
(PIV) and Entrust, to allow integration into the existing cryptographic
frameworks used at LANL.

\begin{figure}[htb]
\begin{center}
\includegraphics[width=8.75cm]{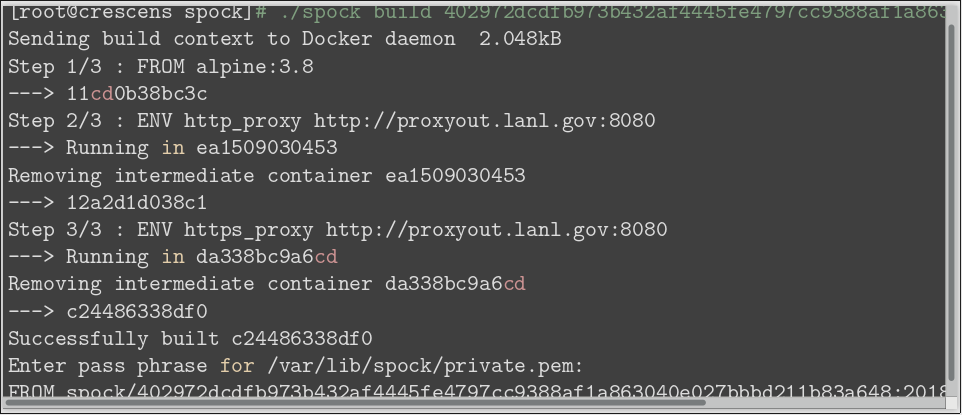}
\end{center}
\caption{Creating a Docker image from a trusted Dockerfile.}
\label{fig:root-build}
\end{figure}

\textbf{Creating a runnable image for Docker:} An image is created through the
\textit{spock build} command, referencing the hash of a signed Dockerfile, as
seen in Figure~\ref{fig:root-build}. The newly created image is uniquely
identified by both the date and time of the image creation and the source
Dockerfile hash (build recipe), which is signed by the creator. The image
identifier is subsequently used by other \textit{child} builds, such that
\textit{spock} only allows running and creating new images from \textbf{both}
signed and valid Dockerfiles and referenced images. This creates provenance,
where future auditing can reveal who created each image, when it was created,
and how it was created, with clear lineage and process.

Furthermore, \textit{spock} will disallow rebuilding images, as that is one of
the vulnerabilities of Docker, in general. Creating an image is a temporal
artifact, reliant on the current state of the network, software repositories,
man-in-the-middle attacks, etc. To control authority and increase reliability,
only one version of an image, resulting from a build recipe, is ever allowed to
exist at any time within the \textit{spock}-ified Docker environment. An image
must be invalidated (removed from the system along with all dependent images)
before it can be rebuilt, thus ensuring a single source of data.

\begin{figure}[htb]
\begin{center}
\includegraphics[width=8.75cm]{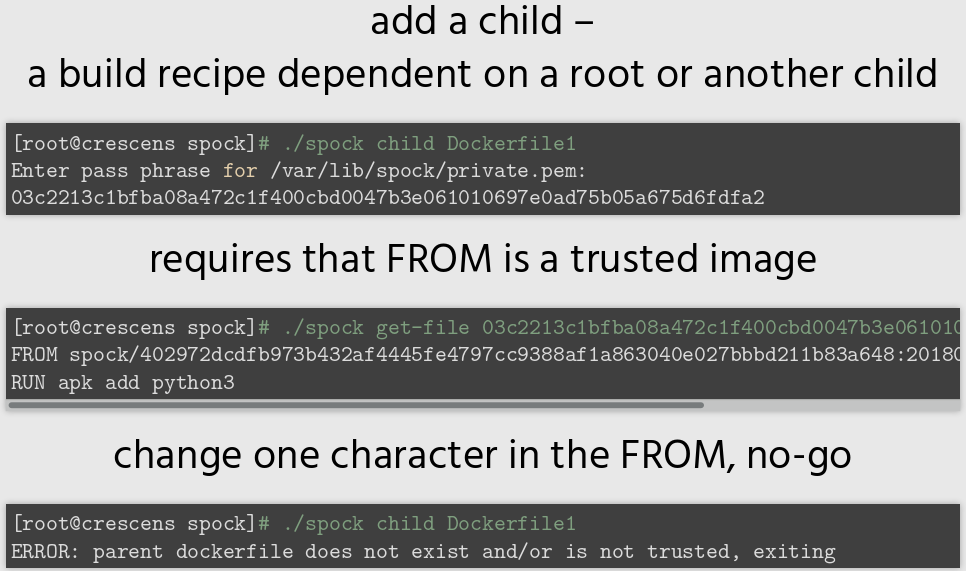}
\end{center}
\caption{Creating a dependent build from an existing build.}
\label{fig:child-build}
\end{figure}

\textbf{Creating a child Docker image:} A \textit{child} Dockerfile in
\textit{spock} is a build recipe that references an already trusted image, such
as a root image or another child tracked by \textit{spock}. It differs from a
root as a child is dependent on an already trusted image, whereas a root is
dependent on a ``foreign'' source such as Docker Hub or an arbitrary tarball.
The reason for this feature is that it isolates a set of images into a pool
that is based on untrusted sources (roots, using data from non-trusted
entities) versus a set of images that have been previously trusted and signed
(i.e., children, using data previously signed by entities in the trusted user group).
This forces more scrutiny and deliberation into creating roots and auditing
them.

\begin{figure}[htb]
\begin{center}
\includegraphics[width=8.75cm]{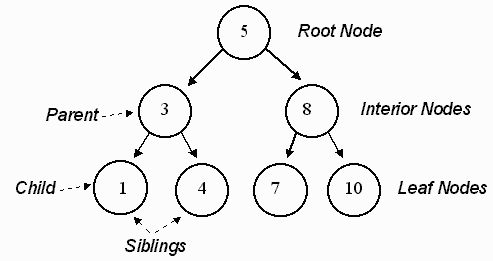}
\end{center}
\caption{A tree of roots and children, where each node is a signed Dockerfile
  and image. The Dockerfile and image at node 5 would reference an outside source
  (FROM), such as from Docker Hub. All other nodes would use FROM on a signed
  source and be descendent builds and images of 5. Thus, we can say the
  lineage of node 1 is ``5 to 3 to 1,'' and investigate the build process used
  to create each image, image contents, and who signed each step.}
\label{fig:tree}
\end{figure}

Figure~\ref{fig:child-build} shows an example of creating a child, where the
command, \textit{spock child}, is the same as the previously shown command,
\textit{spock root}, except that a child Dockerfile requires that the image it
references (i.e., \textit{FROM}) is a trusted image in the \textit{spock}
repository. This means that a child is referencing an image built at a specific
time with a specific Dockerfile, both signed by a trusted entity. The bottom
half of the figure shows \textit{spock} disallowing the creation of a new child
when it does not reference an existing trusted image. The creation of a
trusted image from a child is the same as before, with the \textit{spock build}
command, and it requires signing the newly created image. The combination of
roots and children creates a tree of provenance, like Figure~\ref{fig:tree},
where all children images are dependent in some way on a previous build.

\textbf{Querying \textit{spock}'s database:}

\begin{figure}[htb]
\begin{center}
\includegraphics[width=8.75cm]{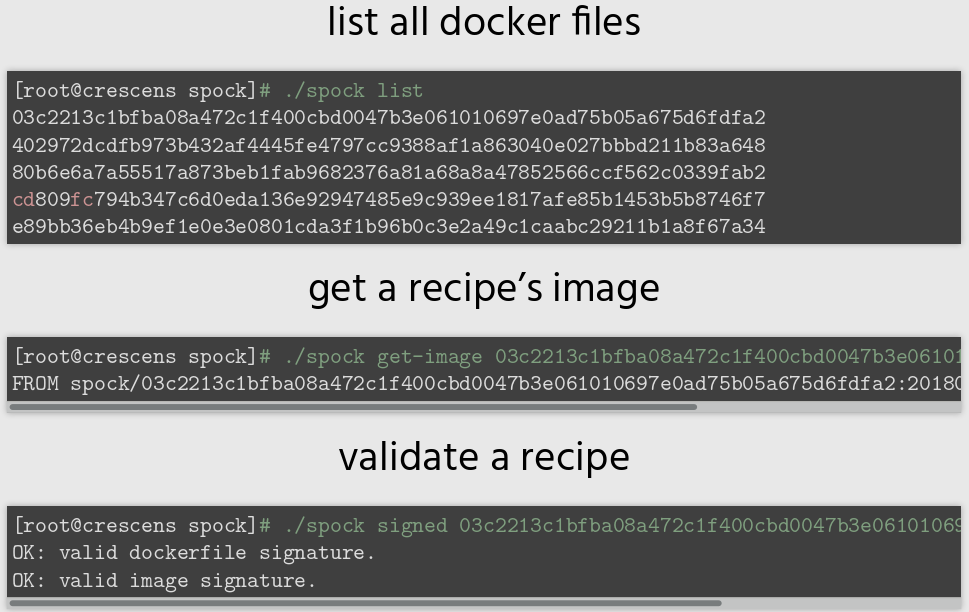}
\end{center}
\caption{Listing, validating, and showing the images of signed Dockerfiles.}
\label{fig:query-1}
\end{figure}

With the information that \textit{spock} stores in its database, it allows
users and administrators multiple ways to query and perform digital forensics
on Dockerfiles and images. In Figure \ref{fig:query-1}, we show examples of
listing various builds, validating that the builds are properly signed, and the
images that were built from each of the Dockerfiles. In future implementations,
\textit{spock} will also contain the ability to store past images that have
been removed from the system, as well as duplicates of files and data that have
copied into a Docker image. This will also allow users and administrators to
do \textit{difference rebuilding} to rebuild images using the established build
processes to find out if there is any temporal variability in images, such as
network and archival differences.

\begin{figure}[htb]
\begin{center}
\includegraphics[width=8.75cm]{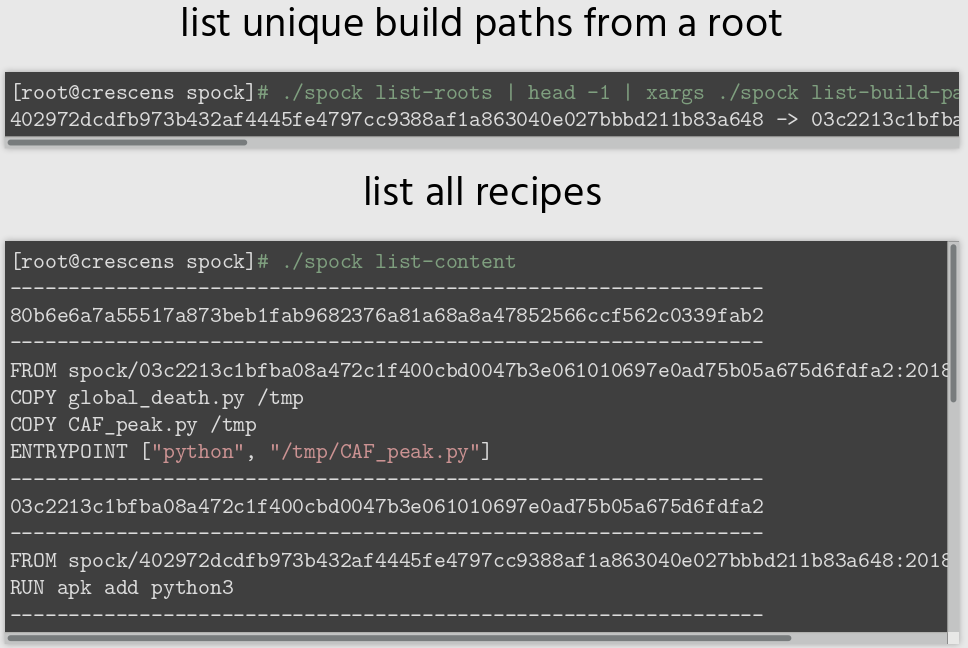}
\end{center}
\caption{Showing dependent builds from a root and listing the content
of all builds. Notice that first one is malicious and will be removed
in Figure~\ref{fig:remove-image}.}
\label{fig:query-2}
\end{figure}

In Figure~\ref{fig:query-2}, we show investigating all of the unique build
paths from a root, as well as listing all of the Dockerfile content for those
paths.  In this scenario, we show a malicious install in one of the Dockerfiles
with the identifying hash of ``80b6e...."

\textbf{Removing a malicious image:}

\begin{figure}[htb]
\begin{center}
\includegraphics[width=8.75cm]{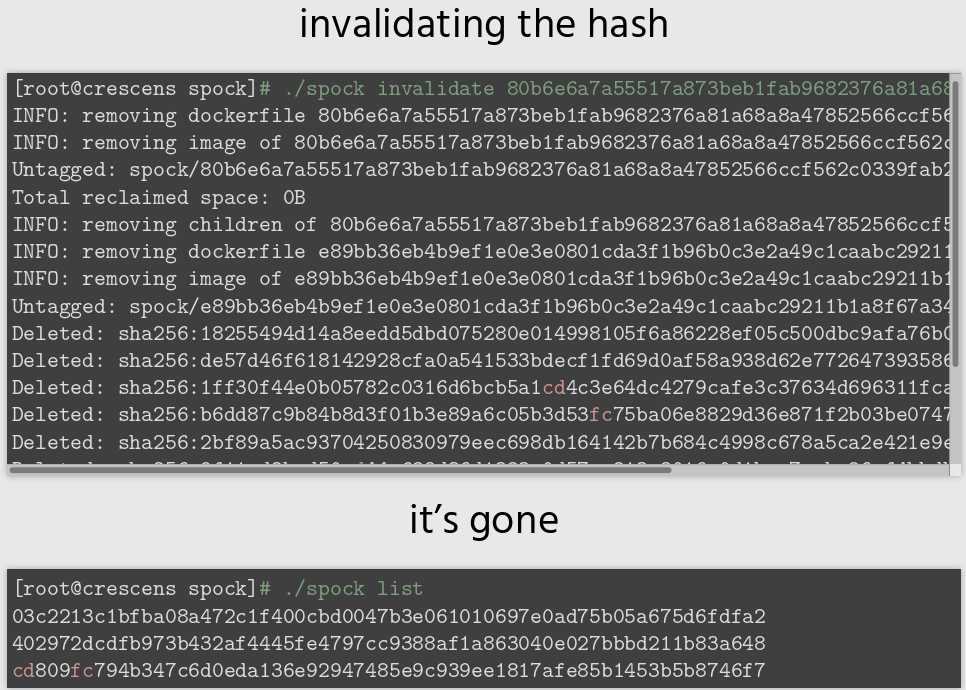}
\end{center}
\caption{Removing a malicious image and automatically removing all
dependent builds and images.}
\label{fig:remove-image}
\end{figure}

In Figure~\ref{fig:remove-image}, we remove the offending Dockerfile, found in
Figure~\ref{fig:query-2}, which will trigger removing the image built by the
Dockerfile, as well as all of the dependent builds and images from that
Dockerfile. As seen in the figure, \textit{spock} will purge the system of the
offending files. Additionally, it creates an archive of those files for later
auditing and inspection.  Also, it will not allow users to build from those
Dockerfiles again or those images because both will be purged from the system.
In future implementations, \textit{spock} will also contain the ability to
distrust a user, and all of their recipes and images will be removed from the
system.

A \textit{spock}-ified Docker environment will run the client on all cloud
nodes, in addition to the development and building nodes. That way, each time a
node attempts to run a Docker image, it will need to verify that it is valid
and signed. This will require additional features added to \textit{spock}, such
as the sharing of trust data, live updates to running containers, and
halting any untrusted content.

\textbf{Differences from Docker's notary:} In a vanilla Docker environment, it
has the capability to introduce image signing and ``only use signed images''
via \textit{notary}~\cite{notary}. Notary introduces identities and signing
keys, such that uploaded images can be identified by signing keys, but it's not
a complete solution. For instance in a public Docker Hub setting, it would be
possible to have and use a malicious image signed by a hacker. For instance,
this was recently the case for Gentoo Linux~\cite{gentoo-hacked}, where the code commits
were signed, but signing does not imply that the signer was a trusted entity.
This can be solved by running an internal notary server and Docker Hub, such
that a limited set of users would have access, which a \textit{spock}
environment would likely use to limit cloud nodes to running signed images.

Though, the largest flaw with notary and Docker, related to trust, is that it
is difficult to perform forensics on existing data, focusing on removing
temporal variance and limiting runnable images to a single source.  Docker and
notary are constructed around image signing, not how images were made.  For
example, \textit{docker history} captures how an image was built, but there
isn't any enforcement of disallowing the rebuilding of an image or running those
same commands again.  Since Docker tags are mutable, in a notary environment,
it would be possible to tag a newly created image with a signed tag, which may
have a completely different lineage and history than the previous image with
the same tag. In the provenance of an image, \textit{docker history} only shows
the linear set of commands that created an image (not its lineage) or a
composed set of commands---such as, if it is the descendant of multiple images,
which may have temporal differences, or who created the build commands
(source code).  \textit{Spock} fixes this by recording the set of build
recipes, like the \textit{ports} system utilized in BSD and Gentoo, in
addition to limiting images to be unique artifacts of build recipes.

Secondly, \textit{spock} supports forensics ``out of the box,'' by recording
all data, not just the build recipes.  For example in a vanilla Docker
environment, removing an image without capturing it does not allow
investigators to determine potential vulnerabilities as the data have been purged
from the system. Thus, \textit{spock} requires that that Dockerfiles are
artifacts that must be captured, in addition to copied data, and past
untrustworthy images and Dockerfiles. This allows investigators to go back and
determine where vulnerabilities may have taken place.  Additionally,
\textit{spock} will have many forensic analysis tools not found in Docker,
such as performing image rebuild for differencing with trusted images. For
instance, Docker can perform differences between a running container and the
base image, but there aren't capabilities to do differences between a trusted
image and a rebuilt image to inspect the temporal differences that may occur.
Furthermore, \textit{spock} has many other concepts and features related to
security and forensics built in, such as differentiating between root builds
versus child builds, which have been previously mentioned.

\subsubsection{Conclusion}

We have developed a prototype tool, \textit{spock}, which is a security and
forensics oriented front-end for \textit{Docker}, an important software tool
for utilizing the cloud. The cloud becomes infinitely more usable with Docker
due to the ease for users to deploy custom software stacks into nodes without
system administrator intervention. Though, that ease comes at a risk of
accidentally deploying malicious code via Docker Hub or other public software
repositories. \textit{Spock} improves trust and reduces the likelihood of
running malicious code by introducing signing of Dockerfile recipes and
creating a provenance of Docker images. This creates the capability to track
source data, along with the users and administrators that created the images
and when they created them: who, when, and how. Also, it minimizes the effects
of temporal variance in Docker images, limiting new images to those that have
already been trusted.

In the future, it is possible that a production version of \textit{spock}
could be developed and deployed into an Amazon Secure C2S environment,
both for the building and running of trustworthy Docker images. 

\clearpage
\section{Acknowledgements}

The organizers of the Information Science \& Technology Institute's ``Exploring Cloud Computing 2018'' would like to thank Terence Joyce and Brady Jones from the Associate Directorate for Business Innovation (ADBI) for their technical support in this effort as well as the Chief Information Officer, Mike Fisk, and Matthew Heavner for their financial support.


LA-UR-18-31581
\end{document}